\documentclass[12pt,a4paper]{article}
\pdfoutput=1
\usepackage[utf8]{inputenc}
\usepackage[T1]{fontenc}
\usepackage{mathrsfs}
\usepackage{tikz}
\usepackage{tikz-feynman}  
\usetikzlibrary{snakes}

\usepackage{soul}
\usepackage[normalem]{ulem}

\usepackage{style}
\usepackage[english]{babel}
\usepackage{url}
\usepackage{footnote}

\usepackage{subcaption}
\usepackage{amsmath}
\usepackage{amssymb}
\usepackage{amsthm}
\usepackage{psfrag}
\usepackage{graphicx}
\usepackage{hyperref}
\usepackage{feynmp}

\newcommand{\be}{\begin{equation}}
\newcommand{\ee}{\end{equation}}
\newcommand{\beqa}{\begin{eqnarray}}
\newcommand{\eeqa}{\end{eqnarray}}

\newcommand\m{\mu}

\newcommand\x{\textbf x}

\newcommand\n{\nu}
\renewcommand\r{\rho}

\renewcommand\l{\lambda}

\newcommand{\SL}{\text{SL}\left(2,\mathbb{R}\right)}

\newcommand{\Lstr}{\text{\sout{$\mathcal{L}$}}}

\def\ba{\begin{aligned}}
\def\ea{\end{aligned}}

\def\e{{\rm e}}
\def\d{\partial}
\newcommand{\bseq}{\begin{subequations}}
\newcommand{\eseq}{\end{subequations}}

\renewcommand{\ln}{\mathop{\rm ln}\nolimits}

\title{Love Symmetry}

\author[a]{Panagiotis Charalambous}
\author[a]{Sergei Dubovsky}
\author[b,a]{Mikhail M. Ivanov}

\affiliation[a]{Center for Cosmology and Particle Physics, Department of Physics,
New York University,\\
New York, NY 10003, USA}
\affiliation[b]{School of Natural Sciences, Institute for Advanced Study,\\1 Einstein Drive, Princeton, NJ 08540, USA}

\emailAdd{pc2560@nyu.edu}
\emailAdd{dubovsky@nyu.edu}
\emailAdd{ivanov@ias.edu}

\abstract{
Perturbations of massless fields in the Kerr-Newman black hole background 
enjoy a (``Love'') SL$(2,\mathbb{R})$ symmetry in the 
suitably defined near zone approximation. 
We present a detailed study of this symmetry 
and show how the intricate behavior
of black hole responses
in four and higher dimensions 
can be understood from the SL$(2,\mathbb{R})$
representation theory.
In particular, 
static perturbations of four-dimensional 
black holes belong to 
highest weight $\SL$ representations. It is this highest weight properety that forces the static 
Love numbers to vanish.
We find that the Love symmetry is tightly connected to the 
enhanced isometries of extremal black holes.
This relation is simplest for extremal 
charged spherically symmetric
(Reissner-Nordstr{\"o}m) solutions, where
the Love symmetry exactly
reduces to the isometry of the near 
horizon AdS$_2$ throat.
For rotating (Kerr-Newman) black holes one is lead to consider an infinite-dimensional $\SL\ltimes \hat U(1)_{\mathcal{V}}
 $
extension of 
the Love symmetry.
It contains  three
physically distinct 
subalgebras: 
the Love algebra,
the Starobinsky near zone algebra,  
and the near horizon algebra 
that becomes the Bardeen-Horowitz 
isometry in the 
extremal limit.
We also discuss other aspects of the Love symmetry,
such as the geometric meaning of its generators for 
spin weighted fields, connection to the no-hair theorems,
non-renormalization 
of Love numbers, its relation to  
 (non-extremal) Kerr/CFT 
correspondence and prospects of its existence in modified theories of gravity.
}

\begin{document}


\maketitle
\flushbottom

\section{Introduction}
\label{sec:int}

The detection of gravitations waves from compact binary mergers
with the LIGO/Virgo
interferometer has opened up a new era
of gravitational wave astronomy~\cite{LIGOScientific:2016aoc}.
One of the goals of this newly emerged science is
to probe the internal dynamics of individual compact objects in the binary. 
This can be achieved by measuring effects of tidal deformations on the emitted gravitational waveforms.
In the low frequency limit these effects are parametrized by (static) tidal Love numbers \cite{Flanagan2008, Binnington:2009bb}.
For neutron stars these measurements serve as a powerful probe of the nuclear matter equation of state under extreme conditions~\cite{Flanagan2008,Yagi:2013bca,Chatziioannou2020}.
The neutron star--neutron star merger 
GW170817 has already 
provided first bounds 
on the neutron star 
Love numbers and on the corresponding equation of state~\cite{LIGOScientific:2017vwq}.

For black holes the role of Love numbers
is somewhat different -- they provide us with
a null test and a possibility to search for exotic signatures. 
This is because
Love numbers of black holes vanish in four dimensional 
general relativity~\cite{Fang2005,Binnington:2009bb}.
A detection of a
compact object with a mass greater
than the Oppenheimer–Volkoff limit $\sim 3~M_{\odot}$~\cite{PhysRev.55.374}
and a non-zero Love number
would be a clear signature of physics beyond
general relativity~\cite{Porto:2016zng}.

Gravitational wave data analysis  
requires efficient tools to produce 
waveform templates.
One of such tools is
the worldline  
effective field theory (EFT), which provides 
a systematic framework 
to compute dynamics of inspiraling binaries
in the post-Newtonian regime~\cite{Goldberger:2004jt,Goldberger:2005cd,Goldberger:2006bd,Porto:2005ac,Rothstein2014,Porto:2016pyg,Levi:2018nxp}.
In this framework each compact member of 
a binary
is approximated as a point particle
and all finite-size effects are captured 
by a set of higher-derivative operators
coupled to the particle's worldline.
This way the EFT describes  
dynamics of a generic compact body
on distances greater than its size. 
In this picture the full general relativity 
plays a role of the UV theory, which is valid 
for scales shorter than the compact object radius.
Determination of the EFT coupling constants (also called Wilson coefficients)
proceeds via matching between the EFT and full general
relativity calculations in the overlapping distance domain.
This matching provides also a manifestly gauge invariant definition of Love numbers. Namely, these are identified 
with the leading EFT finite size 
operators~\cite{Kol:2011vg,Hui:2020xxx,Charalambous:2021mea,Ivanov:2022hlo}.

Following 't Hooft~\cite{tHooft:1979rat},
in the absence of a symmetry in the UV theory, one expects
the Wilson 
coefficients to be order-one numbers
times an appropriate power of the EFT cutoff scale.
Hence, the vanishing of black hole Love numbers 
in four dimensions gives rise to an intriguing naturalness puzzle~\cite{Porto:2016zng}.

In four dimensions, static Love numbers are zero for 
Schwarzschild and Kerr black holes alike~\cite{Chia:2020yla,Charalambous:2021mea,Goldberger2020c,LeTiec2020,LeTiec2020a,Hui:2020xxx,Poisson:2020mdi}.
However, the vanishing of black hole tidal Love numbers in 
four dimensions is not the only peculiar property
that needs to be explained.
One can generalize the notion of black hole 
tidal Love numbers 
to scalar and electromagnetic fields 
and show that the static responses to
their perturbations vanish as well~\cite{Kol:2011vg,Hui:2020xxx,Charalambous:2021mea}.

%

The situation is more intricate
for higher dimensional black holes. 
In the case of higher dimensional 
Schwarzschild solutions,
Love numbers' behavior is parametrized by 
\[
\hat \ell\equiv \ell/(d-3)\;,
\]
where $\ell$ is the multipolar order
and $d$ is the number of spacetime dimensions.
If $\hat \ell$ is an integer, Love numbers vanish for all 
static perturbing fields~\cite{Kol:2011vg,Hui:2020xxx}.
If $\hat \ell$ is a half-integer, Love numbers
exhibit a classical logarithmic Renormalization Group (RG) running.
In all other cases Love numbers do 
not vanish and do not run.

%
%
%

These results suggest  the presence of 
a new symmetry of black holes, whose
selection rules force Love numbers to vanish.
This symmetry, 
which we call ``Love symmetry'',
has been recently presented 
in Ref.~\cite{Charalambous:2021kcz}.
In this paper we study this symmetry in more detail. In particular, we relate it to the enhanced near horizon isometries of extremal black holes, which play a crucial role in the holographic correspondence.
We will see that the vanishing of Love numbers always comes about as a result of a special algebraic property of the corresponding representations of the Love symmetry algebra. Namely, Love numbers vanish whenever the 
corresponding representation exhibits the highest weight property.

The rest of the paper is organized in the following way. In Section~\ref{sec:prel}
we define Love numbers and introduce
some basics of black hole perturbation theory,
such as the Teukolsky equation and 
the near zone approximation.
All important facts about Love numbers
are presented in Section~\ref{sec:facts}.
This section also contains 
Love number calculations 
in higher derivative gravity, illustrating that the logarithmic running is indeed a generic property.

In Section~\ref{sec:love}
we give a detailed 
description of the Love symmetry.
Love symmetry is an approximate $\SL$ symmetry that 
acts on black hole perturbations in a properly
defined near zone.
The Love symmetry is globally well defined 
and hence regular solutions of the 
linearized near zone black hole perturbation equation
(the so called Teukolsky equation~\cite{Teukolsky1972,Teukolsky1973})
form $\SL$ representations.
This is a powerful statement as now
many properties of 
black hole perturbations 
can be extracted from 
the group theory without having to 
solve any differential equations explicitly.
In particular, in four dimensions
the static solutions of the near zone Teukolsky
equation form highest weight $\SL$ representations,
which dictates their polynomial form and hence 
the vanishing of the Love numbers.
Static perturbations of higher dimensional black holes
also belong to highest weight representations 
for integer $\hat \ell$, hence the Love numbers vanish in this case.

There are indications 
that the actual symmetry structure 
of the black hole perturbations 
is greater than the Love symmetry. 
In Section~\ref{sec:gen} we unveil a part
of the full symmetry structure.
Namely, we explicitly construct an extension
of the Love symmetry into an 
$\SL\ltimes \hat U\left(1\right)_{\mathcal{V}}$ algebra,
which captures two 
additional
physically relevant regimes:
the Starobinsky
near zone 
approximation~\cite{Starobinski1973,Starobinski1974},
and the extremal near horizon black hole 
geometry~\cite{Bardeen:1999px,Amsel:2009et}. 
The Starobinsky near horizon algebra addresses properties
of black hole responses 
around the locking frequency.
Another subalgebra of this $\SL\ltimes \hat U\left(1\right)_{\mathcal{V}}$ reduces to the isometry of the 
near horizon black hole AdS$_2$ throat
in the extremal limit. We also present a generalization that allows for a group theory
description of the frequency-dependent 
response solutions in the Starobinsky near zone
approximation.


We give further details of the relationship between the near horizon
symmetries and the Love symmetry in Sec.~\ref{SecNHE}.
We show that the Love symmetry and the extremal near horizon isometries appear to be closely related.
In particular, in the spherically symmetric charged
black hole case, the Love symmetry exactly reduces 
to the $\SL$ isometry of the AdS$_2$ near horizon black hole 
geometry. This gives a clear geometrical interpretation
of the Love symmetry in the Reissner-Nordstr{\"o}m
black hole case as an approximate spacetime isometry.


In the case of rotating Kerr-Newman black holes
the situation is more intricate.
As we discussed before, the Love symmetry is a part of the larger symmetry group
$\SL\ltimes \hat U\left(1\right)_{\mathcal{V}}$.
One particular subalgebra of this group 
reduces to the Kerr-Newman near horizon $\SL$
isometry~\cite{Amsel:2009et} in the properly defined extremal limit.

All in all, the relationship to the extremal near horizon 
$\SL$ isometries suggests a proper 
interpretation of the Love symmetry generators
as approximate Killing vectors of the
Kerr-Newman black hole
geometry. This can be made more precise in the context of subtracted geometries~\cite{Cvetic:2011hp,Cvetic:2011dn}. Subtracted geometries are effective geometries characterized by their property of approximating the environment of the black hole while preserving its internal structure. In this language, the near zone symmetries are realized as isometries of particular subtracted geometries which have the extra feature of preserving properties that extend beyond the vicinity of the event horizon, namely, the location of the ergosphere of the black hole.

This interpretation can be extended to 
Newman-Penrose spin weighted fields that 
encode electromagnetic and gravitational
perturbations.
In Section~\ref{sec:spins} we show that the 
generators of the Love symmetry for 
spin weighted fields have a meaning of the
approximate Geroch-Held-Penrose (GHP) Lie derivatives~\cite{Edgar2000,Ludwig2000,Ludwig2002}
along the (approximate) Killing vectors 
of the Love symmetry.




In Section~\ref{sec:furt} we present 
further possible generalizations of the Love symmetry: 
a new~``middle zone'' symmetry that appears 
in the extremal black hole case
and a continuous family of approximate $\SL$ symmetries
generalizing  the Love symmetry. 
Finally, we discuss requirements
for the Love symmetry to exist 
in modified gravity theories. 

We draw conclusions in Section~\ref{sec:disc}
and sketch future research directions 
in Section~\ref{sec:future}.
Some additional material 
is collected in several appendices.

\textit{Conventions}: We will be working in geometrical units with the speed of light in the vacuum and Newton's gravitational constant set to unity, $c=1$ and $G=1$, except in the beginning of the next section where $G$ will be kept explicit for clarity, and we will adopt the mostly positive Lorentzian metric signature. Small Greek letters will denote spacetime indices running from $0$ up to $d-1$ with $x^{0}$ indicating the time coordinate, while small Latin indices will denote spatial indices running from $1$ up to $d-1$ for a $d$-dimensional spacetime. In addition, we will be using the multi-index notation $\mu_1 \mu_2\dots \mu_{\ell}\equiv L$ with $x^{\mu_1}x^{\mu_2}\dots x^{\mu_{\ell}}\equiv x^{L}$ and $\partial_{\mu_1}\partial_{\mu_2}\dots\partial_{\mu_{\ell}}\equiv\partial_{L}$.
The brackets $\langle ...\rangle$
denote the operation of symmetrization
and removal of the traces. Hence, any tensor 
with the multi-index $\langle L\rangle$
is a symmetric trace-free (STF) tensor of rank $\ell$.

\section{Preliminaries}
\label{sec:prel}

In this section we review some 
material required for the understanding 
of the Love symmetry.
We discuss the definition of Love numbers in 
the Newtonian approximation and EFT, 
and introduce the Teukolsky master equation,
which allows one to extract black hole Love numbers
from perturbative calculations. We will briefly discuss the holographic interpretation of the Love numbers.
We will also introduce 
the near zone approximation.
In addition, we will carry out a calculation of the scalar Love numbers 
in the Riemann-cubed gravity
and show that  
they exhibit a classical RG running 
in the multipolar sectors 
with $\ell>2$.
Finally, we will 
discuss the role 
of black hole 
no-hair 
theorems.

\subsection{Newtonian definition of Love numbers}

Tidal Love numbers were originally defined more than a century ago~\cite{Love1912} as response coefficients of a spherically symmetric body under external tidal forces~\cite{PoissonWill2014}. Consider a spherically symmetric 
body of mass $M$ placed at the origin 
of a Cartesian coordinate system. 
An arbitrary external Newtonian potential
perturbing this body
can be written in terms of its 
STF
tidal moments $\mathcal{E}_{L}$,
\be
	U_{\rm ext}\left(t,\x\right)=-\sum_{\ell = 2}^\infty \frac{(\ell-2)!}{\ell!}\mathcal{E}_{L}\left(t\right)x^L\,, \quad \mathcal{E}_{L}\left(t\right)\equiv - \frac{1}{\left(\ell-2\right)!}\d_{\langle L\rangle} U_{\rm ext}(t,\x)\Big|_{\x=0}\,.
\ee 
The sum over $\ell$ starts from $\ell=2$ because 
the linear $\ell=1$ term affects only the overall center of mass motion of the body.
The tidal force perturbs the body and induces 
mass multipole moments, 
\be
I^{L}\left(t\right)=\int d^3x~\delta\rho(t,\x) x^{\langle L \rangle} \,,
\ee
where $\delta\rho$ is the body's mass density perturbation. 
Then the total potential 
takes the following form ($n^i \equiv x^i/|\x|$),
\be 
U_{\rm tot}\left(t,\x\right)= \frac{G M}{r}-
\sum_{\ell = 2}^\infty 
\frac{(\ell-2)!}{\ell!}\left[
\mathcal{E}_{L}\left(t\right) x^L-\frac{(2\ell-1)!!}{(\ell-2)!}\frac{ G I_L\left(t\right) n^L}{r^{2\ell+1}}
\right]\,.
\ee
It is convenient to switch to spherical coordinates
and
work in the harmonic basis,  
\be
\mathcal{E}_{\ell m}\equiv \mathcal{E}_{L}\int_{\mathbb{S}^2} d^2x~n^L Y^*_{\ell m}\,,\quad 
I_{\ell m}\equiv I_{L}\int_{\mathbb{S}^2} d^2x~n^L Y^*_{\ell m}
\,,
\ee
where $d^2x\equiv \sin\theta d\theta d\phi$ and $Y_{\ell m}$ are spherical harmonics on the $2$-sphere.
Assuming that an external perturbation is 
weak, 
one can use the linear response theory to write down
the following relationship between 
the induced mass multipole moment
and the applied tidal moment in the frequency space\footnote{For a spherically symmetric body, the response coefficients do not depend on the azimuthal number $m$. However, we are including here the case of an axisymmetric body with background symmetries that do not allow mixing between different $\ell$-modes either. This is relevant, for instance, for the Kerr-Newman black holes due to their separability.},
\be
\label{eq:respdef}
G I_{\ell m}\left(\omega\right)
=-\frac{\left(\ell-2\right)!}{(2\ell-1)!!}k_{\ell m}(\omega) R^{2\ell+1} \mathcal{E}_{\ell m}\left(\omega\right)
\,,
\ee
where $R$ is a characteristic length scale of the body's size, e.g. the unperturbed body's radius,
and $k_{\ell m}$ is the dimensionless
response
coefficient. 
Analyticity in the frequency space allows us to write 
the following Taylor expansion
valid for small
frequencies $\omega'$ in the body's rest frame,
\be
\label{eq:klm}
	k_{\ell m}  = \kappa_{\ell m} + i\nu_{\ell m}\omega' + \dots = \kappa_{\ell m} + i\nu_{\ell m}\left(\omega - m \Omega\right) + \dots\,,
\ee
where $\omega$ is the perturbation's frequency 
in the external inertial frame,
$m$ is the azimuthal harmonic number and $\Omega$ is the body's angular velocity.
The appearance of the ``superradiant factor'' $\left(\omega - m \Omega\right)$
is merely a consequence of frame dragging;
it is a purely kinematic effect~\cite{Chia:2020yla,Charalambous:2021mea}.
The real contribution in Eq.~\eqref{eq:klm}
captures the conservative response 
and the corresponding response coefficient $\kappa_{\ell m}$ is called the Love number.
As far as applications to binary inspirals go,
the resulting gravitational waveform is sensitive to the Love numbers starting at the 5PN order
in the Post-Newtonian expansion (for $\ell=2$).

The imaginary contribution $i\nu_{\ell m}$
is purely dissipative, as can be seen from the fact that it is 
odd under the time reversal. For binary inspirals this dissipative response coefficient
captures the tidal heating and the spin exchange 
between a binary's components;
it starts contributing at the 2.5PN order for $\ell=2$.

The distinction between $\kappa_{\ell m}$ and $\nu_{\ell m}$ 
has a simple physical analogy in electromagnetism.
There, the imaginary part of the electric susceptibility gives rise to 
dissipation/amplification of the intensity of propagating plane waves inside the material, while the real part captures conservative effects such as refraction.

\subsection{General relativity and EFT definitions}
\label{sec:GREFT}

In the full general relativity the Love numbers 
can be defined through the
Weyl curvature scalar $\psi_0$~\cite{Newman1962,Teukolsky1972},
\be
	\psi_0 = C_{\mu\n\l\r} l^\m m^\n l^\l m^\r\,,
\ee
where $C_{\mu\n\l\r}$ is the Weyl tensor 
and $l^\m,~m^\n$ are Newman-Penrose null tetrades~\cite{Newman1962,Geroch1973}. The advantage of working
with the Weyl scalar will become clear shortly,
when we discuss the Teukolsky master equation. To define the Love numbers of an isolated gravitating object, one considers equations for linear perturbations around the object and looks for solutions which are regular everywhere, but are allowed to grow at the spatial infinity. Expanding the resulting $\psi_0$ at large distances one finds ~\cite{LeTiec2020a},
\be
\label{eq:psi0as}
	\begin{split}
		\psi_0\Big|_{r\to\infty}=\sum_{\ell=2}\sum_{m=-\ell}^\ell \sqrt{\frac{\left(\ell+2\right)\left(\ell+1\right)}{\ell\left(\ell-1\right)}} r^{\ell-2} \left( \mathcal{E}_{\ell m} \left( \psi_{s,\ell m}+k_{\ell m}^{\cal E}\psi_{r,\ell m} \right) + \dots
		\right)
%
%
		{}_{2}Y_{\ell m}(\theta,\phi)\,,
	\end{split}
\ee
%
%
%
where 
$\mathcal{E}_{\ell m}$
are asymptotic spherical harmonic coefficients of the electric tidal tensor, 
defined as
\be 
	\mathcal{E}_L =\frac{1}{\left(\ell-2\right)!} \nabla_{\langle i_3...i_\ell} C_{0|i_1|0|i_2\rangle}\,.
\ee
Dots in \eqref{eq:psi0as} stand for an analogous (imaginary) contribution associated with the magnetic tidal tensor, which we do not write down explicitly for brevity.
Functions $\psi_{s,\ell m}\left(r\right)$ and $\psi_{r,\ell m}\left(r\right)$ describe growing and decaying solutions of the equations for metric perturbations,
\begin{gather}
	\psi_{s,\ell m}\left(r\right)=1+{\cal O}\left({1\over r}\right)\;,\\
	\psi_{r,\ell m}\left(r\right)=\left(\frac{R}{r}\right)^{2\ell+1}\left(1+{\cal O}\left({1\over r}\right)\right)\;.
\end{gather}
The former is interpreted as the contribution of a tidal source located at the spatial infinity, and the latter as the tidal response. This way coefficients $k^{{\cal E}}$, and their magnetic analogues 
$k^{{\cal B}}$,
provide a relativistic generalization of Love numbers. In the Newtonian gauge, the electric Love numbers defined this way match the Newtonian definition.

However, one may wonder whether this definition is unambiguous under the change of coordinates (such as the switch between the advanced and the Boyer--Lindquist frames for the Kerr metric).
Also, given that the source function $\psi_{s,\ell m}\left(r\right)$ is in general an infinite series in the powers of $1/r$, it is unclear whether the source/response separation in \eqref{eq:psi0as} is well-defined.




These questions are nicely resolved by defining Love numbers in the framework of 
the gravitational point-particle effective field 
theory (EFT)~\cite{Goldberger:2004jt,Porto:2005ac,Porto:2016pyg,Levi:2018nxp}. 
One constructs the EFT by integrating out the body's 
short scale 
degrees of freedom and parameterizing their 
impact on the long distance dynamics through 
effective operators for long wavelength degrees of freedom. The latter include 
the long distance metric perturbation $h_{\mu\nu}$
defined w.r.t. the Minkowski background 
\[
g_{\mu\nu} = \eta_{\mu\nu} + h_{\mu\nu}
\]
and
the 
center of mass position $x_{\rm cm}^\mu$.\footnote{If the body is spinning, one also needs to include rotating degrees of freedom parametrized by a local tetrad on a worldline~\cite{Porto:2005ac,Porto:2016pyg,Levi:2015msa,Levi:2018nxp}.}

This description is valid on length scales longer than the size of the object $R$. For black holes, which is the main object of interest in this paper, $R$ is set by the gravitational 
 radius 
 \[
 R\sim
 r_{s}=2GM\;.
 \]
The point-particle effective action is obtained by 
expanding the bulk action for gravity and
writing down all possible effective worldline 
operators consistent 
with the symmetries of the problem~\cite{Goldberger:2004jt},
\be
	\begin{gathered}
		S_{\text{EFT}}\left[h_{\m\n},x^\m_{\rm cm}\right] = 
		S_{\text{EH}}\left[\eta_{\m\n}+h_{\m\n}\right] 
		+ S_{\text{pp}}\left[h_{\m\n},x_{\rm cm}\right] + S_{\text{finite-size}}\left[h_{\m\n},x^\m_{\rm cm}\right]
	\end{gathered}
\ee
where $S_{\text{EH}}$ is the Einstein-Hilbert action
and $S_{\text{pp}}$ is the usual point-particle
action,
\be
	S_{\text{pp}}
 	= -M\int d\tau \left(g_{\mu \nu}\frac{dx^\mu_{\rm cm}}{d\tau}\frac{dx^\nu_{\rm cm}}{d\tau}\right)^{1/2}\,.
\ee
The action $S_{\text{finite-size}}$ contains higher-derivative corrections 
on the worldline. In the simplest case 
of a spherically symmetric and non-rotating body at the leading order in metric perturbations and for static configurations, we find
\be
\label{eq:Sfs}
\begin{split}
	& S_{\text{finite-size}} = \\
	&\sum_{\ell}\frac{
	\lambda_{\ell}}{2\ell!}\int d\tau\int d^4x\,\delta^{(4)}
	\left(x-x_{\rm cm}\left(\tau\right)\right)
	\nabla_{\langle \mu_3}...\nabla_{\mu_\ell} \left(
	C_{\m_1 |\n |\m_2\rangle\rho} v^\n v^\r\right)
	\nabla^{\langle \mu_3}...\nabla^{\mu_\ell} 
	\left(C^{\m_1 |\alpha |\m_2\rangle \beta} v_\alpha v_\beta\right)
	\\
	& +\text{magnetic}\,,
	\end{split}
\ee
where $v_\mu$ is the body's 4-velocity,
and, as before, we suppressed an analogous series of  magnetic operators.

Using Eq.~\eqref{eq:Sfs}
we can compute the long distance metric of a body
in the presence of an external source. Neglecting gravitational non-linearities, the only non-trivial contribution is associated with local worldline operators,
which can be represented diagrammatically as  
\be\label{eq:1pointResponse}
	h_{\mu\nu} = \vcenter{\hbox{\begin{tikzpicture}
			\begin{feynman}
			\vertex[dot] (a0){};
			\vertex[left=0.4cm of a0] (p0){$\lambda_{\ell}$};
			\vertex[below=1.5cm of a0] (p1);
			\vertex[above=1.5cm of a0] (p2);
			\vertex[below right=2cm of a0] (a1);
			\vertex[above right=2cm of a0] (a2);
			\vertex[above right=2cm of a0] (a22){$\times$};
			\diagram*{
				(p1) -- [double,double distance=0.5ex] (p2),
				(a1) -- [photon] (a0) -- [photon](a2),
			};
			\end{feynman}
			\end{tikzpicture}}} 
			\,,
\ee
where the double line depicts
the worldline of the body, wiggly lines stand for the (retarded) graviton propagator, and  ``$\times$''  indicates that the corresponding propagator is replaced  by a linearized solution 
growing at infinity, corresponding to the background tidal field.

Non-linearity of the Einstein theory implies that additional contributions are present, which may be diagrammatically represented by the following infinite series,
\be\label{eq:1pointSourceT}
	h_{\mu\nu} = \vcenter{\hbox{\begin{tikzpicture}
			\begin{feynman}
			\vertex[dot] (a0);
			\vertex[below=1.5cm of a0] (p1);
			\vertex[above=1.5cm of a0] (p2);
			\vertex[right=1cm of a0, dot] (b0);
			\vertex[below right=2cm of a0] (b1);
			\vertex[above right=2cm of a0] (b2);
			\vertex[above right=1.6cm of a0] (b22){$\times$};
			\diagram*{
				(p1) -- [double,double distance=0.5ex] (p2),
				(a0) -- [photon] (b0),
				(b1) -- [photon] (b0) -- [photon](b2),
			};
			\end{feynman}
			\end{tikzpicture}}} + 
		\vcenter{\hbox{\begin{tikzpicture}
				\begin{feynman}
				\vertex[dot] (a0);
				\vertex[below=1.5cm of a0] (p1);
				\vertex[above=1.5cm of a0] (p2);
				\vertex[above=0.8 cm of a0, dot] (a1);
				\vertex[below=0.8 cm of a0, dot] (a2);
				\vertex[right=0.9cm of a0, dot] (b0);
				\vertex[below right=2cm of a0] (b1);
				\vertex[above right=2cm of a0] (b2);
				\vertex[above right=1.6cm of a0] (b22){$\times$};
				\diagram*{
					(p1) -- [double,double distance=0.5ex] (p2),
					(a1) -- [photon] (b0),
					(a2) -- [photon] (b0),
					(b1) -- [photon] (b0) -- [photon] (b2),
				};
				\end{feynman}
				\end{tikzpicture}}} +
		\vcenter{\hbox{\begin{tikzpicture}
				\begin{feynman}
				\vertex[dot] (a0);
				\vertex[below=1.5cm of a0] (p1);
				\vertex[above=1.5cm of a0] (p2);
				\vertex[above=0.8 cm of a0, dot] (a1);
				\vertex[below=0.8 cm of a0, dot] (a2);
				\vertex[right=0.5cm of a0] (a3);
				\vertex[right=0.7cm of a0, dot] (b0);
				\vertex[below right=2cm of a3] (b1);
				\vertex[above right=2cm of a3] (b2);
				\vertex[above right=1.6cm of a3] (b22){$\times$};
				\vertex[right=0.5cm of b0] (b3);
				\diagram*{
					(p1) -- [double,double distance=0.5ex] (p2),
					(a1) -- [photon] (b0),
					(a2) -- [photon] (b0),
					(b0) -- [photon] (b3),
					(b1) -- [photon] (b3) -- [photon](b2),
				};
				\end{feynman}
				\end{tikzpicture}}} + \dots
\ee
By requiring that the EFT result for the asymptotic field agrees with the full solution, one may determine an infinite series of coupling constants (``Wilson coefficients'') $\lambda_\ell$.
In Newtonian gravity, only the contributions \eqref{eq:1pointResponse} are present, so that the Wilson coefficient $\lambda_\ell$ is equal (up to an overall multiplicative factor) to the corresponding (Newtonian) Love number $k_\ell$. Hence, also in full general relativity, it is natural to identify Love numbers with the corresponding Wilson coefficients  $\lambda_\ell$. 
Contributions of the type \eqref{eq:1pointResponse} correspond then to the response part of the solution $\psi_r$, and contributions of the type \eqref{eq:1pointSourceT} to the source $\psi_s$. We see that the EFT definition resolves the source/response ambiguity and is manifestly reparametrization invariant.
 
However, the EFT definition also makes it clear that an additional irreducible ambiguity is present in the definition of Love numbers, which may be thought of as a remnant of the source/response separation problem. Namely, by the naive power counting one expects many of the diagrams in \eqref{eq:1pointSourceT} to exhibit  UV divergences. Hence, as is always the case, to perform the EFT matching one needs to introduce the UV cutoff scale (or, equivalently, the renormalization scale $\mu$ if one is using dimensional regularization) at which one performs the matching between the UV theory and 
the EFT. As a result, generically one expects Love numbers to depend on the matching scale $\mu$, {\it i.e.} to exhibit the classical RG running.

Somewhat surprisingly, one finds that the RG running of black hole Love numbers is an exceptional rather than a generic phenomenon in the Einstein gravity.  
It never takes place in four dimensions~ \cite{Ivanov:2022hlo}, whilst for higher-dimensional Schwarzschild black holes it occurs only for very special values of the multipole number $\ell$ (such that $\hat{\ell}=\ell/(d-3)$ is integer)~\cite{Kol:2011vg}.

Note that depending on the choice of coordinates, the source/response mixing may be present even in the absence of the classical RG running~\cite{Charalambous:2021kcz,Ivanov:2022hlo}. 
If this is the case, the two contributions can be disentangled bypassing the full EFT calculation via a formal trick of
``analytic continuation,''
i.e. promoting 
the multipolar index $\ell$ 
to an arbitrary 
real 
number during the matching procedure~\cite{Starobinski1973,Starobinski1974,Page:1976df,Charalambous:2021mea}. The equivalence between the analytic continuation of the multipolar index and the full EFT calculation has recently been demonstrated explicitly for the example of a Schwarzschild black hole in \cite{Ivanov:2022hlo} via the EFT computation of the relativistic corrections to the ``source'' profile of the $1$-point function to all PN orders.


Several further comments are in order. First, it is possible 
to account for the black hole spin
within the EFT along the lines of Ref.~\cite{Goldberger:2020wbx,Charalambous:2021mea}
and references therein.
In the body's rest 
frame and using the proper time 
worldline parametrization 
it amounts to promoting 
the Love numbers to 
tensors, i.e. 
\be
	S_{\text{finite-size}} = \sum_{\ell}\frac{1}{2\ell!} \int d\tau ~\Lambda^{j_1...j_\ell}_{i_1...i_\ell}E^{i_1...i_\ell} E_{j_1...j_\ell}\,.
\ee
The local Love tensors
responsible for conservative 
effects are even w.r.t.
the exchange of upper 
and lower indices, $i_a\leftrightarrow j_b$
for any $a,b=1,...,\ell$. 

Second, 
we note that 
it is straightforward to generalize 
the definition of Love numbers to 
scalar and electromagnetic fields~\cite{Kol:2011vg,Hui:2020xxx,Charalambous:2021mea,Ivanov:2022hlo}.
In full general relativity, one can compute the corresponding Newman-Penrose scalars, 
expand them at large distances, 
and read off coefficients 
in front of the relevant decaying power of $r$.
Equivalently,
one can also include long-wavelength scalar 
and electromagnetic perturbations 
in the point-particle EFT
and fix the relevant Wilson 
coefficients through matching.

Third, it is worth noting 
that matching to the full GR calculations
can also be done at the level of the S-matrix elements 
for the graviton-black hole scattering~\cite{Goldberger:2007hy,Cheung:2020sdj,Bern:2020uwk,Creci:2021rkz,Ivanov:2022hlo}. This matching
is performed on-shell and allows to explicitly
avoid ambiguities related to an apparent 
coordinate-dependence of the matching procedure. 




\subsection{Love numbers and holography}
Note that so far we have ignored
dissipative effects in our description of the EFT. 
These are not of the direct interest for the purposes of this paper. On the other hand, black holes are our main object of study. Black holes are perfect absorbers so that dissipation plays a very important role in their dynamics. So, for completeness, let us briefly describe how dissipative effects are incorporated in the EFT. This also gives us an ample opportunity to discuss the interpretation of  Love numbers in the context of holography and in the Kerr/CFT program.

Local worldline 
EFT operators introduced so far can only reproduce 
conservative tidal effects. 
In order to account for the dissipation, 
one needs to introduce 
internal  degrees of freedom $X$
on the black hole horizon 
and to couple them to the tidal 
tensors $E^{i_1...i_\ell}$ through the dissipative
multipole moments $Q^{(E)}_{i_1...i_\ell}$~\cite{Goldberger:2005cd,Goldberger:2019sya,Goldberger:2020fot,Goldberger:2020wbx,Ivanov:2022hlo}.
In the spherically symmetric case,
in the body's rest frame,
and with the worldline 
affine parameter equal to the proper time 
we have
\be 
	S_{\rm dissipative}=\sum_{\ell}\int d\tau ~Q^{(E)}_{i_1...i_\ell}(X)E^{i_1...i_\ell}+\text{magnetic}~\,.
\ee 
$Q$ here is a composite operator
whose exact dependence on $X$
is unknown. A black hole describes a highly excited state of the $X$-system, which can be accounted by using the in-in formalism.
As a result, the black hole response is a sum of two contributions. First, there is a local conservative response given by the diagrams shown in \eqref{eq:1pointResponse}. In addition there is a dynamical internal response which can be represented as 
\be\label{eq:dynamicResponse}
	h_{\mu\nu} = \vcenter{\hbox{\begin{tikzpicture}
			\begin{feynman}
			\vertex[] (a0){};
			\vertex[left=0.2cm of a0] (a00);
			\vertex[above=0.4cm of a00] (a0t){};
			\vertex[below=0.4cm of a00] (a0b){};
			\vertex[left=0.4cm of a0] (p0);
			\vertex[above=0.4cm of p0] (p0t){$Q$};
			\vertex[below=0.4cm of p0] (p0b){$Q$};
			\vertex[below=1.5cm of a0] (p1);
			\vertex[above=1.5cm of a0] (p2);
			\vertex[below=0.6cm of a0] (p11);
			\vertex[above=0.6cm of a0] (p22);
			\vertex[below right=2cm of a0] (a1);
			\vertex[above right=2cm of a0] (a2);
			\vertex[above right=2cm of a0] (a22){$\times$};
			\diagram*{
				(p1) -- [double,double distance=0.5ex] (p2),
				(a1) -- [photon] (a0b) , (a0t) -- [photon](a2),
				(p11) -- [ghost](p22)
			};
			\end{feynman}
			\end{tikzpicture}}} 
			\,,
\ee
where the dotted lined denotes an insertion of the $\langle Q Q\rangle$ two-point function\footnote{We assume that the operators are chosen in such a way that their vev's are zero $\langle Q\rangle=0$.}.
This way correlation functions of $Q$ can be extracted through 
matching to various observables such as
 low-energy graviton
absorption cross-sections. The latter determine the imaginary part of $QQ$ correlators. The real part can be reconstructed by making use of dispersion relations following from analyticity of the correlators.
This reconstruction leaves undetermined a real polynomial piece of the $QQ$ correlators, which corresponds to the Love numbers. For more details on how dissipation is treated in the worldline EFT, see the recent work \cite{Ivanov:2022hlo}, with explicit matching calculations 
for the Schwarzschild black holes, and
also references therein.

Our discussion implies that generically the physical Love numbers is a sum of the Wilson coefficient $\lambda$'s and of the polynomial part of the $QQ$ correlators. However, by making use of the field redefinitions which shift $Q$'s by functions of the bulk metric evaluated at the origin, one may work in the operator basis where Love numbers are entirely associated with $\lambda$'s.

This description of the black hole EFT is holographic in its nature \cite{Goldberger:2005cd}. Namely, the idea of holography is essentially the statement that $X$'s are more than an EFT bookkeeping device, but represent an actual quantum mechanical system describing the black hole dynamics. Remarkably, the AdS/CFT correspondence \cite{Aharony:1999ti}  tells us exactly what this system is for certain classes of black holes in string theory.
For instance,  black $3$-branes in type IIB string theory are described in this way by the maximally supersymmetric Yang--Mills theory. The discovery of this correspondence \cite{Maldacena:1997re} was guided by the matching calculations \cite{Klebanov:1997kc,Gubser:1997yh} of the type we just described.

A promise of the Kerr/CFT correspondence \cite{Guica:2008mu,Castro:2010fd} is that one day a similar success may be achieved for the actual real world black holes. It is natural to ask what is the role of the Love numbers in this story. The discussion above implies that they do not have an intrinsic CFT interpretation. Rather, they describe how the CFT (describing a near horizon throat emerging in the extremal limit) is glued to the rest of the spacetime. It will be interesting to calculate Love numbers for classes of black holes with known holographic duals and contrast them with what one finds in four-dimensional general relativity.

Note that from the holographic viewpoint it is natural to think about black holes with different values of a mass as excitations of a single quantum mechanical system rather than representing genuinely different systems.   
Namely, an extremal black hole of a minimal possible mass at the fixed values of gauge charges (and zero Hawking temperature) corresponds to the ground state. Non-extremal black holes
correspond instead to excited finite temperature states. As a result, symmetries of the system become more manifest as one approaches the extremal limit. Geometrically, a near extremal charged black hole develops a near horizon anti de Sitter throat. Isometries of this region correspond to the conformal symmetry present in the holographic description. This makes it natural to study the fate of the Love symmetry in the near extremal limit, which is one of the main tasks of this paper.    

\subsection{Teukolsky master equation}


In order to extract black hole Love numbers in a systematic and 
gauge-independent fashion, one can match 
results of EFT calculations to solutions of black hole 
perturbation theory. 
In four spacetime dimensions, it is  
convenient to study these perturbations within the Newman-Penrose (NP) spin-coefficient formalism~\cite{Newman1962,Geroch1973}. 
In this approach all independent components of the Maxwell and Weyl tensors
are captured by the so-called Newman-Penrose scalars
that are defined as projections of these tensors onto complex null tetrads. 
The NP scalars have fixed weights under 
local Lorentz rotations which define their spin weights $s$,
and they are also referred to as spin $s$ scalars.
Spin-$0$ ($s=0$) and spin-$1$ ($s=\pm1$) perturbations correspond to test scalar and Maxwell fields in the black hole background respectively, while spin-$2$ ($s=\pm2$) tidal perturbations correspond to perturbations of the black hole geometry itself. 

The advantage of working with the spin weighted scalars is 
that for some of them the relevant dynamical equations fully factorize in the generic  
Kerr-Newman black hole metric. 
The corresponding separable master equation was obtained by Teukolsky in Ref.~\cite{Teukolsky1972}. 
In the Boyer-Lindquist coordinates the Kerr-Newman black hole metric reads
\be\label{eq:KerrNewmanMetricBL}
	\begin{gathered}
	    ds^2 = -dt^2 + \frac{r_{s}r-r_{Q}^2}{\Sigma}\left(dt - a\sin^2\theta\,d\phi\right)^2 + \frac{\Sigma}{\Delta}\,dr^2 \\
		+ \left(r^2+a^2\right)\sin^2\theta\,d\phi^2 + \Sigma\,d\theta^2 \\
		\Delta = r^2 -r_{s}r + a^2 + r_{Q}^2 = \left(r-r_{+}\right)\left(r-r_{-}\right) \,\,,\,\, \Sigma=r^2+a^2\cos^2\theta \,,\\
		r_{\pm} = \frac{1}{2}\left[r_{s}\pm\sqrt{r_{s}^2-4\left(a^2+r_{Q}^2\right)}\right]=
M\pm\sqrt{M^2-\left(a^2+Q^2\right)}
		\,.
	\end{gathered}
\ee
This metric is characterized by three parameters: mass $M$, angular momentum $J$ and electric charge $Q$, encoded in the Schwarzschild radius, spin parameter and charge parameter as 
\[
r_{s}=2GM\,,\;\;a={J\over M}\,,\;\;\mbox{\rm and}\;\; r_{Q}=\sqrt{G}Q
\]
respectively. In what follows we mostly work in the $G=1$ units.
A spin-$s$ NP scalar $\Psi_{s}$
can be factorized as follows
\be
	\Psi_s =  \Phi_s\left(t,r,\phi\right) S\left(\theta\right)= e^{-i\omega t}e^{im\phi}R\left(r\right)S\left(\theta\right) \,.
\ee
It satisfies the Teukolsky 
master equation~\cite{Teukolsky1973,Dudley1977}
which takes the following form 
in the Kinnersley tetrad~\cite{Kinnersley1969}
\be\label{eq:teuk}
	\begin{split}
		\mathbb{O}^{\left(s\right)}\Psi_{s} = \ell(\ell+1) \Psi_{s}\,,\quad \mathbb{P}^{\left(s\right)}\Psi_{s} =  \ell(\ell+1)\Psi_{s}\,,
  \end{split}
\ee
where $\ell(\ell+1)$ is a separation constant
identified with the angular problem eigenvalue. 
The orbital number $\ell$ 
\textit{is not an integer} in general.

The explicit form of the radial and angular differential operators is,
\be\label{eq:teukExplicit}
	\begin{split}
			 & \mathbb{O}^{\left(s\right)} = 
			 \Delta^{-s}\partial_{r}\Delta^{s+1}\partial_{r} 
			 - \frac{a^2}{\Delta}\partial_{\phi}^2  -\frac{\left(r^2+a^2\right)^2}{\Delta}\partial_{t}^2 - 2\frac{a\left(2Mr-Q^2\right)}{\Delta}\partial_{t}\partial_{\phi} \\
			& + s\left(s+1\right) + s\frac{a\Delta^{\prime}}{\Delta}\partial_{\phi}
			 + s\left[\frac{2M\left(r^2-a^2\right)-2{Q}^2r}{\Delta}-2r\right]\partial_{t}
		 \,, \\
			& \mathbb{P}^{\left(s\right)}= -\bigg[ \Delta_{\mathbb{S}^2}^{\left(s\right)} - s\left(s+1\right) + a^2\sin^2\theta\partial_{t}^2 -2isa\cos\theta\partial_{t}\bigg]\,,
	\end{split}
\ee
with $\Delta_{\mathbb{S}^2}^{\left(s\right)}$ the spin weighted Laplacian on the sphere,
\be
	\Delta_{\mathbb{S}^2}^{\left(s\right)} = \frac{1}{\sin\theta}\partial_{\theta}\sin\theta\partial_{\theta}+\frac{\left(\partial_{\phi}+is\cos\theta\right)^2}{\sin^2\theta} + s\,.
\ee
For $s=0$ the spacetime scalar function $\Psi_{s}$ is just the massless scalar field, for $s=+1$ ($s=-1$) this is the transverse ingoing (outgoing) radiation Maxwell-NP scalar and for $s=+2$ ($s=-2$) this is the transverse ingoing (outgoing) radiation Weyl scalar~\cite{Teukolsky1973}. The Weyl scalar $\psi_0$ discussed in Section~\ref{sec:GREFT} corresponds to $\Psi_2$
in this nomenclature.  

\subsection{Near zone approximation(s)}
Let us now introduce the near zone approximation~\cite{Starobinski1973,Starobinski1974,Maldacena1997,Castro:2010fd,Chia:2020yla,Charalambous:2021kcz}.
Its general purpose is to come up with an exactly solvable truncation of the full Teukolsky equation that would be accurate in a certain vicinity of the black hole, which is large enough to overlap also with the asymptotically flat region. 
Let us write the 
full Teukolsky operators in the following form,
\be
\label{eq:NZgen}
	\begin{split}
		& \mathbb{O}^{\left(s\right)} \equiv \mathbb{O}^{\left(s\right)}_{\rm NZ} + \epsilon  \mathbb{O}^{\left(s\right)}_\epsilon = \Delta^{-s}\partial_{r}\Delta^{s+1}\partial_{r} + V_0 +\epsilon V_1 +s\left(s+1\right)\,,\\
		& \mathbb{P}^{\left(s\right)} \equiv \mathbb{P}^{\left(s\right)}_{\rm NZ} +\epsilon \mathbb{P}^{\left(s\right)}_{\epsilon} = -\left[ \Delta_{\mathbb{S}^2}^{\left(s\right)} - s\left(s+1\right) + \epsilon\left(a^2\sin^2\theta\partial_{t}^2 -2isa\cos\theta\partial_{t}\right) \right] \,,
	\end{split}
\ee
where we split the potential into the so-called 
near and far 
contributions $V_0$ and $V_1$,
\be
\label{eq:V0V1}
\ba
	V_0 = &-\frac{\left(r_{+}^2+a^2\right)^2}{\Delta}\left[(\partial_t +\Omega\,\partial_\phi)^2 + 4\Omega\frac{r-r_{+}}{r_{+}-r_{-}}\partial_t \partial_\phi - 2s\,\beta^{-1}\partial_{t}\right] +s\frac{a\Delta^{\prime}}{\Delta}\partial_\phi \,,\\
	V_1 = &-\frac{\left(r+r_{+}\right)\left(r^2 + r_{+}^2 +2a^2\right)}{r-r_{-}}\partial_t^2 + 2\frac{r_{+}^2+a^2}{r-r_{-}}\left(\beta-2M\right)\Omega\,\partial_t\partial_\phi \\
	&+ 2s\left[\frac{M\left(r+r_{+}\right)-Q^2}{r-r_{-}}-r\right]\partial_t \,,
\ea
\ee
and we have introduced the (inverse of the) Hawking temperature,
\be
	\beta \equiv \frac{1}{2\pi T_H} =2\frac{r_{+}^2+a^2}{r_{+}-r_{-}} \,,
\ee
and the black hole's angular velocity,
\be 
	\Omega \equiv \frac{a}{r_+^2+a^2} \,.
\ee
The split~\eqref{eq:NZgen} is accomplished by means of
a formal parameter $\epsilon$
which is equal to unity for the full physical Teukolsky 
equation, while $\epsilon=0$ corresponds to the leading near zone approximation.
The latter
is accurate as long as the following 
conditions are satisfied,
\be
\label{eq:NZ}
	\omega r\ll 1\,,\quad M\omega\ll 1\,. 
\ee
This regime covers the region around
the black hole $r\gtrsim r_+$
and also has an overlap with the
asymptotically far region $r\gg r_+$.

Importantly, the near zone expansion 
\textit{is not equivalent} to the low 
frequency expansion 
because $V_0$
and $V_1$ contain terms 
with equal powers of frequency.
Nevertheless, it 
does provide an 
accurate approximation at low frequencies in the near zone region  \eqref{eq:NZ}
because the $V_1$ corrections are suppressed there. In particular, the near zone approximation is exact 
for $\omega = 0$
modes, which are relevant for static tidal response
calculations.

Related to this, there is an ambiguity in the definition of a near zone expansion associated with a freedom to move $\omega$-dependent terms between $V_0$ and $V_1$ as long as $V_1$ stays finite at the horizon. Other choices of the near zone split can be found in, e.g.~\cite{Starobinski1973,Maldacena1997,Castro:2010fd}.
For future reference, let us explicitly present the Starobinsky choice,
\begin{gather}
	\label{eq:starCas}	V^{\rm Star}_0 = -\frac{\left(r_{+}^2+a^2\right)^2}{\Delta}\left(\partial_{t} + \Omega\,\partial_{\phi}\right)^2 + s\frac{\left(r_{+}^2+a^2\right)\Delta^{\prime}}{\Delta}\left(\partial_{t} + \Omega\,\partial_{\phi}\right) \,, \\
	V^{\rm Star}_1 = -\frac{\left(r+r_{+}\right)\left(r^2 + r_{+}^2 +2a^2\right)}{r-r_{-}}\partial_t^2 - 4\frac{Ma}{r-r_{-}}\partial_t\partial_\phi + 2s\left[M\frac{r-r_{+}}{r-r_{-}}-r\right]\partial_t \,.\label{eq:starCas1}
\end{gather}
As will be explained later, we chose a particular near zone split 
\eqref{eq:NZgen} because in 
this approximation black hole 
pertubations exhibit the enhanced Love symmetry.

As far as the angular problem is concerned, 
the leading near zone approximation 
is chosen to coincide with the static approximation.
The leading angular 
eigenfunctions are then given by
the standard spin weighted spherical harmonics 
(see Ref.~\cite{Charalambous:2021kcz} for our conventions), 
\be
	S\left(\theta\right) = e^{-im\phi} {}_{s}Y_{\ell m}\left(\theta,\phi\right) \,.
\ee
The orbital number $\ell\ge\left|s\right|$ is always an integer in the leading near zone approximation.

\section{Love numbers in general relativity and beyond}
\label{sec:facts}

In this section we summarize a number of facts about black hole Love numbers.
First, we present a master formula 
for Kerr-Newman black hole response coefficients
to perturbing fields of spin $s$
in the near zone approximation.
This formula embodies the facts
that in four dimensional general relativity black hole 
Love numbers are zero but 
dissipative numbers are not.
Then we turn to higher dimensional Schwarzschild 
black holes, which exhibit more a complex 
Love number phenomenology.
After that, we present an explicit example of non-vanishing and running Love numbers. Namely, we will show that black hole Love numbers are not zero and exhibit a logarithmic running in the Riemann-cubed gravity. Finally we will comment on the relation between Love numbers and black hole (no) hair.

\subsection{Master formula for black hole Love numbers in four dimensions}
\label{sec:master}

In order to calculate static black hole Love numbers describing response to 
a perturbing fields of spin $s$, one needs to 
solve the static ($\omega=0$) radial Teukolsky equation
\[
\mathcal{O}^{\left(s\right)}R =\ell(\ell+1)R\;,
\]
where $\mathcal{O}^{\left(s\right)}$ is the operator obtained from  $\mathbb{O}^{\left(s\right)}$ after expanding into monochromatic and harmonic modes, that is, after replacing $\partial_{t}\rightarrow -i\omega$ and $\partial_{\phi}\rightarrow im$. At this point it is instructive to consider an extension 
of the static Teukolsky equation to finite frequencies, by using the Starobinsky near zone approximation for $\mathcal{O}^{\left(s\right)}$.
This will allow us to capture a part of the full finite frequency response associated with frame dragging.
Then we need to find a solution to
\be
	\mathcal{O}^{(s)~\text{Star}}_{\rm NZ}R =\ell(\ell+1)R \,,
\ee
which is regular at the black hole horizon. It is given by
\be
\label{eq:psiS}
	R = \Delta^{-s}\left(\frac{r-r_+}{r-r_-}\right)^{i Z} {}_2F_1\left(-\ell-s,\ell+1-s;1+2iZ-s;\frac{r_{+}-r}{r_{+}-r_{-}}\right) \,,
\ee
where ${}_2F_1(a,b;c;x)$ is the Gauss 
hypergeometric function
and we have defined
\be 
	\begin{split}
		& Z\equiv\frac{r_{+}^2+a^2}{r_+-r_-}\left(m\Omega-\omega\right) \,.
	\end{split}
\ee 

It is  most  straightforward  to interpret this result for a static scalar response in the Schwarzschild case. This corresponds  to setting  
\[
s=0\,,\;Z=0\;,
\]
so that the resulting radial function reduces to a hypergeometric function, which is purely polynomial in $r$ in this case. This indicates that the static scalar response completely vanishes in the Schwarzschild example.

This  conclusion is  not immediately obvious at  $s\neq0$ and/or $Z\neq 0$  due  to the  presence  of a  non-polynomial factor $\Delta^{-s}\left(\frac{r-r_+}{r-r_-}
\right)^{i Z}$ in (\ref{eq:psiS}). However,  this factor does not  affect the extraction of  the Love numbers,  which follows from the fact that it multiplies  the full solution\footnote{By this we mean that if we were to write
down the most general solution, without imposing the regularity condition at the horizon, this overall multiplicative factor would still be present.}, and  hence appears both in front  of the source and of the response. Another way to  see that this factor cannot affect coordinate independent response coefficients is to note that  its presence or  absence 
depends on  the choice  of   coordinates and of the null  tetrad (the Boyer-Lindquist 
frame and the Kinnersley tetrad correspondingly in  our case). For instance,  the form factor $\left(\frac{r-r_+}{r-r_-}
\right)^{i Z}$ can be completely
removed by a transition to the advanced coordinates \cite{Charalambous:2021mea}.  Similarly, the $\Delta^{-s}$ can  be stripped  off  by an appropriate null boost~\cite{Teukolsky1972,Teukolsky1973}.

All in all, the reponse  coefficients are completely  determined by the
hypergeometric function in Eq.~\eqref{eq:psiS}.
In the near zone approximation $\ell+s$ is a non-negative integer
and hence this hypergeometric function 
is a polynomial of order 
$\ell+s$. This implies that the static Love numbers
are zero also in the Kerr-Newman case for general spin $s$ perturbations and $\Omega$.

Note,  that in principle one should be careful to draw the latter conclusion just based  on the polynomial form  of the  solution (up  to an overall multiplicative factor). Indeed,  as  discussed in  section~\ref{sec:prel} to  extract the  response  coefficients one needs  to disentangle the source and  response  contributions  into the full  solution. In the case at hand this is achieved most easily by a formal analytic continuation to non-integer values of $\ell$. As a result, one obtains that the black hole still exhibits a non-vanishing  {\it dissipative} response  in spite of the (quasi)polynomial form of~\eqref{eq:psiS}. It can be summarized 
by the following master formula for the black hole response coefficients
to a perturbing field of integer spin weight $\left|s\right|\le2$ \cite{Charalambous:2021mea}\footnote{In \cite{Charalambous:2021mea}, the master formula 
was derived only for neutral Kerr black holes. 
However, the inclusion of an electric charge~\cite{Berti2005} is straightforward
because it does not alter the form of the perturbation equations.  
},
\be\label{eq:ResponseCoefficients}
	k_{\ell m}^{\left(s\right)} = \frac{i\sinh2\pi Z}{2\pi}\left|\Gamma\left(\ell-2iZ+1\right)\right|^2\left(-1\right)^{s+1}\frac{\left(\ell+s\right)!\left(\ell-s\right)!}{\left(2\ell\right)!\left(2\ell+1\right)!}\left(\frac{r_{+}-r_{-}}{r_{s}}\right)^{2\ell+1}\,.
\ee
These harmonic response coefficients are purely imaginary
and also odd w.r.t. time-reversal transformations.
Thus, they correspond  to dissipative effects. As a consequence of using the leading Starobinsky near zone approximation they depend on the the frequency only through the frame-dragging factor $\omega-m\Omega$.
The conservative 
static responses of black holes, i.e. Love numbers, 
are exactly zero for all spins.

We observe that, in addition to vanishing of the Love numbers, the black hole response exhibits an additional surprising feature. Namely, at non-integer $\ell$ the hypergeometric function in \eqref{eq:psiS} splits into a sum of non-overlapping source and response contributions of the form
\[
{}_2F_1\sim r^\ell\left(1+\dots\right)+k r^{-\ell-1}\left(1+\dots\right)\;,
\]
where dots stand for two infinite series of power law corrections $\sum  a_n/r^n$ with positive integer $n$'s. 
At the (physical) integer values of $\ell$ the two series overlap and conspire to produce a polynomial answer for the full solution, even though both source and response are still given by non-trivial infinite series as reflected by the presence of non-trivial response coefficients \eqref{eq:ResponseCoefficients}. Hence, the actual puzzle to explain is the (quasi)polynomial form of \eqref{eq:psiS}. We will see that it arises as a consequence of the highest weight property of the corresponding representation of the Love symmetry.

From Eq.~\eqref{eq:ResponseCoefficients}
we also see a technical advantage of the Starobinsky 
near zone approximation. It provides a simple frequency-dependent 
extension which allows to 
capture co-rotating 
modes while keeping the complexity of the solution
the same as in the static case.
It is important to remember though that the result~\eqref{eq:ResponseCoefficients}
is \textit{exact} only for static perturbations ($\omega=0$). Its
frequency-dependent part is \textit{approximate} and does not capture the full black hole response at non-zero $\omega$.
A systematic calculation shows that Love numbers receive 
corrections already
at linear order in $\omega$~\cite{Charalambous:2021mea},
\be
\label{eq:ksf}
	\begin{split}
		\delta  k^{(s)}_{\ell m} =
 		& -2(\omega r_{s})m\gamma \ln \left(\frac{r-r_+}{r_{+}-r_{-}}\right)
	\\
		&\times
		\left(\frac{r_{+}-r_{-}}{r_s}\right)^{2\ell+1}(-1)^{s}\frac{(\ell+s)!(\ell-s)!}{(2\ell)!(2\ell+1)!} \prod_{n=1}^\ell\left(n^2+4(m\gamma)^2\right)
		\,,
	\end{split} 
\ee
where 
\[
\gamma\equiv
\frac{a}{r_+-r_-} \,,
\]
and we retained only the scheme-independent 
logarithmic part. This result indicates that non-static (``dynamical'') 
black hole Love numbers are in general non-zero and exhibit logarithmic running, 
in agreement 
with expectations
of Wilsonian naturalness. 

%
%
%

Let us also note that the above analysis does not  
apply for extremal black holes with $a^2+Q^2=M^2$. This happens because  the correction to the potential $V^{Star}_1$, given by \eqref{eq:starCas1} and neglected in the Starobinsky near zone approximation, develops a singularity at the horizon in the extremal limit. 
Nevertheless, there exists 
a consistent near zone truncation 
of the Teukolsky equation
even in the extremal limit,
see Appendix~\ref{app:extr}.
In this approximation, 
the conservative response coefficients
do not vanish for finite frequencies, 
but the static Love numbers 
are still zero,
\be 
	k^{(s)}_{\ell m}= 0\,,\quad a^2+Q^2=M^2 \,.
\ee 
In Section~\ref{sec:furt},
we will provide a somewhat complementary explanation for this fact by 
presenting a symmetry that is valid 
in the so-called ``middle zone'',
interpolating between the near horizon region (but not the horizon itself)
and the asymptotically flat region.


\subsection{Higher dimensions}

Love numbers have also been computed for higher $d$-dimensional spherically symmetric asymptotically flat black hole geometries,
\be
	ds^2 = -f\left(r\right)dt^2 + \frac{dr^2}{f\left(r\right)} + r^2 d\Omega_{d-2}^2 \,,\quad f\left(r\right) = 1 -\left(\frac{r_{s}}{r}\right)^{d-3} \,,
\ee
where $d\Omega_{d-2}^2$ is the line element on the unit $\mathbb{S}^{d-2}$.
The responses to 
scalar and electric-type spin-1 and spin-2 perturbations take the following general form~\cite{Kol:2011vg,Hui:2020xxx},
\be\label{eq:TLNsD}
	k_{\text{el},\ell}^{\left(s\right)} =A_{\hat{\ell}}^{(s)} \tan\pi\hat{\ell} \,,
\ee
where $A_{\hat{\ell}}^{(s)}$ are certain 
non-zero and real constants and
\[
	\hat{\ell}\equiv\frac{\ell}{d-3}\;.
\]
These Love numbers are in general not zero, but they still vanish in
particular situations when $\hat{\ell}\in\mathbb{N}$. The apparent divergences at $\hat{\ell}\in\mathbb{N}+\frac{1}{2}$ signal the classical renormalization group (RG) running 
with a finite scheme-independent coefficient in front 
of the logarithmic corrections~\cite{Kol:2011vg}.

The magnetic-type Love numbers have a very different behavior. Their main features are captured by the following expressions for spin-$1$ and spin-$2$ perturbations~\cite{Hui:2020xxx},
\be
	k_{\text{mag},\ell}^{\left(s\right)} 
	= B_{\hat{\ell}}^{(s)} \frac{\sin\pi\left(\hat{\ell}+\frac{1}{d-3}\right)\sin\pi\left(\hat{\ell}-\frac{1}{d-3}\right)}{\sin2\pi\hat{\ell}}\,,
\ee
with $B_{\hat{\ell}}^{(s)}$ certain non-zero and real constants. Magnetic-type Love numbers are never zero for $d>4$, not even when $\hat{\ell}\in\mathbb{N}$, but they still exhibit a classical RG flow whenever $\hat{\ell}\in\mathbb{N}+\frac{1}{2}$. 
The magnetic- and electric-type Love numbers
coincide only in $d=4$ as a result of the electric-magnetic duality
that takes place only in this spacetime dimensionality~\cite{Porto:2007qi,Hui:2020xxx}.

\subsection{Running scalar Love numbers in $R^3$ gravity}
\label{sec:Rcubed}

A simple important example of non-vanishing and running
Love numbers is given by the black hole scalar response in the Riemann cubed $R^3$ gravity
in four dimensions.

General effective field theory arguments suggest the presence of
higher order curvature corrections 
in the gravity action~\cite{Donoghue:2017pgk}. 
In the pure gravity case the $R^2$ corrections 
to the Einstein-Hilbert action
can be eliminated by means of the leading (GR) equations 
of motion in the vacuum and the Gauss-Bonnet identity. 
Hence the first non-trivial contribution
appears at the cubic order in curvature.
In perturbation theory this contribution
becomes important at the two-loop order~\cite{Goroff:1985th}.
The action of the $R^3$ gravity is given by,
\be
\label{eq:R3act}
	S_{\text{gr}}\left[g\right] = -\frac{1}{16\pi G}\int d^{4}x\,\sqrt{-g}\,\left(R + \alpha\,R^{\mu\nu}_{\,\,\,\,\,\,\,\,\rho\sigma}R^{\rho\sigma}_{\,\,\,\,\,\,\,\,\kappa\lambda}R^{\kappa\lambda}_{\,\,\,\,\,\,\,\,\mu\nu}\right)\,.
\ee
This modified gravitational action leads to the field equations of the form
\be
	G_{\mu\nu} + \alpha K_{\mu\nu}=0 \,,
\ee
where
\[
G_{\mu\nu}=R_{\mu\nu}-\frac{1}{2}Rg_{\mu\nu} \,,
\] 
is the Einstein tensor and
\[
K_{\mu\nu} \equiv -\frac{1}{2}g_{\mu\nu}R^{\alpha\beta}_{\,\,\,\,\,\,\,\,\rho\sigma}R^{\rho\sigma}_{\,\,\,\,\,\,\,\,\kappa\lambda}R^{\kappa\lambda}_{\,\,\,\,\,\,\,\,\alpha\beta} + 3R_{\mu\alpha\kappa\lambda}R^{\kappa\lambda\rho\sigma}R_{\nu\,\,\,\,\rho\sigma}^{\,\,\,\,\alpha} - 6\nabla_{\sigma}\nabla_{\rho}\left(R_{(\mu|}^{\,\,\,\,\,\,\,\,\sigma\kappa\lambda}R_{\kappa\lambda\,\,\,\,|\nu)}^{\,\,\,\,\,\,\,\rho}\right)\,.
\]

The first step in calculating the black hole Love numbers is to find the background geometry. We focus on the modified Schwarzschild solution describing an asymptotically flat and spherically symmetric black hole in the vacuum of this theory of gravity. As such, the general background metric can be written as,
\be
	ds^2 = -f_{t}\left(r\right)dt^2 + \frac{dr^2}{f_{r}\left(r\right)} + r^2 d\Omega_{2}^2\,,
\ee
We perturbatively expand the problem around the Schwarzschild results at $\alpha=0$.
To keep track of the boundary conditions at the horizon, it is convenient 
to introduce the dimensionless variable
\[
x=\frac{r_{h}}{r}\,,
\] 
where $r=r_{h}$ is the radial position of the event horizon. We keep $r_h$ fixed at all values of $\alpha$. Note that in this parameterization the ADM mass of the resulting solutions
is a non-trivial function of $\alpha$.
Up to linear order in $\alpha$, we then find
\be\ba
	f_{t} &= \left(1-x\right)\left(1 + \frac{\alpha}{r_{h}^4}\left(-5x\frac{1-x^5}{1-x} - 5x^6\right) + \mathcal{O}\left(\alpha^2\right) \right) \,, \\
	f_{r} &= \left(1-x\right)\left(1 + \frac{\alpha}{r_{h}^4}\left(-5x\frac{1-x^5}{1-x} + 49x^6\right) + \mathcal{O}\left(\alpha^2\right) \right) \,.
\ea\ee

Let us calculate now the black hole response at the leading order in $\alpha$. For simplicity, we restrict to the scalar field response, so that we need to compute the static scalar field profile by solving the Klein-Gordon equation,
\be
\label{eq:R3KG}
	f_{r} \phi_{\ell m}^{\prime\prime} + \frac{\left(f_{t}f_{r}\right)^{\prime}}{2f_{t}}\phi_{\ell m}^{\prime} = \frac{\ell\left(\ell+1\right)}{x^2}\phi_{\ell m}\,,
\ee
where an expansion into spherical harmonics has been used. 
We look for a perturbative solution of \eqref{eq:R3KG} as a series in $\alpha$,
\be
    \phi_{\ell m}\left(x\right) = \phi_{\ell}^{\left(0\right)} + \frac{\alpha}{r_{h}^4}\phi_{\ell}^{\left(1\right)}\left(x\right) + \mathcal{O}\left(\alpha^2\right)  \,,
\ee
where at zeroth order in $\alpha$ one gets the conventional GR solution, 
\be
	\phi_{\ell}^{\left(0\right)} = \frac{\left(\ell!\right)^2}{\left(2\ell\right)!}P_{\ell}\left(\frac{2-x}{x}\right) \,,
\ee
corresponding to vanishing Love numbers. 
Higher order terms $\phi_{\ell}^{\left(n\right)}$ are all regular at the horizon at $x=1$ and grow at infinity slower than the leading order solution,
\[
\phi_{\ell}^{\left(n\right)}x^{\ell}\to 0\;\mbox{ \rm as }\; x\to 0\;,
\]
for $n>0$.

The leading corrections $\phi_{\ell}^{\left(1\right)}$ are then extracted by solving the resulting linear inhomogeneous ODE. For $\ell=1,2$ we find,
\be\ba
    \phi_{1}^{\left(1\right)} &={7\over 2}+{2\over 5}x^2+{2\over 5}x^3+{9\over 25}x^4+{83\over 25} x^5 \,,
    \\
    \phi_{2}^{\left(1\right)} &= -{5\over x}+{5\over3}+{236\over 75}x^3+{118\over 25}x^4-{83\over 25} x^5 \,.
\ea\ee
We see that these  corrections are no longer polynomial in $r$ and, in particular, terms proportional to $x^{\ell+1}\propto r^{-\ell-1}$ are arising, but there is still  no sign of logarithmic RG running.
This changes for higher mulitpole  numbers  $\ell\ge3$. For example, for $\ell=3$, we find
\begin{gather}
	\phi_{3}^{\left(1\right)}={230385\over 2x^2}-{143994\over x}+{154877\over  4}-480  x+48x^2-5x^3-{1851x^4\over 250}+{249x^5\over  125} \\
	-{1920\left( 11x^2-60x+60\right)\log  x\over x^2}+{5760(x-2)(x^2-10x+10)\left(\log x\log(1-x)+\text{Li}_2\left(x\right)\right)\over x^3}
	\nonumber
\end{gather}
%
%
At face value, the coefficients in 
front of the $x^{\ell+1}$ terms that correspond to the Love numbers read, for $\ell=1,2,3,4$,
\be
\label{eq:varkappas}
	\begin{gathered}
    	\varkappa_1 = \frac{2}{5} \frac{\alpha}{r_{h}^4} \,\,\,,\,\,\, \varkappa_2 = \frac{236}{75} \frac{\alpha}{r_{h}^4} \\
	    \varkappa_3 = \frac{\alpha}{r_{h}^4}\left(\frac{450501}{12250} + \frac{288}{7}\log x\right)\,\,\,,\,\,\,\varkappa_4 = \frac{\alpha}{r_{h}^4}\left(\frac{1540202}{11025} + \frac{736}{7} \log x\right) \,,
	\end{gathered}
\ee
where we denoted them as $\varkappa_{\ell}$ in order to emphasize that it would be premature to identify these with the Love numbers at this stage.

Indeed, as we discussed before, in order to 
rigorously identify the Love numbers
through the matching procedure, 
we need to compute the graviton
corrections to the source term 
and subtract them from the full GR solution.
An alternative to this procedure 
is to perform an analytic 
continuation $\ell\to \mathbb{R}$, which removes an overlap between the source and the response.
Unfortunately, the complexity of the equations of motion in $R^3$ gravity 
does not allow us to construct a solution for generic $\ell$. 
Therefore, a more accurate calculation
involving a systematic expansion 
of the bulk action into interaction
vertices and calculations of the 
corresponding loop corrections 
is required. However, as we will see now, for many purposes, it is possible to bypass a complete calculation by making use of the straightforward dimensional 
analysis alone.

In particular, the situation simplifies  when the logs
appear. 
Logarithmic corrections 
are present both in 
the IR and UV theories. 
In the IR, the logarithms are associated
with divergences
in the EFT loop integrals~\cite{Kol:2011vg}. 
The finite 
logarithmic part is associated with the RG running of
the physical Love number. 
This is exactly the situation
that we have in $R^3$ gravity 
with $\ell\geq3$, and we conclude that the $\ell\geq3$ Love numbers exhibit log running there, and $\varkappa_{\ell}$ are the corresponding leading log coefficients.

\subsection{Power counting for Love numbers}
\label{sec:count}
Let us see now that the presence of this classical RG running for $\ell\geq 3$ scalar Love numbers in $R^3$ gravity 
can be understood based on a power counting argument. In fact, the argument applies for a general theory of gravity with the action of the following schematic form 
\be
    S_{\text{gr}} \sim \frac{1}{16\pi G}\int d^{d}x\,\sqrt{-g}\,\sum_{k=1}^{\infty}\alpha_{k}\left(R_{\mu\nu\rho\sigma}\right)^{k} \,,
\ee
where the coupling constants $\alpha_{k}$ have mass dimensions $2\left(1-k\right)$. As we discussed, the source/response ambiguity and associated running of Love numbers  occurs when gravitational non-linearities caused by the source contribute to the total field at the same order, $r^{-\ell-1}$, as the response.
A typical EFT diagram representing these non-linearities has the following schematic form
\be\label{eq:1pointSource}
	\vcenter{\hbox{\begin{tikzpicture}
			\begin{feynman}
			\vertex[dot] (a0);
			\vertex[below=1.5cm of a0] (p1);
			\vertex[above=1.5cm of a0] (p2);
			\vertex[above=1cm of a0] (a1);
			\vertex[above=0.5cm of a0] (a2);
			\vertex[below=1cm of a0] (a3);
			\vertex[left=0.2cm of a1] (a11);
			\vertex[left=0.2cm of a3] (a31);
			\vertex[right=0.1cm of a0] (a10){$\vdots$};
			\vertex[right=0.7cm of a0, blob] (b0){};
			\vertex[right=0.425cm of a0] (a00);
			\vertex[right=1cm of a00, dot] (b00){};
			\vertex[below right=2cm of a00] (b1);
			\vertex[above right=2cm of a00] (b2);
			\vertex[above right=1.55cm of a00] (b22){$\times$};
			\diagram*{
				(p1) -- [double,double distance=0.5ex] (p2),
				(a1) -- [photon] (b0),
				(a2) -- [photon] (b0),
				(a3) -- [photon] (b0),
				(b1) -- [plain] (b00) -- (b2),
			};
			\draw [decoration={brace}, decorate] (a31.south west) -- (a11.north west)
node [pos=0.5, left] {\(N\)};
			\end{feynman}
			\end{tikzpicture}}}
	\sim \mathcal{E}_{\ell m}r^{\ell}\left(\frac{GM}{r}\right)^{N}\times
	\vcenter{\hbox{\begin{tikzpicture}
			\begin{feynman}
			\vertex[blob] (a0){};
			\vertex[above left=0.37cm and 0.147986cm of a0] (a1){$\times$};
			\vertex[above left=0.24cm and 0.32cm of a0] (a2){$\times$};
			\vertex[below left=0.0cm and 0.5cm of a0] (d1){$\cdot$};
			\vertex[below left=0.15cm and 0.47697cm of a0] (d2){$\cdot$};
			\vertex[below left=0.28612cm and 0.41cm of a0] (d3){$\cdot$};
			\vertex[below left=0.37cm and 0.147986cm of a0] (a3){$\times$};
			\vertex[right=0.38cm of a0, dot] (b0){};
			\diagram*{
			};
			\end{feynman}
			\end{tikzpicture}}}
\ee
where $\mathcal{E}_{\ell m}r^{\ell}$, $M^{N}$ and $\left(G/r\right)^{N}$ come from the asymptotic source insertion, graviton-worldline vertices and graviton propagators attached to the worldline respectively. The remaining blob diagram contributes a factor of 
\be
\label{NDA}
	\left({G\over r^2}\right)^L\prod_k \left(\alpha_k\over r^{2(k-1)}\right)^{n_k} \,,
\ee 
where $L$ is a number of loops in the blob and $n_k$  is the the multiplicity of the $\alpha_k$ vertex. The classical contribution to the response, which is our focus here, corresponds to $L=0$. In these expressions we reconstructed the powers of $r$ using the ``naive dimensional analysis" (NDA). NDA correctly reproduces powers of $r$ (but misses the logarithms, which generically arise whenever the EFT diagrams are divergent) provided one is using the mass independent scheme such as the dimensional regularization to regulate the EFT infinities.
To give rise to a non-trivial contribution into the corresponding Love number, the diagram should scale as $r^{-\ell-1}$, which implies that 
\be
\label{eq:Neq}
	N=2\ell+1-2\sum_{k}n_k(k-1)\;.
\ee
A specific case considered in section~\ref{sec:Rcubed}---the leading order calculation in $R^3$ gravity---corresponds to $n_3=1$ with $n_2=0$ and  $n_{k>3}=0$. In this case one also finds $N\geq 2$, given that the $\alpha_3$ vertex has at least three graviton legs so that for tree level diagrams the number of gravitons attached to the worldline can't be smaller than two (this accounts for the fact that one graviton connects to the bulk scalar field vertex). This leaves as with the condition
\[
\ell\geq  3
\]
for the possibility of the source responce mixing, in agreement with the presence of log running as observed in the microscopic calculation in section~\ref{sec:Rcubed}.
As a byproduct, this argument indicates also that there is no source-response mixing at $\ell = 1,2$, so that the values of $\varkappa_{1,2}$ given by \eqref{eq:varkappas} are the actual Love numbers\footnote{Let us recall, however, that in applications of these expressions it may be necessary to account for the non-trivial relation between $r_h$ and the black hole ADM mass in our conventions.}. It is worth mentioning
that the case of gravitational perturbations
in Riemann-cubed gravity was studied in Ref.~\cite{Cai2019},
which showed that the $\ell=2$ Love numbers are not zero without addressing an issue of the source/response separation. A similar analysis and conclusions were made in the gravity theory with the $R^4$ corrections in Ref.~\cite{Cardoso2018}.


It is instructive to contrast our results in $R^3$ gravity with properties of Love numbers in the Einstein theory in $d$ dimensions. There the analog of \eqref{eq:Neq} is
\be
\label{eq:dNeq}
	N\left(d-3\right)=2\ell+d-3\;,
\ee
where we accounted for the fact that in $d$ dimensions the Newton constant $G$ has dimension $mass^{d-2}$ and that the decaying solution of the Laplace equation scales as 
\[
\phi_\ell\propto r^{-\ell-d+3}\;.
\]
Hence, based on the dimensional analysis presented above, one expects to find logarithmic running of the Love numbers whenever \eqref{eq:dNeq} can be satisfied with integer $N$, {\it i.e.} for integer and half-integer values of $\hat{\ell}$. The latter expectation indeed holds, while the former does not. Love numbers 
exhibit running only for half-integer values of $\hat{\ell}$. At integer values of $\hat{\ell}$ one finds vanishing Love numbers instead. 
In particular, at $d=4$ one expects the Love numbers to exhibit running at all values of $\ell$; instead they all vanish identically.

Technically, 
for spin-0 and spin-2 fields, 
the absence of logarithmic running for integer values of $\hat{\ell}$ can be understood within EFT from the special
structure of nonlinear vertices 
in the Einstein action 
in the conveniently chosen
ADM gauge~\cite{Kol:2011vg,Ivanov:2022hlo}.
However, this still leaves one wonder whether this fact itself is indicative of the additional symmetry structure in the microscopic theory, which turns out to be the case as we will see.

%

\subsection{Love numbers vs no-hair theorems}
Vanishing of Love numbers is often linked to another famous property of black holes---no-hair theorems. Let us briefly compare these phenomena, the main point being that the two are in fact quite distinct and different, at least as
far as the discussion of EFT naturalness and fine-tuning go. To make a proper comparison it is important to be precise about what black hole hair are. 
Note also that the no-hair property is often used to distinguish black holes from conventional objects such as stars, rocky planets or scalar field solutions. Again, we will see that to draw such a distinction it is important to be precise in defining the notion of hair.

A simple observation showing that the vanishing of Love numbers
is not a consequence of no-hair 
theorems is 
based on the Bekenstein 
argument~\cite{Bekenstein:1972ky,Cardoso:2016ryw}.
The no-hair theorem in the 
Bekenstein
sense is the statement 
that one cannot ``anchor'' a regular 
and decaying at infinity scalar field 
profile to the black hole geometry.
This statement is true both 
in four and in higher dimensions. 
However, we know that generically Love 
numbers do not
vanish in higher dimensions.

Let us now define hair 
more systematically. 
Possible black hole hair can be divided into three broad categories.
The first type of hair is a situation when a black hole is a member of a family of solutions which have additional continuous parameters on top of mass, spin and gauge charge(s). 
This type of hair 
is sometimes called ``primary hair.''
This primary hair 
match the perturbative 
notion of Bekenstein, as its
presence implies 
the presence of a zero mode,
i.e. 
the possibility to anchor 
an additional field 
to a black hole, see e.g.~\cite{Dubovsky:2007zi}.

The primary hair have to be distinguished from ``secondary hair'', which refer to a situation when black holes support additional fields (typically, scalar fields), which are not associated to any conserved gauge charges. 
An example of this situation
is a theory with a non-minimally
coupled scalar field (e.g.~\cite{Berti:2015itd,Herdeiro:2015waa}), 
which give new black hole
solutions that are different 
from the ones existing 
in the absence of such 
couplings. 
Black holes do not acquire any new continuous parameters in this case, so that the corresponding field profiles,
if they exist, 
are still uniquely fixed by the values of a black hole mass, spin and charge(s). 

Finally, for a proper comparison of black holes to conventional objects, such as rocky planets, it is important to consider also a possibility of ``discrete hair''. This is a situation when at fixed values of mass, spin and gauge charge(s)
one finds a (potentially very large) 
``discretuum'' 
of distinct black hole solutions. 
Similarly to the secondary hair no new continuum parameter is present in this case.
In the 
context of rocky planets, 
there are new parameters characterizing 
the multipolar structure such as 
e.g. ``mountain height''.
Another example is given by 
higher-dimensional black holes,
which are characterized by a ``discrete'' set of topologies~\cite{Hollands:2012xy}.

Let us see now that these three scenarios have very different properties both from the EFT viewpoint and in the microscopic theory, and also as far as the relation to Love numbers is concerned. 
To check the absence (or presence) of primary hair at the level of the microscopic theory one needs to perform a study which is very similar to the calculation of the Love numbers. Indeed, primary hair correspond to time-independent solutions to the Teukolsky equation, which are regular both at the horizon and at the spatial infinity. 

Importantly, the presence of primary hair would be a fine-tuning and would require a symmetry explanation, similarly to the vanishing of Love numbers. In the EFT description primary hair give rise to additional gapless degrees of freedom on the worldline. Also in this description their presence would be indicative of fine-tuning unless some additional symmetry is present. This implies that the absence of black hole primary hair (contrary to the vanishing of Love numbers) is not surprising, and is not that special to black holes. Primary hair are generically absent also for conventional objects, such as rocky planets and solitons. The only peculiar feature of black holes in this respect, is that, unlike for conventional objects, their continuum parameters can only be {\it gauge} charges, {\it i.e.}, black holes are neutral w.r.t. global charges.

In spite of these differences, the absence of primary hair still has some relevance for the properties of Love numbers. Namely, if primary hair were present in a certain sector, it would be impossible to define the corresponding tidal response. Indeed, Love numbers are defined by the decaying tail of the black hole perturbation, which is regular at the horizon. The presence of primary hair would allow to change the decaying tail arbitrarily by adding the corresponding solution. A physical realization of an object with primary hair would be a planet made of something like plastic, which can be  deformed continuously. An object like this may exhibit a hysteresis property, preventing a definition of the tidal response.

One may be puzzled by this discussion given that the absence of hair is often used to distinguish black holes from conventional objects. We already saw one sense in which this is correct, namely, unlike black holes, conventional objects may carry primary hair corresponding to global charges. In addition, it is the possibility of discrete hair which distinguishes conventional objects from black holes in four-dimensional general relativity. Thus, unlike a black hole, a piece of rock of fixed mass and angular momentum (and other charges) may take many different shapes, which cannot be smoothly deformed into each other without breaking the rock into smaller pieces and gluing them back together. However, the absence or presence of such discrete hair is invisible from the viewpoint of the worldline EFT describing each individual shape, and has no bearing on the properties of  Love numbers.

Finally, secondary hair are associated to tadpole couplings in the worldline EFT. Similarly to the vanishing of Love numbers, their absence also appears as tuning in the worldline EFT unless additional symmetries are present. However, secondary hair again has little to do with the properties of Love numbers. Secondary hair is a property of the background solution, while the Love numbers are determined by the behavior of small perturbations.

\section{Love symmetry}
\label{sec:love}

In this section we present 
a hidden symmetry which 
governs dynamics 
of black hole perturbations in the 
near zone approximation.
We call it ``Love symmetry''
because it is this symmetry 
that forces the black hole Love numbers
to vanish.
We will describe this symmetry 
as follows. 
First, we start with a simple case 
of a massless scalar field equation 
in the Schwarzschild 
black hole background, which is known 
to posses an enhanced $\SL$ symmetry in 
the near zone approximation~\cite{Bertini2012}.
We will explicitly show that 
static scalar Schwarzschild 
black hole perturbations belong to highest weight
representations of $\SL$, which 
implies the vanishing of 
scalar Love numbers. 
We will extend this argument
to a massless scalar field in the Kerr-Newman black hole
background,
which posses a similar $\SL$ structure.
Then, we generalize our results 
to generic spin-$s$ fields.
We will present a general
$\SL$ symmetry which addresses the vanishing of
scalar ($s=0$), electromagnetic ($s=\pm1$), 
and gravitational ($s=\pm2$) Love numbers 
of Kerr-Newman black holes in four dimensions. 
Finally,
we will show how the Love symmetry
of a scalar wave equation
can be generalized to higher dimensions.

\subsection{Scalar perturbations of Schwarzschild black holes}

Consider a massless scalar field in
the Schwarzschild black hole background,
\be
	\varphi = \Phi\left(t,r,\phi\right) S\left(\theta\right) =  e^{-i\omega t}e^{im\phi} R\left(r\right) S\left(\theta\right) \,.
\ee
The radial wavefunction $\Phi$ satisfies the following 
Klein-Gordon (equivalently, spin-$0$ Teukolsky) equation
\be
	\left(\d_r \Delta \d_r +V_0+\epsilon V_1\right)\Phi = \ell(\ell+1)\Phi\,,
\ee
where $V_0$ and $V_1$ are given by \eqref{eq:V0V1} with $s=0$ and $a=0$.
Consider now the following vector fields~\cite{Bertini2012,Charalambous:2021kcz},
\be\label{eq:SL2RSchwarzschild4}
	\begin{split}
		L_0 = -\beta\,\partial_{t} \,,\quad L_{\pm1} = e^{\pm t/\beta}\left[\mp\sqrt{\Delta}\,\partial_{r} + \partial_{r}\left(\sqrt{\Delta}\right)\beta\,\partial_{t}\right] \,,
	\end{split}
\ee
where $\beta$ is the inverse Hawking temperature of the Schwarzschild black hole,
\be
	\beta = \frac{1}{2\pi T_{H}} = 2r_{s} \,.
\ee
By transforming 
these vectors into advanced/retarded Eddington-Finkelstein coordinates, one can check that they are regular both at the future 
and at the past event horizons (see Appendix~\ref{app:adv} for more details).
Also, they obey the $\SL$ algebra commutation relations,
\be\label{eq:SL2RAlgebra}
	\left[L_{m},L_n\right] = \left(m-n\right)L_{m+n}\,,\quad m,n=-1,0,+1\,.
\ee
The quadratic Casimir of this algebra reproduces a differential operator of the Klein-Gordon 
equation in the near zone ($\epsilon=0$),
\be
	\mathcal{C}_2 = L_0^2 - \frac{1}{2}\left(L_{+1}L_{-1}+L_{-1}L_{+1}\right) = \partial_{r}\Delta\partial_{r} - \frac{r_{s}^4}{\Delta}\,\partial_{t}^2 = \mathbb{O}^{\left(s=0\right)}_{\text{NZ}}\,.
\ee
The solution of this equation $\Phi$
is an eigenvector of both $L_0$ and $\mathcal{C}_2 $\,,
\be
\begin{split}
	L_0 \Phi = h\Phi= i\omega\beta \Phi\,,\quad \mathcal{C}_2 \Phi = \ell(\ell+1)\Phi\,,
\end{split}
\ee
where $\ell$ is an integer number by virtue of the static angular eigenvalue problem, namely, $\ell$ is the orbital number label of the spherical harmonics.
Therefore, all solutions $\Phi$ form representations of $\SL$. Moreover, as a consequence of the regularity of the vector fields $L$'s, solutions regular at the horizon are closed under the $\SL$ action.  

This is a powerful statement, because now we can 
derive many properties of black hole perturbations
from group theory arguments.
In particular, let us demonstrate
how the Love symmetry implies the vanishing 
of static scalar Love numbers $k_{\ell}^{\left(0\right)}$. These are independent of the azimuthal number $m$ due to 
spherical symmetry of the background, so in this section we set $m=0$ without loss of generality.
They are extracted from the static solution, which is 
a null-vector of $L_0=-\beta\,\d_t$. This solution belongs 
to a highest weight representation of $\SL$\footnote{See Appendix~\ref{app:sl2r} for a brief review of the indecomposable $\SL$ representations.}. To see this, let us 
explicitly construct this representation. 
It is generated by a highest weight (primary) vector $\upsilon_{-\ell,0}$ with weight $h=-\ell$,
\be
	L_{+1}\upsilon_{-\ell,0} = 0\,,\quad L_0\upsilon_{-\ell,0} = -\ell\upsilon_{-\ell,0}\,.
\ee
Up to an overall normalization factor, this solution is given by
\be
	\upsilon_{-\ell,0} = e^{\ell t/\beta}\Delta^{\ell/2}\,.
\ee
This function solves the near zone massless Klein-Gordon equation for multipolar order $\ell$ with an imaginary frequency $\omega_{-\ell,0}=i\ell/\beta$ and is regular both at the future and at the past event horizons. 
Since the generators $L_{\pm1}$ are regular on the horizon, all the descendants,
\be
	\upsilon_{-\ell,n} = \left(L_{-1}\right)^{n}\upsilon_{-\ell,0}
\ee
are also regular solutions of the massless Klein-Gordon equation, now with frequency $\omega_{-\ell,n}=i\left(\ell-n\right)/\beta$. In particular, we immediately see that the physical static solution with zero frequency is an element of this highest weight representation,
\be
	\Phi \left(\omega=0\right) \propto \upsilon_{-\ell,\ell} = \left(L_{-1}\right)^{\ell}\upsilon_{-\ell,0}\,.
\ee
As such, it must be annihilated by $\left(L_{+1}\right)^{\ell+1}$. From the explicit action of the vector fields \eqref{eq:SL2RSchwarzschild4}, we observe that,
for any arbitrary radial function $F\left(r\right)$,
\be
	\left(L_{+1}\right)^{n}F\left(r\right) = \left(-e^{t/\beta}\sqrt{\Delta}\right)^{n}\frac{d^{n}}{dr^{n}}F\left(r\right)\,.
\ee
For the static solution $\upsilon_{-\ell,\ell} = F\left(r\right)$, the annihilation condition then reads,
\be
	\left(L_{+1}\right)^{\ell+1}\upsilon_{-\ell,\ell} = \left(-e^{t/\beta}\sqrt{\Delta}\right)^{\ell+1}\frac{d^{\ell+1}}{dr^{\ell+1}}\upsilon_{-\ell,\ell} = 0\,,
\ee
implying that the physical static solution must be a degree-$\ell$ polynomial in $r$,
\be
	\upsilon_{-\ell,\ell}
	 = \sum_{n=0}^{\ell}c_{n}r^{n} = c_{\ell}r^{\ell} + \dots + c_1r + c_0\,.
\ee
Clearly, this solution does not have terms with decaying powers $\propto r^{-\ell-1}$, which is precisely the condition for the vanishing of static Love numbers.

\begin{figure}[!ht!]
	\centering
	\begin{tikzpicture}
		\node at (0,0) (uml4) {$\upsilon_{-\ell,2\ell}$};
		\node at (0,1) (uml3) {$\upsilon_{-\ell,\ell}$};
		\node at (0,2) (uml2) {$\upsilon_{-\ell,2}$};
		\node at (0,3) (uml1) {$\upsilon_{-\ell,1}$};
		\node at (0,4) (uml0) {$\upsilon_{-\ell,0}$};
		
		\draw [snake=zigzag] (1,-0.1) -- (5,-0.1);
		\draw (1,0) -- (5,0);
		\draw (1,1) -- (5,1);
		\node at (3,1.5) (up) {$\vdots$};
		\node at (3,0.5) (um) {$\vdots$};
		\draw (1,2) -- (5,2);
		\draw (1,3) -- (5,3);
		\draw (1,4) -- (5,4);
		\draw [snake=zigzag] (1,4.1) -- (5,4.1);
		
		\draw[blue] [->] (2.5,2) -- node[left] {$L_{+1}$} (2.5,3);
		\draw[blue] [->] (2,3) -- node[left] {$L_{+1}$} (2,4);
		\draw[red] [<-] (4,3) -- node[right] {$L_{-1}$} (4,4);
		\draw[red] [<-] (3.5,2) -- node[right] {$L_{-1}$} (3.5,3);
	\end{tikzpicture}
	\caption{The finite-dimensional highest weight representation of $\SL$ whose elements solve the near zone equations of motion for a massless scalar field in the Schwarzschild black hole background with multipolar index $\ell$ and contains the regular static solution.}
	\label{fig:HWSL2RSchwarzschild}
\end{figure}
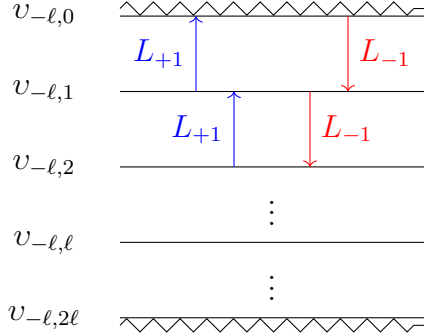

Note that in the case of a scalar field in the 
four-dimensional Schwarzschild black hole background one can arrive
at the same conclusion starting 
from a lowest weight vector.
Indeed, we can construct the lowest weight vector of weight $h=+\ell$,
\be
	L_{-1}\bar{\upsilon}_{+\ell,0} = 0\,,\quad L_0\bar{\upsilon}_{+\ell,0} = +\ell\bar{\upsilon}_{+\ell,0} \quad \Rightarrow \quad \bar{\upsilon}_{+\ell,0} = e^{-\ell t/\beta}\Delta^{\ell/2} \,,
\ee
which is also a solution of the massless Klein-Gordon equation with multipolar index $\ell$ that is regular on both the future and past event horizons, but this time with the frequency $\bar{\omega}_{+\ell,0}=-i\ell/\beta$. A regular static solution would then be identified as the particular ascendant with zero $L_0$-eigenvalue,
\be
	\Phi \left(\omega=0\right) \propto \bar{\upsilon}_{+\ell,-\ell} = \left(L_{+1}\right)^{\ell}\bar{\upsilon}_{+\ell,0}
\ee
By the uniqueness of the regular solution, this implies that the highest and lowest weight representations are actually the same, i.e. this is a finite $\left(2\ell+1\right)$-dimensional representation of $\SL$, and consequently
\be
	\bar{\upsilon}_{+\ell,0} = \upsilon_{-\ell,2\ell}\,.
\ee
Our construction of the highest weight representation of $\SL$ in the near zone equations of motion for the scalar perturbations of Schwarzschild black hole is summarized in Fig.~\ref{fig:HWSL2RSchwarzschild}.

The fact that the highest weight and lowest weight representations of $\SL$
coincide is unique to the Schwarzschild background and to spin-$0$ fields. 
We will see momentarily that
for the Kerr black holes, and also 
for spin-$s$ fields in the Schwarzschild metric, 
this is not the case.
In these more general cases the highest weight representation contains solutions that are regular on the physical future event horizon, 
while the lowest weight representation contains physically irrelevant solutions that are regular on the past event horizon and singular on the future event horizon.

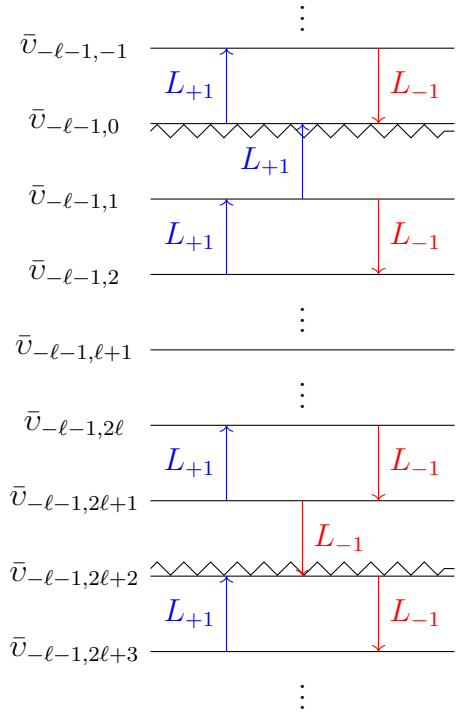
\begin{figure}[!ht!]
	\centering
	\begin{tikzpicture}
		\node at (0,-1) (uml8) {$\bar{\upsilon}_{-\ell-1,2\ell+3}$};
		\node at (0,0) (uml7) {$\bar{\upsilon}_{-\ell-1,2\ell+2}$};
		\node at (0,1) (uml6) {$\bar{\upsilon}_{-\ell-1,2\ell+1}$};
		\node at (0,2) (uml5) {$\bar{\upsilon}_{-\ell-1,2\ell}$};
		\node at (0,3) (uml4) {$\bar{\upsilon}_{-\ell-1,\ell+1}$};
		\node at (0,4) (uml3) {$\bar{\upsilon}_{-\ell-1,2}$};
		\node at (0,5) (uml2) {$\bar{\upsilon}_{-\ell-1,1}$};
		\node at (0,6) (uml1) {$\bar{\upsilon}_{-\ell-1,0}$};
		\node at (0,7) (uml0) {$\bar{\upsilon}_{-\ell-1,-1}$};
		
		\node at (3,-1.5) (umm) {$\vdots$};
		\draw (1,-1) -- (5,-1);
		\draw [snake=zigzag] (1,0.1) -- (5,0.1);
		\draw (1,0) -- (5,0);
		\draw (1,1) -- (5,1);
		\draw (1,2) -- (5,2);
		\node at (3,2.5) (um) {$\vdots$};
		\draw (1,3) -- (5,3);
		\node at (3,3.5) (up) {$\vdots$};
		\draw (1,4) -- (5,4);
		\draw (1,5) -- (5,5);
		\draw (1,6) -- (5,6);
		\draw [snake=zigzag] (1,5.9) -- (5,5.9);
		\draw (1,7) -- (5,7);
		\node at (3,7.5) (upp) {$\vdots$};
		
		\draw[blue] [->] (2,-1) -- node[left] {$L_{+1}$} (2,0);
		\draw[red] [<-] (4,-1) -- node[right] {$L_{-1}$} (4,0);
		\draw[red] [<-] (3,0) -- node[right] {$L_{-1}$} (3,1);
		\draw[blue] [->] (2,1) -- node[left] {$L_{+1}$} (2,2);
		\draw[red] [<-] (4,1) -- node[right] {$L_{-1}$} (4,2);
		\draw[blue] [->] (2,4) -- node[left] {$L_{+1}$} (2,5);
		\draw[red] [<-] (4,4) -- node[right] {$L_{-1}$} (4,5);
		\draw[blue] [->] (3,5) -- node[left] {$L_{+1}$} (3,6);
		\draw[blue] [->] (2,6) -- node[left] {$L_{+1}$} (2,7);
		\draw[red] [<-] (4,6) -- node[right] {$L_{-1}$} (4,7);
	\end{tikzpicture}
	\caption{The infinite-dimensional representation of $\SL$ of whose elements solve the near zone equations of motion for a massless scalar field in the Schwarzschild black hole background with multipolar index $\ell$ and contain the singular static solution.}
	\label{fig:SingSL2RSchwarzschild}
\end{figure}

In the Schwarzschild $s=0$ case, instead, a static solution that is singular on the event horizon belongs to the representation shown in Fig. \ref{fig:SingSL2RSchwarzschild}. This is spanned by vectors $\tilde{\upsilon}_{-\ell,m}$, $m\in\mathbb{Z}$. The upper part of the ladder is constructed by ascendants of the lowest weight state $\tilde{\upsilon}_{-\ell-1,0}$ with weight $h=-\ell-1$,
\be
	L_{-1}\tilde{\upsilon}_{-\ell-1,0} = 0\,,\,\, L_0\tilde{\upsilon}_{-\ell-1,0} 
	= -\left(\ell+1\right)\tilde{\upsilon}_{-\ell-1,0} \,\, \Rightarrow \,\, \tilde{\upsilon}_{-\ell-1,0} = e^{\left(\ell+1\right) t/\beta}\Delta^{-\frac{\ell+1}{2}} \,.
\ee
All the elements in this upper ladder are in fact regular at the event horizon with frequencies $\tilde{\omega}_{-\ell-1,-n} = i\left(\ell+1+n\right)/\beta$. Nevertheless, singular solutions exist in the middle part of the ladder, constructed by requiring that $\tilde{\upsilon}_{-\ell-1,0}$ is itself an ascendant of $\tilde{\upsilon}_{-\ell-1,1}$, a condition that gives an inhomogeneous first-order differential equation to solve,
\be
	\tilde{\upsilon}_{-\ell-1,0} \propto L_{+1}\tilde{\upsilon}_{-\ell-1,1} \ne 0 \quad \Rightarrow \quad \tilde{\upsilon}_{-\ell-1,1} \propto e^{\ell t/\beta}\Delta^{-\ell/2} \ln\frac{r-r_{s}}{r} \,.
\ee
Clearly, $\tilde{\upsilon}_{-\ell-1,1}$ is singular at the horizon\footnote{We are ignoring here an irrelevant additive piece which is regular at the event horizon and is annihilated by $L_{+1}$. This freedom reflects the fact that, starting from one particular singular solution, we can always construct another (linearly dependent) singular solution by adding to the profile a regular solution.}. The subsequent descendants will then also be singular, up until $\tilde{\upsilon}_{-\ell-1,2\ell+2}$ beyond which we enter the lower part of the ladder. At that step, $\tilde{\upsilon}_{-\ell-1,2\ell+2}$ becomes a highest weight vector of weight $h=+\ell+1$. Then, all $\tilde{\upsilon}_{-\ell-1,2\ell+2+n}$, $n\ge0$ are regular descendants with frequencies $\tilde{\omega}_{-\ell-1,2\ell+2+n} = -i\left(\ell+1+n\right)/\beta$. The region of interest of course is the middle part of the ladder, spanned by the singular at the horizon vectors $\tilde{\upsilon}_{-\ell-1,n+1}$, $n=0,\dots,2\ell$ which have frequencies $\omega_{-\ell-1,n+1} = i\left(\ell-n\right)/\beta$. The singular static solution is then identified as the state $\tilde{\upsilon}_{-\ell-1,\ell+1}$ and the structure of the representation implies,
\be
	L_{-1}\left(L_{+1}\right)^{\ell+1}\tilde{\upsilon}_{-\ell-1,\ell+1} = 0 \Rightarrow \frac{d}{dr}\left(\Delta^{\ell+1}\frac{d^{\ell+1}}{dr^{\ell+1}}\tilde{\upsilon}_{-\ell-1,\ell+1}\right) = 0 \,.
\ee
This is indeed the condition satisfied by the singular at the horizon static solution as can be checked by its explicit expression in terms of Legendre polynomials of the second kind.

Finally, let us briefly comment
on the massive wave equation
with mass $\mu$. 
In the regime 
$\m\left(r-r_s\right)\ll 1$, and 
$\mu M\ll 1$,
it has the
following form in the 
Schwarzschild black hole 
near zone approximation~\cite{Bertini2012},
\be
	\left[\d_r  \Delta \d_r  -\frac{r_{s}^4}{\Delta}\,\d_t^2 \right] \Phi = \left[\ell(\ell+1)+\mu^2 r_s^2\right] \Phi \,.
\ee
The mass changes the eigenvalue of the Love symmetry
Casimir such that the physical 
static solution $\Phi\left(\omega=0\right)$
regular at the horizon
does not belong to the 
highest weight $\SL$ representation anymore.
A direct calculation shows that Love numbers
are not zero. At the leading order in  $\mu^2r_s^2$, 
they take constant values without any running.
This example illustrates that the presence of the Love symmetry alone is not enough to ensure 
the vanishing of Love numbers. The crucial role is played by the highest weight property of the corresponding representations.

\subsection{Scalar perturbations of Kerr-Newman black holes}

The generators of the Love symmetry in the Kerr-Newman black hole background 
have the form
\be\label{eq:SL2RKerr}
	L_0 = -\beta\,\partial_{t} \,,\quad L_{\pm1} = e^{\pm t/\beta}\left[\mp\sqrt{\Delta}\,\partial_{r} + \partial_{r}\left(\sqrt{\Delta}\right)\beta\,\partial_{t} + \frac{a}{\sqrt{\Delta}}\,\partial_{\phi}\right]\,,
\ee
where the inverse Hawking temperature of the Kerr black hole is now given by
\be
	\beta = \frac{1}{2\pi T_{H}} = 2\frac{r_{+}^2+a^2}{r_{+}-r_{-}}\;.
\ee
Note that
these  generators are singular in the extremal limit 
$r_{-}\to r_{+}$, which also corresponds to the vanishing Hawking temperature. This limit will be discussed separately.
The new generators also satisfy the $\SL$ algebra \eqref{eq:SL2RAlgebra} while the resulting Casimir operator again coincides with a near zone Teukolsky differential operator
\be
	\mathcal{C}_2 = \partial_{r}\Delta\partial_{r} - \frac{\left(r_{+}^2+a^2\right)^2}{\Delta}\left[\left(\partial_{t}+\Omega\,\partial_{\phi}\right)^2 + 4\Omega\frac{r-r_{+}}{r_{+}-r_{-}}\partial_{t}\partial_{\phi}\right] = \mathbb{O}_{\text{NZ}}^{\left(s=0\right)}\,.
\ee
As in the Schwarzschild case, the Love vector fields are regular on both future and past event horizons as can be seen by transforming them to the advanced/retarded Kerr coordinates  (see Appendix~\ref{app:adv} for more details).

In what follows we assume that the scalar
black hole perturbation $\Phi$
carries an 
arbitrary azimuthal quantum number
$m$, which is an eigenvalue of
the $U\left(1\right)$ axial rotation symmetry,
\be
	J_0\Phi
	=-i\partial_{\phi}\Phi
	=m \Phi
	\,.
\ee
Note that the azimuthal $U\left(1\right)$ commutes with the Love symmetry.
Now we can show that Kerr-Newman black hole Love numbers 
also vanish as a result of the Love symmetry. 
Most of the argument can be straightforwardly 
borrowed from the Schwarzschild case.
The highest weight vector with $h=-\ell$ is defined by,
\be
\label{eq:HWkerr}
	L_{+1}\upsilon_{-\ell,0}^{\left(m\right)}=0\,,\quad L_0\upsilon_{-\ell,0}^{\left(m\right)} = -\ell\upsilon_{-\ell,0}^{\left(m\right)}\,.
\ee
Solving Eq.~\eqref{eq:HWkerr}, we obtain (up to an arbitrary normalization constant)
\be
	\upsilon_{-\ell,0}^{\left(m\right)} = \left(\frac{r-r_{+}}{r-r_{-}}\right)^{im\gamma}e^{im\phi}e^{\ell t/\beta}\Delta^{\ell/2} \,,\quad \gamma\equiv\frac{a}{r_{+}-r_{-}} \,.
\ee
This solves the (radial) near zone Teukolsky equation for multipolar order $\ell$, with imaginary frequency $\omega_{-\ell,0}=i\ell/\beta$. 
An important difference w.r.t. the Schwarzschild case is the regularity condition. 
The solution  $\upsilon_{-\ell,0}^{\left(m\right)}$ is still regular on the physical future event horizon, but it is now singular on the past event horizon. As a result all its descendants,
\be
	\upsilon_{-\ell,n}^{\left(m\right)}=\left(L_{-1}\right)^{n}\upsilon_{-\ell,0}^{\left(m\right)}\,,
\ee
are also solutions of equations of motion regular on the future event horizon. On the other hand, for a lowest weight vector with $h=+\ell$ we have
\be
	\begin{gathered}
		L_{-1}\bar{\upsilon}_{+\ell,0}^{\left(m\right)}=0
		\,,\quad 
		L_0\bar{\upsilon}_{+\ell,0}^{\left(m\right)} = +\ell\bar{\upsilon}_{+\ell,0}^{\left(m\right)}\,, \\
		\Rightarrow \bar{\upsilon}_{+\ell,0}^{\left(m\right)} = \left(\frac{r-r_{+}}{r-r_{-}}\right)^{-im\gamma}e^{im\phi}e^{-\ell t/\beta}\Delta^{\ell/2}\,.
	\end{gathered}
\ee
This solution is regular at the past event horizon, but singular at the future horizon. It gives rise to an ascending tower of solutions that are regular on the past event horizon.

\begin{figure}
	\centering
	\begin{subfigure}[b]{0.49\textwidth}
		\centering
		\begin{tikzpicture}
		\node at (0,1) (uml3) {$\upsilon_{-\ell,\ell}$};
		\node at (0,2) (uml2) {$\upsilon_{-\ell,2}$};
		\node at (0,3) (uml1) {$\upsilon_{-\ell,1}$};
		\node at (0,4) (uml0) {$\upsilon_{-\ell,0}$};
		
		\draw (1,1) -- (5,1);
		\node at (3,1.5) (up) {$\vdots$};
		\node at (3,0.5) (um) {$\vdots$};
		\draw (1,2) -- (5,2);
		\draw (1,3) -- (5,3);
		\draw (1,4) -- (5,4);
		\draw [snake=zigzag] (1,4.1) -- (5,4.1);
		
		\draw[red] [<-] (2.5,2) -- node[left] {$L_{-1}$} (2.5,3);
		\draw[red] [<-] (2,3) -- node[left] {$L_{-1}$} (2,4);
		\draw[blue] [->] (4,3) -- node[right] {$L_{+1}$} (4,4);
		\draw[blue] [->] (3.5,2) -- node[right] {$L_{+1}$} (3.5,3);
		\end{tikzpicture}
		\caption{A highest weight $\SL$ representation that contains perturbations of Kerr-Newman black holes that are regular on the future event horizon.}
	\end{subfigure}
	\hfill
	\begin{subfigure}[b]{0.49\textwidth}
		\centering
		\begin{tikzpicture}
		\node at (0,3) (upll) {$\bar{\upsilon}_{+\ell,\ell}$};
		\node at (0,2) (upl2) {$\bar{\upsilon}_{+\ell,2}$};
		\node at (0,1) (upl1) {$\bar{\upsilon}_{+\ell,1}$};
		\node at (0,0) (upl0) {$\bar{\upsilon}_{+\ell,0}$};
		
		\draw [snake=zigzag] (1,-0.1) -- (5,-0.1);
		\draw (1,0) -- (5,0);
		\draw (1,1) -- (5,1);
		\draw (1,2) -- (5,2);
		\node at (3,2.5) (up) {$\vdots$};
		\draw (1,3) -- (5,3);
		\node at (3,3.5) (um) {$\vdots$};
		
		\draw[blue] [->] (2.5,0) -- node[left] {$L_{+1}$} (2.5,1);
		\draw[blue] [->] (2,1) -- node[left] {$L_{+1}$} (2,2);
		\draw[red] [<-] (4,1) -- node[right] {$L_{-1}$} (4,2);
		\draw[red] [<-] (3.5,0) -- node[right] {$L_{-1}$} (3.5,1);
		\end{tikzpicture}
		\caption{A lowest weight $\SL$ representation containing perturbations of Kerr-Newman black holes that are regular on the past event horizon.}
	\end{subfigure}
	\caption{The infinite-dimensional highest and lowest weight representations of $\SL$ whose elements solve the near zone equations of motion for a massless scalar field in the Kerr-Newman black hole background with multipolar index $\ell$ and contain the static solutions.}
	\label{fig:HWSL2RKerr}
\end{figure}
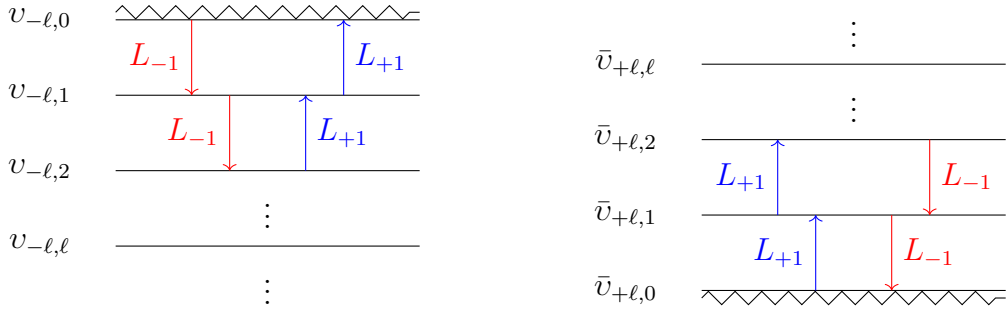

A physical black hole formed as a result of a collapse does not exhibit the past event horizon. Hence, we are interested in the solutions which are regular at the future event horizon, and their singularity at the past horizon does not pose a problem. This singles out the highest-weight representation. As follows from the above discussion, it is now infinite-dimensional, falling into the general category of Verma modules, see Fig.~\ref{fig:HWSL2RKerr}.

As before, the physical static solution is identified with the null $L_0$-eigenstate in the highest weight representation,
\be
	\Phi\left(\omega=0\right) \propto \upsilon_{-\ell,\ell}^{\left(m\right)} = \left(L_{-1}\right)^{\ell}\upsilon_{-\ell,0}^{\left(m\right)}\;.
\ee
Observing that,
\be
	\left(L_{+1}\right)^{n}\left[ \left(\frac{r-r_{+}}{r-r_{-}}\right)^{im\gamma}e^{im\phi}F\left(r\right)\right] = \left(\frac{r-r_{+}}{r-r_{-}}\right)^{im\gamma}e^{im\phi} \left(-e^{t/\beta}\sqrt{\Delta}\right)^{n} \frac{d^{n}}{dr^{n}}F\left(r\right)
\ee
for arbitrary $F\left(r\right)$ and $n$, the annihilation condition for the static case ($n=\ell+1$) implies a polynomial form for $F\left(r\right)$,
\be
	\begin{gathered}
		\upsilon_{-\ell,\ell}^{\left(m\right)} =\left(\frac{r-r_{+}}{r-r_{-}}\right)^{im\gamma}e^{im\phi}F\left(r\right) \,,\\
		\left(L_{+1}\right)^{\ell+1}\upsilon_{-\ell,\ell}^{\left(m\right)} = \left(\frac{r-r_{+}}{r-r_{-}}\right)^{im\gamma}e^{im\phi}\left(-e^{t/\beta}\sqrt{\Delta}\right)^{\ell+1}
		\frac{d^{\ell+1}}{dr^{\ell+1}}F\left(r\right) = 0 \,,\\
		\Rightarrow \quad 
		\Phi\left(\omega = 0\right) = \left(\frac{r-r_{+}}{r-r_{-}}\right)^{im\gamma}e^{im\phi}\sum_{n=0}^{\ell}c_{n}r^{n}\,.
	\end{gathered}
\ee
While this does not look like a pure polynomial solution due the overall $r$-dependent factor, as discussed in section \ref{sec:master}, this form-factor can be attributed to frame-dragging and does not affect Love numbers. A polynomial form of the solution apart of this factor indicates that the highest weight property ensures the vanishing of the Love numbers also in this case.

It is natural to ask whether other solutions which enter in the same multiplets with the static ones have any physical interpretation. 
In the Schwarzschild case the corresponding non-vanishing imaginary frequencies are given by
\[
	\omega_n= i\,2\pi T_H n \;.
\]
Interestingly, the spacing between these modes matches the spacing of 
highly-damped quasinormal modes (QNMs) given by~\cite{Nollert:1993zz}
\be
	\omega_k =\frac{i}{4M}\left(k+{1}/{2}\right) = i\,2\pi T_H (k +1/2) \,. 
\ee
However, while the spacing of states 
in the Kerr-Newman Love highest weight multiplet 
is also $2\pi T_H$, 
the spacing of highly-damped 
QNMs actually does not 
match $2\pi T_H$~\cite{Berti:2004um,Berti:2009kk}.\footnote{In the Kerr case, 
early works
suggested that
the spacing of the imaginary part 
of the highly-damped 
quasinormal modes 
matches $2\pi T_H$, but only 
for modes with the positive magnetic 
number $m>0$~\cite{Hod:2003jn,Berti:2003jh}. Later studies refuted these results~\cite{Berti:2004um,Berti:2009kk}.}
On the other hand, the spacing of the so-called
anomalous total transmission modes
is indeed equal to $2\pi T_H$
for the Kerr black holes~\cite{Cook:2016fge,Cook:2016ngj}. It will be interesting to explore whether this is more than a sheer coincidence.
%


\subsection{General perturbations of Kerr-Newman black holes}
Let us now present the most general case of the four-dimensional Kerr-Newman black hole perturbations with 
an arbitrary integer spin weight $s$. The generators of the Love $\SL$ algebra 
are not spacetime vectors anymore: they acquire a scalar part proportional to the spin weight of the perturbation~\cite{Charalambous:2021kcz},
\be\label{eq:SL2RsKerrNewman}
	\begin{gathered}
		L_0^{(s)} = -\beta\,\partial_{t} + s \\
		L_{\pm1}^{(s)} = e^{\pm t/\beta}\left[\mp\sqrt{\Delta}\,\partial_{r} + \partial_{r}\left(\sqrt{\Delta}\right)\beta\,\partial_{t} + \frac{a}{\sqrt{\Delta}}\,\partial_{\phi} - s\left(1\pm1\right)\partial_{r}\left(\sqrt{\Delta}\right) \right]\,.
	\end{gathered}
\ee
The Casimir of the algebra spanned by the above generators
is given in Eq.~\eqref{eq:NZgen}-\eqref{eq:V0V1}.
These additional $s$-dependent
parts are singular at the horizon. 
However, this singular behavior is an artifact 
related to 
the use of the Kinnersley tetrads~\cite{Kinnersley1969} 
that are singular at the horizon themselves.
The relevant spin-$s$ scalars
are regular at the horizon 
for the appropriate well-behaved 
choice of coordinates and tetrads~\cite{Teukolsky1972}.
Another important difference w.r.t. the spin-$0$ case is that the static solution now has a non-zero $L_0$ eigenvalue,
\be
	L_0 \Psi_{s}\left(\omega=0\right)= s\,\Psi_{s}\left(\omega=0\right)\,.
\ee
Let us show now that this state belongs 
to a highest weight representation 
with weight $-\ell$ (recall that  $\ell\geq |s|$).
We start with a highest weight vector $\upsilon_{-\ell,0}^{\left(m\right)}$ of azimuthal number $m$ that satisfies
\be 
	L_{+1}\upsilon_{-\ell,0}^{\left(m\right)}=0\,\,\,,\,\,\,L_0\upsilon_{-\ell,0}^{\left(m\right)} = -\ell\upsilon_{-\ell,0}^{\left(m\right)} \,.
\ee
It is straightforward to integrate these equations
and to obtain
\be
	\upsilon_{-\ell,0}^{\left(m\right)} = \left(\frac{r-r_{+}}{r-r_{-}}\right)^{im\gamma}e^{im\phi}e^{\left(\ell+s\right) t/\beta}\Delta^{\left(\ell-s\right)/2}\,.
\ee
By transforming this vector into the advanced coordinates we see that it is regular on the future horizon, but singular on the past one. Now it is instructive to rewrite $L^{(s)}_{+1}$ as
\be
	L^{(s)}_{+1}= L_{+1}-2s e^{t/\beta}\d_r(\Delta^{1/2})\,,
\ee
where $L_{+1}$ is the vector in the scalar field case,
which is manifestly regular at the horizon.


A Newman-Penrose scalar $\Psi_s$ that is regular at the future event horizon must have the form 
$\Delta^{-s}f$ with $f$ being a function that is regular on the future event horizon~\cite{Teukolsky1972,Teukolsky1973}. For any function $f$ we have
\be
	L^{(s)}_{+1} \Delta^{-s}f  = \Delta^{-s}L_{+1}f \,,\quad L^{(s)}_{-1} \Delta^{-s}f = \Delta^{-s} \left[L_{-1}-2s\,e^{-t/\beta}\partial_{r}\left(\sqrt{\Delta}\right)\right]f \,.
\ee
As $L_{\pm1} f$ is regular for regular $f$, any vector of the form 
$\Delta^s(L^{(s)}_{\pm 1})^n[\Delta^{-s}f]$
is regular too. 
This way, acting on 
$\upsilon_{-\ell,0}^{\left(m\right)}$
with $L^{(s)}_{\pm1}$, we will 
get new regular solutions
thereby
generating a 
multiplet containing
solutions to the Teukolsky equation 
that are regular at the future event horizon.
The descendant with $L_0$-eigenvalue equal to $s$ is then identified with the regular static solution,
\be
	\Psi_{s}\left(\omega=0\right) \propto \upsilon_{-\ell,\ell+s}^{\left(m\right)} = \left(L^{(s)}_{-1}\right)^{\ell+s}\upsilon_{-\ell,0}^{\left(m\right)} \;.
\ee
This state satisfies the annihilation condition 
$(L_{+1}^{(s)})^{\ell+s+1}\upsilon_{-\ell,\ell+s}
=L_{+1}^{(s)}\upsilon_{-\ell,0}=0$. Noting that, for any $n$ and $F\left(r\right)$,
\be
	\begin{split}
		&\left(L^{(s)}_{+1}\right)^{n}\left[\left(\frac{r-r_{+}}{r-r_{-}}\right)^{im\gamma}e^{im\phi} \Delta^{-s} F\left(r\right)\right] \\
		&= \left(\frac{r-r_{+}}{r-r_{-}}\right)^{im\gamma}e^{im\phi} \Delta^{-s}\left(-e^{t/\beta}\sqrt{\Delta}\right)^{n} \frac{d^{n}}{dr^{n}}F\left(r\right) \,,
	\end{split}
\ee
the physical static solution 
$\upsilon_{-\ell,\ell+s}^{\left(m\right)}=\left(\frac{r-r_{+}}{r-r_{-}}\right)^{im\gamma}e^{im\phi}\Delta^{-s}F\left(r\right)$ 
satisfies
\be
	\left(L_{+1}\right)^{\ell+s+1}\upsilon_{-\ell,\ell+s}^{\left(m\right)} = \left(\frac{r-r_{+}}{r-r_{-}}\right)^{im\gamma}e^{im\phi} \Delta^{-s}\left(-e^{t/\beta}\sqrt{\Delta}\right)^{\ell+s+1} \frac{d^{\ell+s+1}}{dr^{\ell+s+1}}F\left(r\right) = 0\,,
\ee
which implies that $F\left(r\right)$ is polynomial of order $\ell+s$. Consequently, the static solution is given by
\be
	\Psi_{s}\left(\omega=0\right) = \left(\frac{r-r_{+}}{r-r_{-}}\right)^{im\gamma}e^{im\phi}\Delta^{-s}\sum_{n=0}^{\ell+s}c_{n}r^{n}\,.
\ee
From this expression we conclude
that black hole Love numbers w.r.t. general 
spin-$s$ perturbations vanish.

\subsection{Scalar perturbations of higher dimensional Schwarzschild black holes}

The Love symmetry is also present in 
higher dimensions, which was first pointed out in Ref.~\cite{Bertini2012}. The $d$-dimensional Schwarzschild black hole background geometry is given by,
\be
	\begin{split}
		ds^2 = -f\left(r\right)dt^2 + \frac{dr^2}{f\left(r\right)} + r^2d\Omega_{d-2}^2 \,,\quad f\left(r\right) = 1 - \left(\frac{r_{s}}{r}\right)^{d-3}\,,
	\end{split}
\ee
where $d\Omega_{d-2}^2$ is the line element on the hypersphere $\mathbb{S}^{d-2}$ and the Schwarzschild radius is related to the ADM mass $M$ of the black hole as,
\be
	r_{s}^{d-3} = \frac{16\pi GM}{\left(d-2\right)\Omega_{d-2}} \,.
\ee
A massless scalar field propagating in this background geometry obeys the following Klein-Gordon equation,
\be
\label{eq:KGhigh}
	g^{\mu\nu}\nabla_{\mu}\nabla_{\nu}\varphi = \left[\frac{1}{r^{d-2}}\partial_{r}\left(r^{d-2}f\partial_{r}\right) - \frac{1}{f}\partial_{t}^2 + \frac{1}{r^2}\Delta_{\mathbb{S}^{d-2}}\right]
	\varphi = 0 \,.
\ee
We separate the variables as follows,
\be
	\varphi  = \Phi\left(t,r\right) S\left(\Theta\right)= e^{-i\omega t}R
	\left(r\right) S\left(\Theta\right)\,,
\ee
where $\Theta$ is a shorthand for the angular coordinates.
Then, Eq.~\eqref{eq:KGhigh}
splits into the radial and angular problems,
\be\label{eq:SchwarzschildD_KG}
	\begin{gathered}
		\left[\partial_{\rho}\Delta\partial_{\rho} - \frac{r^{2\left(d-2\right)}}{\left(d-3\right)^2\Delta}\partial_{t}^2\right]\Phi = \hat{\ell}\left(\hat{\ell}+1\right)\Phi\,, \quad
		\left[-\frac{1}{\left(d-3\right)^2}\Delta_{\mathbb{S}^{d-2}}\right] S= \hat{\ell}\left(\hat{\ell}+1\right)S\,,
	\end{gathered}
\ee
where we introduced the new radial variable $\rho \equiv r^{d-3}$ and 
\be
	\Delta\left(\rho\right) \equiv \rho^2f = \rho\left(\rho-\rho_{s}\right)\,,\quad \rho_{s}\equiv r_{s}^{d-3} \,.
\ee
The solutions of the angular problem are the
higher-dimensional spherical harmonics
$Y_{\ell m}\left(\Omega\right)\equiv Y_{\ell,m_1m_2\dots m_{n-1}}\left(\theta_1,\theta_2,\dots\theta_{n}\right)$~\cite{Hui:2020xxx}\footnote{In $n+1$ spatial dimensions, the azimuthal index $m$ becomes a multi-index $m \rightarrow m_1m_2\dots m_{n-1}$ with $\left|m_1\right| \le m_2 \le m_3 \dots m_{n-1} \le \ell$. In addition, there are $n$ angular variables $\theta_{A}$, $A=1,\dots,n$.}.
In order to solve the radial problem, one can use 
the following near zone approximation,
\be
	\left[\partial_{\rho}\Delta\partial_{\rho} - \frac{r^{2\left(d-2\right)}}{\left(d-3\right)^2\Delta}\partial_{t}^2\right] \approx \partial_{\rho}\Delta\partial_{\rho} - \frac{r_{s}^{2\left(d-2\right)}}{\left(d-3\right)^2\Delta}\partial_{t}^2\,,
\ee
which is indeed valid in the near zone region $\omega\left(r-r_{s}\right)\ll 1$. Similar to the $4$-dimensional case, there exists an $\SL$ Love symmetry generated by the vector fields~\cite{Bertini2012,Charalambous:2021kcz},
\be\label{eq:SL2RSchwarzschildD}
	L_0 = -\beta\,\partial_{t} \,,\quad L_{\pm1} = e^{\pm t/\beta}\left[\mp\sqrt{\Delta}\,\partial_{\rho} + \partial_{\rho}\left(\sqrt{\Delta}\right)\beta\,\partial_{t}\right] \,,
\ee
with $\beta=\left(2\pi T_H\right)^{-1}={2r_{s}}/\left({d-3}\right)$ standing for the inverse Hawking temperature of the higher-dimensional Schwarzschild black hole.

Using the Love symmetry, one can explain now 
the properties of higher dimensional black hole Love numbers.
An important difference w.r.t. the four-dimensional case 
is that the
parameter $\hat\ell$ defining 
the Casimir is not generically 
an integer. Instead, the allowed values of $\hat{\ell}$ are
\[
\hat{\ell}={\ell\over d-3},
\]
where $\ell$ is a non-negative integer.
This produces a 
different representation structure.
If $\hat\ell$ is an integer, the static solution
belongs to a highest weight representation as before.
In this case one can run an argument 
completely identical to the four-dimensional case, which proves that the 
static solution is a finite polynomial in $\rho$,
\be
	\text{If $\hat{\ell}\in\mathbb{N}$} \Rightarrow 
	\Phi\left(\omega=0\right) \propto \upsilon_{\hat{\ell},\hat{\ell}} = \sum_{n=0}^{\hat{\ell}}c_{n}\rho^{n} = c_{\hat{\ell}}r^{\ell} 
	+ \dots + c_0\,.
\ee
The absence of terms $\propto r^{-\left(\ell+d-3\right)}$ implies that
Love numbers are zero, which reproduces the 
result based on the explicit regular solution of the 
static Klein-Gordon equation according to which the series in the r.h.s. of the above equation 
corresponds to the Legendre polynomial of degree $\hat \ell$.

Let us also briefly comment on the role of the $\SL$ Love symmetry regarding the running of Love numbers. In $d=4$, for either Schwarzschild or Kerr-Newman black holes, the static Love numbers vanish at all scales. In $d>4$ on the other hand, the scalar Love numbers for the Schwarzschild black hole have a more intricate structure (\cite{Kol:2011vg,Hui:2020xxx}),
\be
	k_{\ell}^{\left(0\right)} =
	\begin{cases}
		\alpha_{\ell} & \text{for $2\hat{\ell}\notin\mathbb{N}$} \\
		\alpha_{\ell} - \beta_{\ell}\log \frac{r}{L} & \text{for $\hat{\ell}\in\mathbb{N}+\frac{1}{2}$} \\
		0 & \text{for $\hat{\ell}\in\mathbb{N}$}
	\end{cases} \,,
\ee
where $\alpha_{\ell}$ are the renormalized static scalar Love numbers at length scale $L$ and $\beta_{\ell}$ are the associated $\beta$-functions, which turn out to be non-vanishing only in the case of half-integer $\hat{\ell}$. As we have shown in this section, Love symmetry explains the vanishing of the Love numbers whenever $\hat{\ell}$ is an integer. The absence of running in these cases is also evident from the fact that regular and singular static solution belong to different, locally distinguishable, representations; these are displayed in Fig. \ref{fig:HWSL2RSchwarzschild} and \ref{fig:SingSL2RSchwarzschild} for Schwarzschild and Fig. \ref{fig:HWSL2RKerr} for the Kerr-Newman black hole. In the nomenclature of Refs.~\cite{nla.cat-vn2139291,doi:10.1137/0501037}, Fig. \ref{fig:HWSL2RSchwarzschild} and Fig. \ref{fig:SingSL2RSchwarzschild} are the representations $D(\,2\hat{\ell}\,)$ and $D^{+-}(\,2\hat{\ell}\,)$ respectively, while, in the notation of Ref.~\cite{Howe1992}, these are the type ``$[\circ]$'' and ``$\circ]\circ[\circ$'' representations $U(-\hat{\ell},-\hat{\ell}\,)$ and $U(\hat{\ell}+1,\hat{\ell}+1)$ respectively\footnote{We refer to Appendix \ref{app:sl2r} for more details on the $\SL$ modules we write here and our notation.}. For the Kerr-Newman case, Fig. \ref{fig:HWSL2RKerr}, solutions regular at the future and past event horizon also belong to the locally distinguishable representations $V_{-2\hat{\ell}}$ and $\bar{V}_{2\hat{\ell}}$ of types ``$[\circ[\circ$'' and ``$\circ]\circ]$'' (\cite{Howe1992}) respectively or, in the notation of Refs.~\cite{nla.cat-vn2139291,doi:10.1137/0501037}, in the $D^{-}(\,2\hat{\ell}\,)$ and $D^{+}(\,2\hat{\ell}\,)$ representations respectively. Except for the finite-dimensional highest weight representation (Fig. \ref{fig:HWSL2RSchwarzschild}), all other $\SL$ representations written here (Figs. \ref{fig:SingSL2RSchwarzschild}, \ref{fig:HWSL2RKerr}) are reducible, but not decomposable.

In all other cases with $\hat{\ell}\notin\mathbb{N}$ however, regular and singular static solutions belong to the same standard $\SL$ representations $D(\hat{\ell},0)$ (\cite{nla.cat-vn2139291,doi:10.1137/0501037}) or $W(4\hat{\ell}(\hat{\ell}+1),0)$ (\cite{Howe1992}) and the Love symmetry does not offer any local criteria from which to infer the absence of running. While this is consistent with the case of half-integer $\hat{\ell}$, it fails to capture the vanishing RG flow for the generic case $2\hat{\ell}\notin\mathbb{N}$. In other words, the Love symmetry $\SL$ representation theory implies a necessary, but not sufficient, condition that an RG flow is expected whenever $\hat{\ell}$ is not an integer. 
In addition, one also needs the external input of power counting arguments (see Section \ref{sec:count}) which independently implies that a necessary condition for running Love numbers is that $2\hat{\ell}\in\mathbb{N}$. These two necessary conditions together reduce to the prediction of a non-vanishing RG flow only for half-integer $\hat{\ell}$ which is indeed what is found by explicit computations (\cite{Kol:2011vg,Hui:2020xxx}).
This is somewhat analogous to the appearance of logarithmic running in conformal perturbation theory,
which takes place only if certain 
resonant conditions are satisfied, c.f.~\cite{zamolodchikov1989integrable,Konechny:2003yy}.
Curiously, the appearance 
of logarithms in the degenerate hypergeometric 
function case 
is also known as~``resonance'', see e.g.~\cite{Opdam:2001wrq}.

Finally, we would like to point out that so 
far we have only studied scalar perturbations 
of higher dimensional Schwarzschild black holes.
We believe that the Love symmetry 
can be extended to higher spin perturbations
as well, and that it can also be used 
for an algebraic proof of
peculiar features of Love numbers
in the corresponding sectors \cite{Hui:2020xxx}.
We leave this question for future work.

\section{Generalizations}
\label{sec:gen}


The Love symmetry admits a number of generalizations. In this section, we first present a second $\SL$ structure arising within the Starobinsky near zone approximation \eqref{eq:starCas}-\eqref{eq:starCas1}. While this particular $\SL_{\text{Star}}$ fails to capture the vanishing of static Love numbers for non-axisymmetric perturbations, the Starobinsky near zone approximation has another peculiar property.
Namely, the corresponding Love numbers vanish for all frequencies $\omega$. We show here that this behavior is appropriately captured by a ``finite-frequency'' generalization of the Love symmetry with the highest weight property playing an analogous role to the static case. Lastly, we construct an infinite dimensional extension of the $\SL$ Love symmetry into $\SL\ltimes \hat{U}\left(1\right)_{\mathcal{V}}$ algebra and show that it contains both the Love and the Starobinsky near zone $\SL$ symmetries as two particular subalgebras. This infinite extension will prove central in the connection of near zone $\SL$ symmetries with the enhanced isometries of near horizon geometries of extremal black holes, which will be discussed in the next section.

\subsection{Starobinsky near zone algebra}
Consider now the following $\SL_{\text{Star}}$ generators,
\be\label{eq:SL2RStarobinsky}
	\begin{gathered}
		L_0^{\text{Star}} = -\beta\left(\partial_{t}+\Omega\partial_{\phi}\right) \,,\\
		L_{\pm1}^{\text{Star}} = e^{\pm t/\beta}\left[\mp\sqrt{\Delta}\partial_{r} + \partial_{r}\left(\sqrt{\Delta}\right)\beta\left(\partial_{t}+\Omega\partial_{\phi}\right) \mp s\frac{r-r_{\mp}}{\sqrt{\Delta}}\right] \,.
	\end{gathered}
\ee
The Casimir of this algebra is given by the Teukolsky 
differential equation
in the Starobinsky near zone
approximation Eq.~\eqref{eq:starCas}.
The $L_0^{\rm Star}$ eigenvalue is given by
\be
	L_0^{\text{Star}} \Psi_s = i\beta\left(\omega-m\Omega\right)\Psi_{s}\,,
\ee
from which we see that in the Starobinsky near zone approximation
black hole perturbations at the locking frequency,
\[
	\omega=m\Omega\,,
\]
play a special role. Namely, these black hole perturbations form 
highest weight $\SL_{Star}$ representations as can be shown completely analogously to the static case for the Love $\SL$ symmetry.
As far as static Love numbers are concerned, the Starobinsky
near zone algebra allows us to make only one exact statement: Love numbers vanish for axisymmetric static 
perturbations, $m=0$, $\omega=0$. 


Another interesting property of the Starobinsky near zone regime is the 
emergence of a modified rotational symmetry,
produced by the following generators
\be
	J_0 = -i\,\partial_{\phi} \,,\quad J_{\pm1} = e^{\pm i\left(\phi-\Omega t\right)}\left[\partial_{\theta} \pm i\cot\theta\,\partial_{\phi} \mp \frac{s}{\sin\theta}\right]\,,
\ee
such that the full near zone symmetry gets enhanced to $\SL_{\text{Star}}\times SO\left(3\right)$. This can be contrasted with the usual Love near zone 
$\SL$ which does not commute with the $SO\left(3\right)$ of the angular equations of motion in the general Kerr-Newman background.
The emergence of the modified
spherical symmetry is an intriguing fact hinting at a more 
general symmetry structure of the near zone Teukolsky
equation.

\subsection{A finite frequency generalization}
An intriguing feature of the Starobinsky near zone approximation \eqref{eq:starCas} is that it leads to vanishing Love numbers at all frequencies when we perform the microscopic computation. We remind here that the explicit expression for the response coefficients in the Starobinsky near zone approximation is given in \eqref{eq:ResponseCoefficients}.
These coefficients are purely imaginary and depend on the frequency only through the frame-dragging factor $\omega-m\Omega$, so that the corresponding conservative part is exactly zero at all frequencies in this approximation. 
This observation calls for a symmetry explanation analogous to the static case.
         
This is achieved by the following generators
\be\label{eq:SL2RStarobinskyOmega}
	\begin{gathered}
		L_0^{\left(\tilde{\omega}\right)} = -\beta\left(\partial_{t}+i\tilde{\omega}\right) \,,\\
		L_{\pm1}^{\left(\tilde{\omega}\right)} = e^{\pm t/\beta}\left[\mp\sqrt{\Delta}\partial_{r} + \partial_{r}\left(\sqrt{\Delta}\right)\beta\,\partial_{t} + \frac{a}{\sqrt{\Delta}}\,\partial_{\phi} + i\beta \tilde{\omega}\sqrt{\frac{r-r_{+}}{r-r_{-}}}\right] \,, 
	\end{gathered}
\ee
defined
for some arbitrary parameter $\tilde{\omega}$.
These generators are regular 
in advanced Kerr coordinates.
The modified generators form an $\SL$ algebra which we dub $\SL_{\tilde{\omega}}$. Its quadratic Casimir is given by,
\be
\label{eq:wtilde}
	\mathcal{C}_2^{\left(\tilde{\omega}\right)} = \partial_{r}\Delta\partial_{r} - \frac{\left(r_{+}^2+a^2\right)^2}{\Delta}\left[\left(\partial_{t}+\Omega\partial_{\phi}\right)^2 + 4\frac{r-r_{+}}{r_{+}-r_{-}}\left(\partial_{t}+i\tilde{\omega}\right)\left(\Omega\partial_{\phi}-i\tilde{\omega}\right)\right] \,.
\ee
A scalar black hole perturbation $\Phi$ of frequency $\omega$
is an eigenstate of $L_0^{\left(\tilde \omega\right)}$ 
with the eigenvalue given by
\be
	L_0^{\left(\tilde \omega\right)} \Phi = i\beta (\omega - \tilde{\omega})\Phi\,
\ee
so that the null $L_0^{\left(\tilde \omega\right)}$ eigenstate is a monochromatic 
black hole perturbation with frequency $\omega=\tilde\omega$. By comparing 
\eqref{eq:wtilde} to \eqref{eq:starCas}, we observe that it satisfies the Teukolsky differential equation
in the Starobinsky near zone approximation.
Hence, analogously to the earlier arguments, one concludes that any regular 
black hole perturbation 
of finite frequency $\tilde\omega$ 
in the Starobinsky near zone
approximation
belongs to a highest weight representation of the $\SL_{\tilde \omega}$ algebra \eqref{eq:SL2RStarobinskyOmega}.
The corresponding highest weight vector, satisfying
\[
	L_{+1}^{\left(\tilde\omega\right)}v_{-\ell,0}=0\,,\,\, L_0v_{-\ell,0}=-\ell v_{-\ell,0}\;,
\]
has the following form
\be
	v_{-\ell,0} = \left(\frac{r-r_{+}}{r-r_-}\right)^{i {\tilde Z}}e^{im\phi}e^{\left(\ell/\beta -i\tilde\omega\right)t}\Delta^{\ell/2} \,, \quad \text{where} \quad {\tilde Z} \equiv \frac{r_+^2+a^2}{r_{+}-r_{-}}\left(m\Omega- \tilde\omega\right) \,.
\ee
It describes a regular at the future event horizon solution with frequency $\tilde\omega +i\beta \ell$. The $L^{\left(\tilde\omega\right)}_0$-null vector $v_{-\ell,\ell}$ is then a regular solution of the Teukolsky equation at frequency $\tilde\omega$ in the Starobinsky near zone approximation. It satisfies 
\be
	\left(L^{\left(\tilde \omega\right)}_{+1}\right)^{\ell+1}v_{-\ell,\ell} = 0 \,,
\ee
which implies the quasipolynomial form
of the solution,
\be
	\Phi\left(\omega=\tilde\omega\right) = \left(\frac{r-r_{+}}{r-r_{-}}\right)^{i {\tilde Z}} e^{im\phi}e^{-i\tilde{\omega}t}\sum_{n=0}^{\ell}c_{n}r^{n} \,.
\ee
As before, the overall form-factor can be removed by a transition into the advanced coordinates and
plays no role in
the EFT matching calculation of Love numbers. 
The remaining polynomial does not contain negative 
powers in $r$ and hence the frequency-dependent 
Love numbers vanish identically.

The extension of the $\SL_{\tilde \omega}$ algebra 
to spin $s$ fields is straightforward. One has to start 
with the usual spin-$s$ Love generators~\eqref{eq:SL2RsKerrNewman}
and add the same $\tilde \omega$-dependent pieces
as in Eq.~\eqref{eq:SL2RStarobinskyOmega}. 
One could check that for $\omega=\tilde \omega$
the corresponding Casimir reduces to the 
spin-$s$ Teukolsky equation 
in the Starobinsky near zone approximation.

This proves the very intriguing property of the Starobinsky near zone approximation of vanishing linear conservative response at all frequencies. It is important to stress that, unlike for static Love numbers, this property does not hold for the full response (i.e., beyond the leading order near zone approximation). Indeed, using the full solution, one can see that the conservative black hole response does not vanish already at the linear order in $\omega$~\cite{Charalambous:2021mea}.

Also, it is unclear to us whether this algebraic argument
is a qualitatively new piece of information or
simply a restatement of the result
obtained by a direct solution 
of the corresponding differential equation. 
The problem is the interpretation
of the hyperparameter $\tilde\omega$ in the general 
case of non-monochromatic solutions,
and the geometric meaning of the scalar 
$\tilde\omega$-dependent 
generators appearing in Eq.~\eqref{eq:SL2RStarobinskyOmega}.
We leave a better understanding of the $\SL_{\tilde \omega}$
symmetry for future work.

\subsection{An infinite-dimensional extension}
\label{sec:InfiniteExtension}

We just saw that, in addition to the near zone approximation leading to the Love symmetry,  the Starobinsky near zone approximation also has very special and interesting properties.
We will show now that these two near zone approximations can be naturally combined into a single algebraic structure, which will also turn out to contain $\SL$ enhancements of effective geometries which preserve the internal structure of the black hole.
The construction is based on the observation that, 
for any regular 
at the horizon 
$\SL$ representation
$\mathcal{V}$,
the Love algebra can be extended 
into a semidirect 
product $\SL\ltimes U\left(1\right)_{\mathcal{V}}$,
where $U\left(1\right)_{\mathcal{V}}$
is generated by vector fields of the form
$\upsilon\,\beta\Omega\,\d_\phi$ with $\upsilon\in \mathcal{V}$.

\begin{figure}
	\centering
	\begin{tikzpicture}
		\node at (0,0) (um3) {$\upsilon_{0,-3}$};
		\node at (0,1) (um2) {$\upsilon_{0,-2}$};
		\node at (0,2) (um1) {$\upsilon_{0,-1}$};
		\node at (0,3) (u0) {$\upsilon_{0,0}$};
		\node at (0,4) (up1) {$\upsilon_{0,+1}$};
		\node at (0,5) (up2) {$\upsilon_{0,+2}$};
		\node at (0,6) (up3) {$\upsilon_{0,+3}$};

		\node at (3,-0.4) (um) {$\vdots$};
		\draw (1,0) -- (5,0);
		\draw (1,1) -- (5,1);
		\draw (1,2) -- (5,2);
		\draw [snake=zigzag] (1,2.9) -- (5,2.9);
		\draw (1,3) -- (5,3);
		\draw [snake=zigzag] (5,3.1) -- (1,3.1);
		\draw (1,4) -- (5,4);
		\draw (1,5) -- (5,5);
		\draw (1,6) -- (5,6);
		\node at (3,6.4) (up) {$\vdots$};

		\draw[blue] [->] (2,0) -- node[left] {$L_{+1}$} (2,1);
		\draw[blue] [->] (2.5,1) -- node[left] {$L_{+1}$} (2.5,2);
		\draw[blue] [->] (3,2) -- node[left] {$L_{+1}$} (3,3);
		\draw[blue] [->] (2.5,4) -- node[left] {$L_{+1}$} (2.5,5);
		\draw[blue] [->] (2,5) -- node[left] {$L_{+1}$} (2,6);
		\draw[red] [<-] (4,0) -- node[right] {$L_{-1}$} (4,1);
		\draw[red] [<-] (3.5,1) -- node[right] {$L_{-1}$} (3.5,2);
		\draw[red] [<-] (3,3) -- node[right] {$L_{-1}$} (3,4);
		\draw[red] [<-] (3.5,4) -- node[right] {$L_{-1}$} (3.5,5);
		\draw[red] [<-] (4,5) -- node[right] {$L_{-1}$} (4,6);
	\end{tikzpicture}
	\caption{A representation $\mathcal{V}$ of $\SL$ used to construct the $\SL\ltimes U\left(1\right)_{\mathcal{V}}$ extension of the Love algebra.}
	\label{fig:VSL2R}
\end{figure}
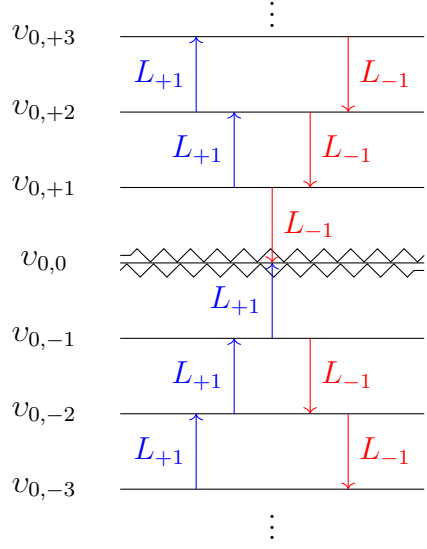

A representation $\mathcal{V}$, which leads to a unified description of the Love symmetry and the Starobinsky near zone $\SL$,
can be constructed as follows. 
We start with $\upsilon_{0,0}=-1$,
which is a single element of the one-dimensional 
unitary $\SL$ representation (``singleton'')~\cite{Barut1965}.
It has $h=0$ and also satisfies $L_{\pm1}\upsilon_{0,0} = 0$.
Then we solve for 
the $h=\mp 1$ vectors $\upsilon_{0,\pm 1}$ satisfying 
\be 
	L_{0}\upsilon_{0,\pm1} = \mp\upsilon_{0,\pm1} \,,\quad L_{\mp 1}\upsilon_{0,\pm 1}=\mp \upsilon_{0,0} \,.
\ee 
These give us
\be
	\upsilon_{0,\pm1} = e^{\pm t/\beta}\sqrt{\frac{r-r_{+}}{r-r_{-}}} \,, \quad \upsilon_{0,0}=-1\,.
\ee
All these functions are regular at the future and past event horizons. 
By definition, starting with $\upsilon_{0,\pm1}$ one can reach $\upsilon_{0,0}$ by acting on them with $L_{\pm1}$. However, one cannot reach $\upsilon_{0,\pm1}$ starting from $\upsilon_{0,0}$. 
A (reducible) representation which contains all three vectors $\upsilon_{0,0}$, $\upsilon_{0,\pm1}$ is spanned by the following vectors $\upsilon_{0,n}$,
\be
	\upsilon_{0,\pm n} = \left(L_{\pm1}\right)^{n-1}\upsilon_{0,\pm1} = \left(\pm1\right)^{n-1}\left(n-1\right)! \,e^{\pm n t/\beta}\left(\frac{r-r_{+}}{r-r_{-}}\right)^{n/2} \,,
\ee
with $n\in\mathbb{N}$. 
We present a graphical representation of this construction in Fig.~\ref{fig:VSL2R}.

This way, we obtain the general $U\left(1\right)_{\mathcal{V}}$ 
representation that extends
the Love symmetry into $\SL\ltimes \hat U\left(1\right)_{\mathcal{V}}$,
\be\label{eq:U1V}
	\upsilon = \sum_{n=-\infty}^\infty \alpha_n v_{0,n}\,\beta\Omega\,\d_\phi \,.
\ee 

A one-parameter 
family of the $\SL$
subalgebras from this $\SL\ltimes U\left(1\right)_{\mathcal{V}}$ is of a
particular interest,
\be
\label{eq:StarLove}
	L_{m}\left(\alpha\right) = L_{m} + \alpha\,\upsilon_{0,m}\,\beta\Omega\,\partial_{\phi}\,,\quad m=0,\pm1 \,.
\ee
The Casimir of this algebra is given by 
\be 
\begin{split}
	\mathcal{C}_2\left(\alpha\right) = \partial_{r}\Delta\partial_{r} - \frac{\left(r_{+}^2+a^2\right)^2}{\Delta}\left(\partial_{t}+\Omega\,\partial_{\phi}\right)^2 + 2\frac{r_{+}^2+a^2}{r-r_-}\left(\partial_{t}+\alpha\,\Omega\,\partial_{\phi}\right)\left(\alpha-1\right)\beta\Omega\,\partial_{\phi} \,.
\end{split}
\ee 
In general, this Casimir 
does not capture the 
physical near zone limit: its static part 
does not match that of the Teukolsky
equation unless $\alpha=0$ or $\alpha = 1$. 
The first choice, $\alpha=0$, corresponds to the Love symmetry. The second special 
choice $\alpha = 1$
reduces this Casimir to the one matching
the Starobinsky 
near zone split.

Nevertheless, it does capture the near horizon characteristic exponents of the Teukolsky equation. In fact, as we show explicitly in Appendix \ref{ApGenerators}, this one-parameter family of $\SL$ subalgebras precisely contains \textit{all} possible globally defined and time-reversal symmetric\footnote{Time-reversal here refers to the simultaneous time-reversal transformation $t\rightarrow-t$ and the flip of the angular momentum of the black hole, $a\rightarrow-a$.} approximations with an $\SL$ enhancement that preserve these near horizon characteristic exponents. This property will prove crucial in the next section where a connection with enhanced isometries of near horizon geometries of extremal black holes will be discussed.


\section{Relation to extremal near horizon isometries}
\label{SecNHE}

The Love vectors are formally singular in
the extremal limit, as the Hawking temperature
approaches zero. In this section we show how 
to take an appropriate extremal limit of
these vectors, and demonstrate 
the relationship between them 
and the extremal near horizon isometries.
In what follows we focus on scalar perturbations.
General spin-$s$ perturbations will be discussed
in detail in the next section.

\subsection{Review of near horizon geometries}

The extremal black hole limit 
has many peculiarities. Let us start with the extremal 
Reissner-Nordstr\"om spherically symmetric charged black holes.
The near horizon geometry is obtained by applying the 
scaling transformations\footnote{The near horizon radial coordinate $\rho$ used here should not be confused with the radial coordinate transformation $\rho=r^{d-3}$ employed in the study of higher-dimensional Schwarzschild black holes in Section \ref{sec:love}.} 
\be
\label{eq:scalRN}
	r = M +  \lambda \rho \,,\quad t = \frac{ \tau}{\lambda} \,,
\ee
and taking the $\lambda\to 0$ limit.
The resulting $\text{AdS}_2\times \mathbb{S}^2$ near horizon metric 
\be
	ds^2 M^{-2}=  -\frac{\rho^2}{M^4} d\tau^2+\frac{d\rho^2}{\rho^2}+d\Omega^2_2 \,,
\ee
acquires the following Killing vectors
\be\label{eq:isoRN}
	\xi_0 = \tau\,\partial_{\tau} - \r\,\partial_{\r} \,, \quad 
	\xi_{+1} = \partial_{\tau}\,, \quad 
	\xi_{-1} = \left(\frac{M^4}{\rho^2} + \tau^2\right)\partial_{\tau} -2\tau\rho\,\partial_{\rho}  \,,
\ee
which satisfy the $\SL$ algebra. 
In the original Boyer-Lindquist coordinates
they take the following form,
\be\label{eq:isoRN2}
	\begin{split}
		\xi_0 &= t\,\partial_{t} - \left(r-M\right)\partial_{r} \,, \quad \xi_{+1} = \lambda^{-1}\partial_{t}\,, \\ 
		\xi_{-1} & = \lambda \left[\left( \frac{M^4}{(r-M)^2} + t^2\right)\partial_{t} - 2t\left(r-M\right)\partial_{r} \right] \,.
	\end{split}
\ee
The Casimir of the $\SL$ near horizon isometry is given by,
\be
	\mathcal{C}_2=\partial_{r}\left(r-M\right)^2\partial_{r} -\frac{M^4}{\left(r-M\right)^2}\,\d_t^2\,,
\ee
and turns out to coincide with the Klein-Gordon
differential operator in the near zone. The Klein-Gordon equation in the near horizon scaling limit can then be written as 
\be
	C_2 \Phi =  \ell\left(\ell+1\right)\Phi\,,
\ee
with integer $\ell$. From the perspective of the full geometry, only static perturbations survive in the throat and this enhanced $\SL$ isometry fully constrains their wave dynamics.

The situation is more intricate in the extremal Kerr-Newman case~\cite{Bardeen:1999px,Amsel:2009et}. 
One obtains the near horizon extremal geometry by taking the $\lambda\rightarrow0$ limit of the following rescaled co-rotating coordinates,
\be
\label{eq:limKerr}
	r =M + \lambda\rho \,,\quad t = \frac{\tau}{\lambda}\,,\quad \phi = \tilde{\phi} + \frac{a}{M^2+a^2}\frac{\tau}{\lambda} \,,
\ee
which generates the metric 
\be 
	ds^2 = \left(1-\frac{a^2}{\rho^2_0}\sin^2\theta\right) \left[ -\frac{\rho^2}{\rho_0^2}d\tau^2 + \frac{\rho_0^2}{\r^2}d\rho^2 + \rho_0^2 d\theta^2 \right] + \frac{\rho_0^2\sin^2\theta}{1-\frac{a^2}{\rho^2_0}\sin^2\theta} \left( d\tilde{\phi} + \frac{2aM\rho}{\rho_0^4}d\tau\right)^2\,,
\ee
with $\rho_0^2\equiv M^2 + a^2$.
This metric possesses the 
azimuthal symmetry $U\left(1\right)$ generated 
by the vector field $J_0 = -i\,\partial_{\tilde{\phi}}$,
as well as additional $\SL$ Killing vectors of the AdS$_2$
factor,
\be\ba\label{eq:SL2RNHEKerrNewman}
	\xi_0 &= \tau\,\partial_{\tau} - \rho\,\partial_{\rho} \,,\quad \xi_{+1} = \partial_{\tau} \,, \\
	\xi_{-1} &= \left( \frac{\left(M^2+a^2\right)^2}{\rho^2} + \tau^2\right)\partial_{\tau} -2\tau\rho\,\partial_{\rho} - \frac{4M a}{\rho}\,\partial_{\tilde{\phi}} \,.
\ea\ee
The full isometry
group is $\SL\times U\left(1\right)$.
The Casimir of the $\SL$ factor
expressed in the original Boyer-Lindquist coordinates is given by
\be
\label{eq:KNNHE}
	\begin{split}
		\mathcal{C}_2 &= \partial_{r}\left(r-M\right)^2\partial_{r} - \left(\frac{M^2+a^2}{r-M}\right)^2\left(\partial_{t}+\Omega\,\partial_{\phi}\right)^2 + \frac{4Ma}{r-M}\left(\partial_{t}+\Omega\,\partial_{\phi}\right)\partial_{\phi} \,,
	\end{split}
\ee
and the radial part of the 
near horizon $s=0$ Teukolsky equation for an extremal Kerr-Newman black hole reads,
\be\label{eq:C2SL2RExtremalKerr}
	\mathcal{C}_2\Phi = \left[\ell\left(\ell+1\right)-2\left(3M^2+a^2\right)m^2\Omega^2\right]\Phi \,,
\ee
where the angular eigenvalues $\ell\left(\ell+1\right)$ are no longer integers, unless $m=0$. The explicit range of values of $\ell$ for non-zero azimuthal numbers is known only numerically and can be found, for example, in \cite{Bardeen:1999px}. Similarly to the Reissner-Nordstr\"{o}m case, not all frequencies of scalar perturbations propagating in the full geometry survive in the throat. In the rotating case, solutions which can be extended to smooth perturbations of the full Kerr-Newman geometry can only have the locking frequency $\omega = m\Omega$.

\subsection{Reissner-Nordstr{\"o}m case}

Now let us take the 
extremal limit of the Love generators 
in the spherically symmetric Reissner-Nordstr{\"o}m case
$a=0$. We parameterize the near-extremal expansion via the Hawking temperature $T_{H}$ of the black hole. Then,
\be
    Q^2=\frac{1-8\pi T_{H}M}{8\pi^2T_{H}^2}\left[\frac{1-4\pi T_{H}M}{\sqrt{1-8\pi T_{H}M}} - 1\right] = M^2\left(1-4\pi^2 T_{H}^2M^2\right) + \mathcal{O}\left(T_{H}^3\right) \,.
\ee
Taylor expanding the Love 
symmetry generators \eqref{eq:SL2RKerr} 
up to $\mathcal{O}(T_{H}^2)$ we obtain
\be
	\begin{split}
		{}& L_0 = -\frac{1}{2\pi T_{H}}\,\partial_{t} \\
		& L_{\pm1} = -\left(r-M\right)\left(\pm1 + t \,2\pi T_{H}\right)\partial_{r} + \left[\frac{1}{2\pi T_{H}} \pm t +\left(\frac{M^4}{\left(r-M\right)^2} + t^2\right)2\pi T_{H}\right] \partial_{t} \,.
	\end{split}
\ee
These generators are singular at $T_H=0$, so that taking the extremal limit of the Love symmetry is somewhat non-trivial. 
To achieve this we consider three linear combinations of generators 
with $T_H$-dependent coefficients which have a finite non-trivial $T_H=0$ limit. 
These linear combinations are 
\be
	\begin{split}
		\xi_{+1} &= \lambda^{-1}\lim_{T_{H}\to 0}\left(-2\pi T_{H} L_0\right) = \lambda^{-1} \partial_{t} \,,\\
		\xi_{0} &= \lim_{T_{H}\to 0}\frac{L_{+1}-L_{-1}}{2} = t\,\partial_{t} -\left(r-M\right)\partial_{r} \,,\\
		\xi_{-1} &= \lambda\lim_{T_{H}\to 0}\frac{L_1+L_{-1}+2L_0}{2\pi T_{H}} = \lambda\left[ \left(t^2 +\frac{M^4}{(r-M)^2}\right)\partial_{t} - 2t\left(r-M\right)\partial_{r} \right] \,.
	\end{split}
\ee
They exactly reproduce the generators of the $\SL$ near horizon isometry of extremal charged black holes \eqref{eq:isoRN2} after identifying the parameter $\lambda$ with the scaling parameter associated with the near horizon limit. 
This limiting procedure is somewhat similar to the Wigner contraction. Unlike in the latter case though, here the limiting algebra of the near horizon isometries is again $\SL$, so it is isomorphic to the original Love symmetry algebra. However, the limit itself is algebraically non-trivial. Namely, the resulting mapping of the generators does not correspond to the algebra automorphism.

To see this it is instructive to investigate the algebraic structure of solutions of the Klein-Gordon equation for extremal Reissner--Nordstr\"{o}m black holes as dictated by this $\SL$ near horizon isometry. In particular, we are interested in the representation that the static solution belongs to. By definition, the static solution is annihilated by $\xi_{+1}$ and is, therefore, a {\it primary} vector $\upsilon_0$ of the highest weight representation,
\be 
	\xi_{+1}v_0=0\,,\quad \xi_0v_0 =-\ell v_0 \,,\quad \Rightarrow 
	\quad v_0=\left(r-M\right)^\ell\,.
\ee
Other solutions in the highest-weight $\SL$ multiplet are then obtained as descendants,
\be
	\begin{split}
		& v_1 = \xi_{-1} v_0 = -2\lambda\ell\,t \left(r-M\right)^\ell\,, \\
		& v_2 = \xi_{-1} v_1 =   -2\lambda^{2}\ell\left[ \left(1-2\ell\right) t^2 +\frac{M^4}{\left(r-M\right)^2} \right] \left(r-M\right)^\ell\,,\\
	\end{split}
\ee
and so on. From all of these elements of the highest weight representation,  $\upsilon_0$ and $\upsilon_1$ are exact solutions of the full Klein-Gordon equation for an extremal Reissner-Nordstr\"{o}m black hole, while $v_n$ with $n>1$ 
are only approximate solutions valid in the near horizon region.
A change of the algebraic structure manifests itself in the fact that the static solution is a primary $\SL$ vector now, which is not the case before taking the extremal limit.
The corresponding solution still has a polynomial dependence on $r$, indicative of the vanishing Love numbers. However, in the extremal limit this polynomial form does not link directly to the highest weight property of the corresponding representation.

\subsection{Kerr-Newman case}
\label{sec:NHEKerrNewman}

Analogously to the Reissner-Nordstr\"{o}m black hole, the extremal limit is realized as the $T_{H}\rightarrow0$ limit of vanishing Hawking temperature, in terms of which,
\be
	a^2+Q^2 = M^2\left(1 - 4\pi^2T_{H}^2M^2\right) + \mathcal{O}\left(T_{H}^3\right) \,.
\ee
However, an attempt to take linear combinations of the Love symmetry generators analogously to the non-rotating case does not recover the near horizon  $\SL$ Killing vectors 
in this case.

However, these Killing vector can still be recovered in the rotating case as well by making use of 
the infinite extension $\SL\ltimes U\left(1\right)_{\mathcal{V}}$ constructed in Section \ref{sec:InfiniteExtension}. Namely, consider a family of $\SL$ subalgebras corresponding to the choice
\be\label{eq:NHa}
	\alpha = 1+4\pi T_{H}M + \mathcal{O}\left(T_{H}^2\right)
\ee
in Eq.~\eqref{eq:StarLove}, for arbitrary subleading contributions in the extremal limit. The corresponding generators are given by,
\be\label{eq:SL2RNHne}
	\begin{gathered}
		L_0^{\text{NHE}} = -\frac{1}{2\pi T_{H}}\left[\partial_{t}+\left(1+4\pi T_{H}M\right)\Omega\,\partial_{\phi}\right] + \mathcal{O}\left(T_{H}\right) \,, \\
		\ba
			L_{\pm 1}^{\text{NHE}} = e^{\pm 2\pi T_{H}t} &\bigg[\mp\sqrt{\Delta}\,\partial_{r} + \partial_{r}\left(\sqrt{\Delta}\right)\frac{1}{2\pi T_{H}}\left(\partial_{t}+\Omega\,\partial_{\phi}\right) \\
			&+ 2M\sqrt{\frac{r-r_{+}}{r-r_{-}}} \Omega\,\partial_{\phi} + \mathcal{O}\left(T_{H}\right) \bigg] \,,
		\ea
	\end{gathered}
\ee
and correspond to the following Casimir,
\be\label{eq:C2nhe2}
	\mathcal{C}_2^{\text{NHE}} = \partial_{r}\Delta\partial_{r} - \frac{\left(r_{+}^2+a^2\right)^2}{\Delta}\left(\partial_{t}+\Omega\,\partial_{\phi}\right)^2 + \frac{4Ma}{r-r_{-}}\left(\partial_{t}+\Omega\,\partial_{\phi}\right)\partial_{\phi} + \mathcal{O}\left(T_{H}\right) \,,
\ee
which indeed reduces to the appropriate near horizon Teukolsky differential operator \eqref{eq:KNNHE} in the extremal limit.

Reshuffling the vector fields $L_{m}^{\text{NHE}}$ and taking the $T_{H}\to 0$ limit
we arrive at the exact near horizon 
extremal Kerr-Newman $\SL$ Killing vectors
\eqref{eq:SL2RNHEKerrNewman},
\be
	\begin{split}
		\xi_{+1} &= \lambda^{-1}\lim_{T_{H}\to 0} \left(-2\pi T_{H} L_0^{\text{NHE}}\right) = \lambda^{-1}\left(\partial_{t} +\Omega\,\partial_{\phi}\right) \,, \\
		\xi_0 &= \lim_{T_{H}\to 0} \frac{L_{+1}^{\text{NHE}}-L_{-1}^{\text{NHE}}}{2}= t\left(\partial_{t}+\Omega\,\partial_{\phi}\right) - \left(r-M\right) \partial_{r} \,, \\
		\xi_{-1} &= \lambda\lim_{T_{H}\to 0} \frac{L_{+1}^{\text{NHE}}+L_{-1}^{\text{NHE}}+2 L_0^{\text{NHE}}}{2\pi T_{H}} \\
		&= \lambda\left[ \left( \frac{\left(M^2+a^2\right)^2}{\left(r-M\right)^2}+t^2 \right)\left(\partial_{t} + \Omega\,\partial_{\phi}\right) - 2t\left(r-M\right)\partial_{r} - \frac{4M a}{r-M}\partial_{\phi} \right] \,,
	\end{split}
\ee
after identifying $\lambda$ with the near horizon scaling parameter as before.

A test field solution $\Phi$ must belong to a certain representation of $\SL$, which is characterized by eigenvalues 
of the Casimir \eqref{eq:C2SL2RExtremalKerr} and the $\xi_0$ operator. For the static axisymmetric ($m=0$) mode $v_0$ we have an integer $\ell$ and
\be
	\mathcal{C}_2 v_0 = \ell\left(\ell+1\right)\,,\quad \xi_{+1} v_0 = 0\,.
\ee
This mode is clearly a primary state of a highest weight representation
with weight $-\ell$. Integrating $\xi_0 v_0 = -\ell v_0$ we obtain,
\be
	v_0 = \left(r-M\right)^{\ell}\,.
\ee
The rest of the representation is then built by acting with $\xi_{-1}$,
\be
	\begin{split}
		& v_1 = \xi_{-1} v_0 = -2\lambda\ell\,t \left(r-M\right)^\ell\,, \\
		& v_2 = \xi_{-1} v_1 =   -2\lambda^{2}\ell\left[ \left(1-2\ell\right) t^2 + \frac{\left(M^2+a^2\right)^2}{\left(r-M\right)^2} \right] \left(r-M\right)^\ell \,,
	\end{split}
\ee
and so on. Similarly to the of the extremal Reissner-Nordstr\"{o}m black hole, only the static solution $\upsilon_0$ satisfies the full extremal $s=0$ Teukolsky equation and can be extended beyond the near horizon approximation; its polynomial form implies the vanishing of the Love numbers for static axisymmetric perturbations. However, this polynomial form is related to the special form of the generator $\xi_{0}$, rather than the highest weight property.

For a generic perturbation with magnetic number $m$ the corresponding frequency is fixed to be $\omega=m\Omega$,
in which case the Casimir eigenvalue is not an integer. These solutions do not belong to highest weight $\SL$ representations and their response coefficients are not zero.

\section{Teukolsky equation from spin weighted Lie derivatives}
\label{sec:spins}

In the previous section we saw that 
the extremal near horizon isometries 
allow one to constrain 
wave dynamics in the corresponding extremal AdS$_2$
factors. In particular, the Casimir made of
the Killing vectors of the 
AdS$_2$ isometry group $\SL$ exactly 
reproduces the appropriate Klein-Gordon 
equation.

A similar picture takes place for general spin-$s$
perturbations, including the gravitational ones.
In this section we review 
an appropriate generalization of AdS$_2$
Killing vectors through 
spin weighted Lie derivatives, which takes into account 
the geometric structure behind spin weighted 
Newman--Penrose scalars. The Casimir 
constructed from these generalized 
vectors reproduces
the corresponding spin-$s$
Teukolsky equation in the near horizon region.

It is natural to ask whether a similar construction can be applied to the Love symmetry generators.
We will argue that the spin-$s$ Love generators
in the non-extremal case
are indeed natural to interpret as approximate 
spin weighted 
Lie derivatives of the Kerr-Newman spacetime.
This interpretation works most nicely for the Starobinsky $\SL$ generators.
Additionally, we discuss the 
geometric meaning of the Love symmetry
in the context of black hole 
subtracted geometries. 

\subsection{Elements of the NP formalism}
\label{sec:NPbasics}

Let us briefly review some basics of the
Newman-Penrose (NP) formalism that will be important for our purposes~\cite{Newman1962}. 
The NP formalism is a particular case of the tetrad formalism, with the local tetrads chosen to be null. The relevant four local tetrads are composed of two real vectors, $\ell^{\mu}$ and $n^{\mu}$, and two complex valued vectors, $m^{\mu}$ and its complex conjugate $\bar{m}^{\mu}$, normalized according to,
\be
	\ell_{\mu}n^{\mu} = -1\,,\quad m_{\mu}\bar{m}^{\mu} = 1 \,.
\ee
These tetrads are related to the metric tensor through
\be
	g_{\mu\nu} = 2\left(-\ell_{(\mu}n_{\nu)} + m_{(\mu}\bar{m}_{\nu)}\right) \,.
\ee
Within the NP formalism,
non-scalar degrees of freedom are encoded into complex spacetime scalars obtained by projecting the relevant tensors 
on the local null tetrads. 
In particular, the $10$ independent components of the Weyl tensor are cast into the $5$ complex Weyl scalars,
\be
	\begin{split}
		\psi_0 &=  C_{\mu\nu\rho\sigma}\ell^{\mu}m^{\nu}\ell^{\rho}m^{\sigma} \,,\quad 
		\psi_1 =  C_{\mu\nu\rho\sigma}\ell^{\mu}n^{\nu}\ell^{\rho}m^{\sigma} \,, \quad 
		\psi_2 =  C_{\mu\nu\rho\sigma}\ell^{\mu}m^{\nu}\bar{m}^{\rho}n^{\sigma} \,, \\ 
		\psi_3 & = C_{\mu\nu\rho\sigma}\ell^{\mu}n^{\nu}\bar{m}^{\rho}n^{\sigma} \,,\quad 
		\psi_4 =  C_{\mu\nu\rho\sigma}n^{\mu}\bar{m}^{\nu}n^{\rho}\bar{m}^{\sigma} \,,
	\end{split}
\ee
while the $6$ independent components of the Maxwell field strength tensor are rearranged into the $3$ complex Maxwell-NP scalars,
\be
	\phi_0 =  F_{\mu\nu}\ell^{\mu}m^{\nu} \,,\quad 
	\phi_1 =  \frac{1}{2}F_{\mu\nu}\left(\ell^{\mu}n^{\nu} + \bar{m}^{\mu}m^{\nu}\right) \,,\quad 
	\phi_2 =  F_{\mu\nu}\bar{m}^{\mu}n^{\nu}\,.
\ee
The NP scalar of a scalar field is the scalar field itself. 
All of the above NP scalars transform homogeneously under two particular local Lorentz transformations known as local rotations and local boosts\footnote{These are also known as type III transformations in the traditional language of the NP formalism.},
\begin{itemize}
	\item \underline{Local rotations}: These are complex local $SO\left(2\right)$ rotations of the tetrad $m^{\mu}$, keeping the real vectors $\ell^{\mu}$ and $n^{\mu}$ invariant,
	\be
		\begin{pmatrix}
			\ell^{\mu} \\
			n^{\mu} \\
			m^{\mu} \\
			\bar{m}^{\mu}
		\end{pmatrix} \xrightarrow{\chi}
		\begin{pmatrix}
			\ell^{\mu} \\
			n^{\mu} \\
			e^{i\chi}m^{\mu} \\
			e^{-i\chi}\bar{m}^{\mu}
		\end{pmatrix} \,.
	\ee
	
	\item \underline{Local boosts}: These are real local rescalings of the real tetrads $\ell^{\mu}$ and $n^{\mu}$,  keeping the complex vector $m^{\mu}$ invariant,
	\be
		\begin{pmatrix}
			\ell^{\mu} \\
			n^{\mu} \\
			m^{\mu} \\
			\bar{m}^{\mu}
		\end{pmatrix} \xrightarrow{\lambda}
		\begin{pmatrix}
			\lambda\ell^{\mu} \\
			\lambda^{-1}n^{\mu} \\
			m^{\mu} \\
			\bar{m}^{\mu}
		\end{pmatrix} \,.
	\ee
\end{itemize}

NP scalars are labeled according to their transformation properties w.r.t. local rotations and boosts via a spin weight $s$ and a boost weight $b$. In particular, an NP scalar $\Psi$ transforming as
\be
	\Psi \xrightarrow{\lambda,\chi}\lambda^{b}e^{is\chi}\Psi
\ee
is said to have weights $\left\{b,s\right\}$. These are related to the GHP weights $p$ and $q$ \cite{Geroch1973} according to,
\be
	b = \frac{p+q}{2}\,\,\,\,,\,\,\,\,s=\frac{p-q}{2} \,.
\ee
The weights associated with the scalar field, Maxwell-NP scalars and Weyl scalar can be found in Table \ref{tbl:NPweights}.

\begin{table}
	\centering
	\begin{tabular}{|c|c|c|}
		\hline
		NP scalar & $\left\{b,s\right\}$ & $\left\{p,q\right\}$ \\
		\hline
		\hline
		$\psi_0$ & $\left\{+2,+2\right\}$ & $\left\{+4,0\right\}$ \\
		\hline
		$\psi_1$ & $\left\{+1,+1\right\}$ & $\left\{+2,0\right\}$ \\
		\hline
		$\psi_2$ & $\left\{0,0\right\}$ & $\left\{0,0\right\}$ \\
		\hline
		$\psi_3$ & $\left\{-1,-1\right\}$ & $\left\{-2,0\right\}$ \\
		\hline
		$\psi_4$ & $\left\{-2,-2\right\}$ & $\left\{-4,0\right\}$ \\
		\hline
		\hline
		$\phi_0$ & $\left\{+1,+1\right\}$ & $\left\{+2,0\right\}$ \\
		\hline
		$\phi_1$ & $\left\{0,0\right\}$ & $\left\{0,0\right\}$ \\
		\hline
		$\phi_2$ & $\left\{-1,-1\right\}$ & $\left\{-2,0\right\}$ \\
		\hline
		\hline
		$\Phi$ & $\left\{0,0\right\}$ & $\left\{0,0\right\}$ \\
		\hline
	\end{tabular}
	\caption{Spin weights, boost weights and GHP weights of the Weyl scalars $\psi_{a}$, the Maxwell-NP scalars $\phi_{a}$ and the scalar field $\Phi$.}
	\label{tbl:NPweights}
\end{table}

In the tetrad formalism, the Christoffel symbols are repackaged into the Ricci rotation coefficients, known as spin coefficients when the tetrad vectors are null as in the NP formalism. In total there are $12$ spin coefficients. These can be categorized into the $8$ so-called ``good'' spin coefficients,
\be\ba\label{eq:NPSpinCoefficientsGood}
	& \kappa = - m^{\mu}D\ell_{\mu} \,,\quad \tau = - m^{\mu}\triangle\ell_{\mu} \,,\quad \sigma = - m^{\mu}\delta\ell_{\mu} \,,\quad \rho = - m^{\mu}\bar{\delta}\ell_{\mu} \,, \\
	& \pi =  \bar{m}^{\mu}Dn_{\mu} \,,\quad \nu =  \bar{m}^{\mu}\triangle n_{\mu} \,,\quad \mu =  \bar{m}^{\mu}\delta n_{\mu} \,,\quad \lambda = \bar{m}^{\mu}\bar{\delta}n_{\mu} \,,
\ea\ee
which transform covariantely under local rotation and boost transformations, and the $4$ so-called ``bad'' spin coefficients,
\be\label{eq:NPSpinCoefficientsBad}
	\begin{split}
		\varepsilon &= -\frac{1}{2}\left(n^{\mu}D\ell_{\mu}-\bar{m}^{\mu}D m_{\mu}\right) \,,\quad \gamma = -\frac{1}{2}\left(n^{\mu}\triangle\ell_{\mu}-\bar{m}^{\mu}\triangle m_{\mu}\right) \,, \\
		\beta &= -\frac{1}{2}\left(n^{\mu}\delta\ell_{\mu}-\bar{m}^{\mu}\delta m_{\mu}\right) \,,\quad \alpha = -\frac{1}{2}\left(n^{\mu}\bar{\delta}\ell_{\mu}-\bar{m}^{\mu}\bar{\delta}m_{\mu}\right) \,,
	\end{split}
\ee
which do not have definite spin and boost weights. In the above expressions, the NP directional derivatives $D$, $\triangle$, $\delta$ and $\bar{\delta}$ are the covariant derivatives projected onto the null tetrads,
\be
	D \equiv \ell^{\mu}\nabla_{\mu} \,,\quad \triangle \equiv n^{\mu}\nabla_{\mu} \,,\quad \delta \equiv m^{\mu}\nabla_{\mu} \,,\quad \bar{\delta} \equiv \bar{m}^{\mu}\nabla_{\mu} \,,
\ee
whose action on a NP scalar also does not transform homogeneously (see Appendix \ref{ApLieDerivative}).

\subsection{Spin weighted generators of isometries}

For the special case of a Killing vector field $\xi^{\mu}$, there exists a canonical generalization $\Lstr_{\xi}$ of the usual Lie derivative $\mathcal{L}_{\xi}$ that acts on a NP scalar of spin weight $s$ and boost weight $b$~\cite{Ludwig2000,Ludwig2002},
\be\label{eq:LieLudwig}
	\Lstr_{\xi} = \mathcal{L}_{\xi} + \left(b\,n_{\mu}\mathcal{L}_{\xi}\ell^{\mu} - s\,\bar{m}_{\mu}\mathcal{L}_{\xi}m^{\mu}\right) \,.
\ee
These are improvements of the traditional GHP Lie derivatives such that the information that $\xi^{\mu}$ is a Killing vector can be read off from the tetrad vectors themselves, namely, if there exists a null tetrad vector which is 
annihilated by $\Lstr_{\xi}$, then $\xi$ is a Killing vector field \cite{Ludwig2000,Ludwig2002}. We will refer to this generalized Lie derivative as the spin weighted Lie derivative in what follows. The spin weighted Lie derivative operator is covariant w.r.t. spin and boost transformations and does not change the spin or boost weights of the NP scalar it acts on.

As an explicit application, let us compute these derivatives along the Killing vectors of the extremal near horizon isometry group \eqref{eq:SL2RNHEKerrNewman}. We will use the fact that $b=s$ for the NP scalars appearing in the Teukolsky equation and the Kinnersley tetrad~\cite{Kinnersley1969},
\be\ba
	\ell &= \frac{r^2+a^2}{\Delta}\left(\partial_{t}+\frac{a}{r^2+a^2}\,\partial_{\phi}\right) + \partial_{r}\,, \\
	n &=
	\frac{\Delta}{2\Sigma}\left[\frac{r^2+a^2}{\Delta}\left(\partial_{t}+\frac{a}{r^2+a^2}\,\partial_{\phi}\right) - \partial_{r} \right]\,, \\
	m &= \frac{1}{\sqrt{2}\left(r+ia\cos\theta\right)}\left[ia\sin\theta\,\partial_{t} + \partial_{\theta} + \frac{i}{\sin\theta}\,\partial_{\phi}\right]\,,
\ea\ee
which we simplify in the scaling limit Eq.~\eqref{eq:limKerr}. Using the notation
\be\ba
	\xi_{m}^{\left(s\right)} \equiv \Lstr_{\xi_{m}}
	 \,\,\,,\,\,\, m=0,\pm1 \,,
\ea\ee
we obtain the following generators corrected by $s$-dependent scalar pieces,
\be\label{eq:SL2RNHEKerr_s}
	\xi_0^{\left(s\right)} = \xi_0- s \,,\quad \xi_{+1}^{\left(s\right)} = \xi_{+1}\,, \quad \xi_{-1}^{\left(s\right)} = \xi_{-1} - 2s\left(\frac{M^2}{\rho}+\tau\right) \,.
\ee
These generators satisfy the $\SL$ commutation relations. The corresponding Casimir reproduces the relevant part of the radial spin-$s$ Teukolsky differential operator for an extremal black hole in the near horizon limit,
\be\label{eq:KNNHE_s}
	\begin{split}
		\mathcal{C}_{2,~\SL}^{\text{NHE},~(s)} & = \rho^{-2s}\partial_{\rho}\rho^{2\left(s+1\right)}\partial_{\rho} - \frac{\left(M^2+a^2\right)^2}{\rho^2} \partial_{\tau}^2 + \frac{4 Ma}{\rho}\partial_{\tau}\partial_{\varphi} + 2s\frac{M^2+a^2}{\rho}\partial_{\tau} + s\left(s+1\right) \\
		& = \left(r-M\right)^{-2s}\partial_{r}\left(r-M\right)^{2(s+1)}\partial_{r} - \frac{\left(M^2+a^2\right)^2}{\left(r-M\right)^2}\left(\partial_{t}+\Omega\,\partial_{\phi}\right)^2 \\
		&~~+ \frac{4Ma}{r-M}\left(\partial_{t}+\Omega\,\partial_{\phi}\right)\partial_{\phi} + 2s\frac{M^2+a^2}{r-M}\partial_{t} + s(s+1) \,.
	\end{split}
\ee
The full near horizon Teukolsky equation is obtained by adding the appropriate $U\left(1\right)$ Casimir,
\be
	\mathcal{C}_{2,\text{full}}^{\text{NHE},~(s)}\Psi_s = \left[\mathcal{C}_{2,~\SL}^{\text{NHE},~(s)} -2\left(3M^2+a^2\right)\Omega^2\,\partial_{\phi}^2\right]\Psi_s=\ell\left(\ell+1\right)\Psi_s \,.
\ee

We see that, in order to describe the symmetry structure of the extremal black hole geometries relevant for spin-$s$ perturbations, we need to introduce the spin weighted Lie derivatives. In a coordinate basis, these derivatives modify the isometry generators by scalar $s$-dependent pieces, see Eq.~\eqref{eq:SL2RNHEKerr_s}. The modified generators inherit the 
algebraic structure of the underlying extremal black hole geometry.

\subsection{Near zone symmetries as isometries of subtracted geometries}

Unfortunately, the generalized Lie derivative can be uniquely defined only for exact Killing vectors, see Appendix \ref{ApLieDerivative}. The non-extremal Love symmetry \eqref{eq:SL2RsKerrNewman} or the Starobinsky near zone $\SL$ symmetry \eqref{eq:SL2RStarobinsky} are not exact and in this situation the additional $s$-dependent scalar pieces in the generators cannot be directly assigned a strict geometric meaning. However, there exists a framework where the vector fields generating the Love or the Starobinsky near zone $\SL$'s are realized as Killing vectors of effective geometries, which are in turn realized as relatives of subtracted geometries of the Kerr-Newman black hole geometry (\cite{Cvetic:2011hp,Cvetic:2011dn}).

Let us briefly review how the construction of subtracted geometries is performed, focusing to the geometry of interest of a Kerr-Newman black hole. In the notation of \cite{Cvetic:2011dn}, the full Kerr-Newman geometry is written as,
\be
	ds^2 = -\Delta_0^{-1/2}G\left(dt+\mathcal{A}\right)^2 +\Delta_0^{1/2}\left(\frac{dr^2}{X}+d\theta^2+\frac{X}{G}\sin^2\theta\,d\phi^2\right) \,,
\ee
where
\be
	G=X-a^2\sin^2\theta \,.
\ee
In the standard notations of \eqref{eq:KerrNewmanMetricBL}, the warp factor $\Delta_0$, the discriminant $X$ and the angular potential $\mathcal{A}$ are given by,
\be\ba
	\Delta_0 = \Sigma^2 \,,\quad X=\Delta \,,\quad \mathcal{A} = \frac{a\sin^2\theta}{G}\left(2Mr-Q^2\right)d\phi \,.
\ea\ee
The main observation then is that the thermodynamic variables of the black hole are completely independent of the warp factor $\Delta_0$, a fact suggesting that $\Delta_0$ encodes information about the environment around the black hole rather than its interior
\cite{Cvetic:2011dn}. Furthermore, the location of the ergosphere only depends on $G$, corresponding to $G=0$. A subtracted geometry as introduced in \cite{Cvetic:2011hp} then corresponds to modifying the warp factor $\Delta_0$, which preserves the internal structure of the black hole. We will more loosely refer to a subtracted geometry of the Kerr-Newman black hole, to whatever geometry which preserves its thermodynamic properties. In other words, besides arbitrary modifications of the warp factor $\Delta_0$, we will also allow alterations of the angular potential $\mathcal{A}$ and the function $G$ such that its near horizon behavior is preserved,
\be\label{eq:SubtractedGeometry}
	ds^2_{\text{sub}} = -\Delta_{0,\text{sub}}^{-1/2}G_{\text{sub}}\left(dt+\mathcal{A}_{\text{sub}}\right)^2 +\Delta_{0,\text{sub}}^{1/2}\left(\frac{dr^2}{\Delta}+d\theta^2+\frac{\Delta}{G_{\text{sub}}}\sin^2\theta\,d\phi^2\right) \,,
\ee
with arbitrary $\Delta_{0,\text{sub}}$ and,
\be\ba
	{}&\lim\limits_{r\rightarrow r_{+}}\mathcal{A}_{\text{sub}} = \lim\limits_{r\rightarrow r_{+}}\mathcal{A} = -\left(r_{+}^2+a^2\right)d\phi \,,
	&\lim\limits_{r\rightarrow r_{+}}G_{\text{sub}} = \lim\limits_{r\rightarrow r_{+}}G = -a^2\sin^2\theta \,.
\ea\ee

Within this framework, both the Love symmetry as well as the Starobinsky near zone $\SL$ vector fields can be realized as isometries of the relevant subtracted geometries and this in turn allows to apply the uniquely defined spin weighted Lie derivative for isometries to attempt to infer the $s\ne0$ pieces.

\subsubsection{Starobinsky near zone symmetry}
We start with the $s=0$ Starobinsky near zone $\SL$ generators \eqref{eq:SL2RStarobinsky}. The relevant subtracted geometry that ensures separability of the massless Klein-Gordon operator is given by \eqref{eq:SubtractedGeometry} with,
\be
	\Delta_{0,\text{Star}} = \left(r_{+}^2+a^2\right)^2 \,,\quad \mathcal{A}_{\text{Star}} = \frac{a\sin^2\theta}{G}\left(r_{+}^2+a^2\right)d\phi \,\quad G_{\text{Star}} = G \,.
\ee
This geometry is what \cite{Hui:2022vbh} refers to as the ``effective near zone geometry'' of the Kerr-Newman black hole, here connected to the earlier notion of subtracted geometries (\cite{Cvetic:2011hp,Cvetic:2011dn}). Let us see if the spin weighted Lie derivative \eqref{eq:LieLudwig} captures the correct $s\ne0$ extensions in \eqref{eq:SL2RStarobinsky}. We first need to identify a proper set of tetrad vectors. This is achieved by,
\be\ba
	\ell_{\text{Star}} &= \frac{r_{+}^2+a^2}{\Delta}\left(\partial_{t}+\Omega\partial_{\phi}\right) + \partial_{r} \,, \\
	n_{\text{Star}} &= \frac{\Delta}{2\left(r_{+}^2+a^2\right)}\left[\frac{r_{+}^2+a^2}{\Delta}\left(\partial_{t}+\Omega\partial_{\phi}\right) - \partial_{r}\right] \,, \\
	m_{\text{Star}} &= \frac{1}{\sqrt{2}\left(r_{+}+ia\right)}\left[\partial_{\theta} + \frac{i}{\sin\theta}\partial_{\phi}\right] \,,
\ea\ee
which preserves the algebraic classification of the Kerr-Newman black hole, i.e. the above subtracted geometry is still a Petrov type-D spacetime. A direct calculation then reveals that the $s\ne0$ pieces in the Starobinsky near zone $\SL$ generators \eqref{eq:SL2RStarobinsky} are exactly reproduced by the spin weighted Lie derivative \eqref{eq:LieLudwig},
\be
	\begin{gathered}
		\Lstr_{L_{0}^{\text{Star}}} = L_{0}^{\text{Star}} = L_{0}^{\text{Star},~\left(s\right)} \,, \\
		\Lstr_{L_{\pm1}^{\text{Star}}} = L_{\pm1}^{\text{Star}} \mp s\,e^{\pm t/\beta}\frac{r-r_{\mp}}{\sqrt{\Delta}} = L_{\pm1}^{\text{Star},~\left(s\right)} \,,
	\end{gathered}
\ee
where $L_{m}^{\text{Star}}$ are the $s=0$ vector fields generating the Starobinsky near zone $\SL$ symmetry of the massless Klein-Gordon equation. This shows that the $s\ne0$ pieces in the Starobinsky near zone $\SL$ generators can indeed be assigned a geometric interpretation in terms of spin weighted Lie derivatives in a particular subtracted geometry.

\subsubsection{Love symmetry}
For the Love symmetry $s=0$ vector fields \eqref{eq:SL2RKerr}, the corresponding subtracted geometry is given by \eqref{eq:SubtractedGeometry} with,
\be
	\begin{gathered}
		\Delta_{0,\text{Love}} = \left(r_{+}^2+a^2\right)^2\left(1+\beta^2\Omega^2\sin^2\theta\right) \,, \\
		\mathcal{A}_{\text{Love}} = \frac{a\sin^2\theta}{G}\left(r_{+}^2+a^2+\beta\left(r-r_{+}\right)\right)d\phi \,,\quad G_{\text{Love}} = G \,.
	\end{gathered}
\ee
One can then attempt to infer the $s\ne0$ generators \eqref{eq:SL2RsKerrNewman} using the spin weighted Lie derivative \eqref{eq:LieLudwig}. The associated tetrad vectors are, up to local boosts and rotations,
\be\ba
	\ell_{\text{Love}} &= \frac{r_{+}^2+a^2}{\Delta}\left(\partial_{t}+\frac{\Delta^{\prime}}{r_{+}-r_{-}}\Omega\partial_{\phi}\right) + \partial_{r} \,, \\
	n_{\text{Love}} &= \frac{\Delta}{2\Delta_{0,\text{Love}}^{1/2}}\left[\frac{r_{+}^2+a^2}{\Delta}\left(\partial_{t}+\frac{\Delta^{\prime}}{r_{+}-r_{-}}\Omega\partial_{\phi}\right) - \partial_{r}\right] \,, \\
	m_{\text{Love}} &= \frac{1}{\sqrt{2}\mathcal{M}_{0,\text{Love}}}\left[\partial_{\theta} + i\sqrt{\frac{1}{\sin^2\theta}+\beta^2\Omega^2}\,\partial_{\phi}\right] \,,
\ea\ee
with $\left|\mathcal{M}_{0,\text{Love}}\right|^2 = \Delta_{0,\text{Love}}^{1/2}$. These would imply the following $s\ne0$ extensions of the Love symmetry vector fields,
\be
	\begin{gathered}
		\Lstr_{L_{0}^{\text{Love}}} = L_{0}^{\text{Love}} = L_{0}^{\text{Love},~\left(s\right)}-s \,, \\
		\Lstr_{L_{\pm1}^{\text{Love}}} = L_{\pm1}^{\text{Love}} \mp s\,e^{\pm t/\beta}\frac{r-r_{\mp}}{\sqrt{\Delta}} = L_{\pm1}^{\text{Love},~\left(s\right)} + s\,e^{\pm t/\beta}\sqrt{\frac{r-r_{+}}{r-r_{-}}} \,,
	\end{gathered}
\ee
which do not reproduce the actual $s\ne0$ pieces involved in the Love symmetry generators \eqref{eq:SL2RsKerrNewman}. As a result a geometric interpretation in terms of spin weighted Lie derivatives for the subtracted geometry does  not  seem to be possible in this case\footnote{As a disclaimer here, there actually exists a possible choice of tetrads that ensures that the spin weighted Lie derivative does indeed capture the correct $s\ne0$ pieces of the Love symmetry generators in \eqref{eq:SL2RsKerrNewman}. This is given by a particular local boost,
\be
	\ell_{\text{Love}} \rightarrow \lambda \ell_{\text{Love}} \,,\quad n_{\text{Love}}\rightarrow \lambda^{-1}n_{\text{Love}} \,,\quad m_{\text{Love}}\rightarrow m_{\text{Love}} \,,
\ee
with $\lambda = e^{\left(t-\phi/\Omega\right)\beta}$. However, this local boost is not globally defined and does not have a smooth spinless limit.}. The corresponding Casimir associated with this spin weighted Lie derivative along the Love symmetry vector fields is,
\be
	\text{\sout{$\mathcal{C}$}}_2^{\text{Love}} = \mathcal{C}_2^{\text{Love},~\left(s\right)} + s \frac{r_{+}-r_{-}}{r-r_{-}}\left(\partial_{t}+\Omega\,\partial_{\phi}\right) \,,
\ee
which fails to be a valid near zone truncation due to the additional static contribution for $\Omega\ne0$, while, for $\Omega=0$, this is just the Starobinsky near zone approximation \eqref{eq:starCas}-\eqref{eq:starCas1}.

Another peculiar property of the Love symmetry generators is that the corresponding $s$-dependent pieces can be completely gauged away through a particular globally defined local boost transformation of the NP scalars involved in the Teukolsky equation,
\be
	\Psi_{s} \rightarrow \tilde{\Psi}_{s} = \left(e^{-t/\beta}\sqrt{\Delta}\right)^{s}\Psi_{s} \Rightarrow L_{m}^{\text{Love},~\left(s\right)}\Psi_{s} = \left(e^{-t/\beta}\sqrt{\Delta}\right)^{-s}L_{m}^{\text{Love}}\tilde{\Psi}_{s} \,.
\ee
In other words, the Love near zone truncation of the spin-$s$ Teukolsky equation \eqref{eq:NZgen}-\eqref{eq:V0V1} is effectively a near zone truncation of the $s=0$ Teukolsky equation for the boosted NP scalars $\tilde{\Psi}_{s}$. Note, however, that the required boost has a non-trivial time dependence, so that it turns static perturbations turn into time-dependent ones.


\subsection{Infinite-dimensional extension and relation to extremal near horizon isometries}
We will finish this section by demonstrating how one can recover the spin weighted Killing vectors of the near horizon geometry of extremal Kerr-Newman black holes from non-extremal $\SL$ algebras. As in the scalar example analyzed in Section \ref{SecNHE}, these non-extremal $\SL$ algebras are subalgebras of the infinite extension $\SL\ltimes U\left(1\right)_{\mathcal{V}}$ found in Section \ref{sec:InfiniteExtension}. Here, the $U\left(1\right)$ vector field in \eqref{eq:U1V} is supplemented by a scalar piece,
\be
	\beta\Omega\,\partial_{\phi} \rightarrow \beta\Omega\,\partial_{\phi} + s \,.
\ee
The $\SL_{\left(\alpha\right)}$ subalgebras of this $s$-extended $\SL\ltimes U\left(1\right)_{\mathcal{V}}$ are then generated by,
\be\label{eq:SL2Ralpha_s}
	L_{m}^{\left(s\right)}\left(\alpha\right) = L_{m}^{\left(s\right)} + \alpha\,\upsilon_{0,m}\left(\beta\Omega\,\partial_{\phi} + s\right) \,,\quad m=0,\pm1 \,,
\ee
where $L_{m}^{\left(s\right)}$ are the $s$-extended Love generators \eqref{eq:SL2RsKerrNewman} and $\upsilon_{0,m}$ belong to the representation $\mathcal{V}$ constructed from $L_{m}^{\left(0\right)}$ (see Fig. \ref{fig:VSL2R}). The associated Casimir is given by,
\be
	\begin{aligned}
		\mathcal{C}_2^{\left(s\right)}\left(\alpha\right) &= \Delta^{-s}\partial_{r}\Delta^{s+1}\partial_{r} - \frac{\left(r_{+}^2+a^2\right)^2}{\Delta}\left(\partial_{t}+\Omega\,\partial_{\phi}\right)^2 \\
		&+ s\frac{\left(r_{+}^2+a^2\right)\Delta^{\prime}}{\Delta}\left(\partial_{t}+\Omega\,\partial_{\phi}\right) + s\left(s+1\right) \\
		&+ 2\frac{r_{+}^2+a^2}{r-r_-}\left(\partial_{t}+\alpha\,\Omega\,\partial_{\phi} + \alpha\,\frac{s}{\beta}\right)\left(\alpha-1\right)\beta\Omega\,\partial_{\phi} \,.
	\end{aligned}
\ee
Again, even though all of these operators preserve the near horizon characteristic exponents of the non-extremal Teukolsky equation, only the choices $\alpha=0$ and $\alpha=1$ give rise to valid near zone approximations, corresponding to the $s$-extended Love and Starobinsky near zones respectively.

Choosing $\alpha$ as in equation \eqref{eq:NHa} for arbitrary sub-extremal contributions, we obtain the $\SL$ algebra,
\be\label{eq:SL2RNHne_s}
	\begin{gathered}
		L_0^{\text{NHE},~\left(s\right)} = -\frac{1}{2\pi T_{H}}\left[\partial_{t}+\left(1+4\pi T_{H}M\right)\Omega\,\partial_{\phi}\right] + \mathcal{O}\left(T_{H}\right) \,, \\
		\ba
			L_{\pm 1}^{\text{NHE},~\left(s\right)} = e^{\pm 2\pi T_{H}t} &\bigg[\mp\sqrt{\Delta}\,\partial_{r} + \partial_{r}\left(\sqrt{\Delta}\right)\frac{1}{2\pi T_{H}}\left(\partial_{t}+\Omega\,\partial_{\phi}\right) + s\frac{r-r_{\mp}}{\sqrt{\Delta}} \\
			&+ 2M\sqrt{\frac{r-r_{+}}{r-r_{-}}} \Omega\,\partial_{\phi} + \mathcal{O}\left(T_{H}\right) \bigg] \,.
		\ea
	\end{gathered}
\ee
The Casimir of this algebra is explicitly given by
\be\label{eq:CasNH1_s}
	\ba
		\mathcal{C}_2^{\text{NHE}} &= \Delta^{-s}\partial_{r}\Delta^{s+1}\partial_{r} - \frac{\left(r_{+}^2+a^2\right)^2}{\Delta}\left(\partial_{t}+\Omega\,\partial_{\phi}\right)^2 + 2\frac{r_{s}a}{r-r_{-}}\left(\partial_{t}+\Omega\,\partial_{\phi}\right)\partial_{\phi} \\
		&+ s\frac{\left(r_{+}^2+a^2\right)\Delta^{\prime}}{\Delta}\left(\partial_{t}+\Omega\,\partial_{\phi}\right) + s\left(s+1\right)  \\
		&+ \frac{r_{+}-r_{-}}{r-r_{-}}\left(2M\Omega\,\partial_{\phi}+s\right)2M\Omega\,\partial_{\phi} + \mathcal{O}\left(T_{H}\right) \,,
	\ea
\ee
which reduces to the appropriate near horizon extremal Teukolsky operator \eqref{eq:KNNHE_s} in the extremal limit.

We can then follow the same prescription we used in \ref{sec:NHEKerrNewman} to recover the spin weighted Killing vectors \eqref{eq:SL2RNHEKerr_s},
\be
	\begin{split}
		\xi_{+1}^{\left(s\right)} &= \lambda^{-1}\lim_{T_{H}\to 0} \left(-2\pi T_{H} L_0^{\text{NHE},~\left(s\right)}\right) \,, \\
		\xi_0^{\left(s\right)} &= \lim_{T_{H}\to 0} \frac{L_{+1}^{\text{NHE},~\left(s\right)}-L_{-1}^{\text{NHE},~\left(s\right)}}{2} \,, \\
		\xi_{-1}^{\left(s\right)} &= \lambda\lim_{T_{H}\to 0} \frac{L_{+1}^{\text{NHE},~\left(s\right)}+L_{-1}^{\text{NHE},~\left(s\right)}+2 L_0^{\text{NHE},~\left(s\right)}}{2\pi T_{H}} \,.
	\end{split} 
\ee

An important remark here that was not relevant in the scalar example in Section \ref{SecNHE} is that the infinite-dimensional extension $\SL\ltimes U\left(1\right)_{\mathcal{V}}$ is now needed not only in the Kerr-Newman case, but also in the spherically symmetric Reissner-Nordstr\"{o}m case in order to correctly reproduce the full spin weighted Killing vectors \eqref{eq:SL2RNHEKerr_s}.

Last, it is instructive to ask whether the $\SL_{\left(\alpha\right)}$ subalgebras of the infinite extension $\SL\ltimes U\left(1\right)_{\mathcal{V}}$ admit similar subtracted geometry analyses as with the near zone $\SL$'s. The subtracted geometry associated with $\SL_{\left(\alpha\right)}$ is given by \eqref{eq:SubtractedGeometry}, now with,
\be\ba
	\Delta_{0,\left(\alpha\right)} &= \left(r_{+}^2+a^2\right)^2\left(1+\left(\alpha-1\right)^2\beta^2\Omega^2\sin^2\theta\right) \\
	\mathcal{A}_{\left(\alpha\right)} &= \frac{a\sin^2\theta}{G_{\left(\alpha\right)}}\left[r_{+}^2+a^2-\left(\alpha-1\right)\beta\left(r-r_{+}\right)\right] \,, \\
	G_{\left(\alpha\right)} &= G + 4\,\alpha\left(\alpha-1\right)\frac{r-r_{+}}{r_{+}-r_{-}}a^2\sin^2\theta \,.
\ea\ee
The $\SL_{\left(\alpha\right)}$ generators \eqref{eq:SL2Ralpha_s} are then Killing vectors of this geometry. Even though this effective geometry preserves the thermodynamic properties of the Kerr-Newman black hole, it fails to capture properties that extend beyond the near horizon behavior. For instance, the ergosphere of this black hole geometry is now located at $G_{\left(\alpha\right)}=0$ which does not match the original locus condition $G=0$ unless $\alpha=0$ or $\alpha=1$. 

Let us finally see whether the $s\ne0$ pieces can be inferred from the spin weighted Lie derivative \eqref{eq:LieLudwig}. The corresponding tetrad vectors are, up to local boosts and rotations,
\be\ba
	\ell_{\left(\alpha\right)} &= \frac{r_{+}^2+a^2}{\Delta}\left(\partial_{t}+\frac{\Delta^{\prime}-2\alpha\left(r-r_{+}\right)}{r_{+}-r_{-}}\Omega\partial_{\phi}\right) + \partial_{r} \,, \\
	n_{\left(\alpha\right)} &= \frac{\Delta}{2\Delta_{0,\left(\alpha\right)}^{1/2}}\left[\frac{r_{+}^2+a^2}{\Delta}\left(\partial_{t}+\frac{\Delta^{\prime}-2\alpha\left(r-r_{+}\right)}{r_{+}-r_{-}}\Omega\partial_{\phi}\right) - \partial_{r}\right] \,, \\
	m_{\left(\alpha\right)} &= \frac{1}{\sqrt{2}\mathcal{M}_{0,\left(\alpha\right)}}\left[\partial_{\theta} + i\sqrt{\frac{1}{\sin^2\theta}+\left(\alpha-1\right)^2\beta^2\Omega^2}\,\partial_{\phi}\right] \,,
\ea\ee
with $\left|\mathcal{M}_{0,\left(\alpha\right)}\right|^2=\Delta_{0,\left(\alpha\right)}^{1/2}$. The spin weighted Lie derivative \eqref{eq:LieLudwig} then outputs,
\be
	\begin{gathered}
		\Lstr_{L_{0}\left(\alpha\right)} = L_{0}\left(\alpha\right) = L_{0}^{\left(s\right)}\left(\alpha\right) - \left(1-\alpha\right)s \,, \\
		\Lstr_{L_{\pm1}\left(\alpha\right)} = L_{\pm1}\left(\alpha\right) \mp s\,e^{\pm t/\beta}\frac{r-r_{\mp}}{\sqrt{\Delta}} = L_{\pm1}^{\left(s\right)}\left(\alpha\right) +\left(1-\alpha\right)s\,e^{\pm t/\beta}\sqrt{\frac{r-r_{+}}{r-r_{-}}} \,,
	\end{gathered}
\ee
and the corresponding spin weighted Casimir is given by,
\be
	\text{\sout{$\mathcal{C}$}}_2^{\left(s\right)}\left(\alpha\right) = \mathcal{C}_2^{\left(s\right)}\left(\alpha=1\right) + 2\frac{r_{+}^2+a^2}{r-r_{-}}\left(\partial_{t}+\alpha\,\Omega\,\partial_{\phi}+\frac{s}{\beta}\right)\left(\alpha-1\right)\beta\Omega\,\partial_{\phi} \,,
\ee
where $\mathcal{C}_2^{\left(s\right)}\left(\alpha=1\right)$ is the Starobinsky near zone $\SL$ Casimir.

Just like in the analysis for the Love symmetry generators, the spin weighted Lie derivative does not agree with the actual $s\ne0$ pieces of $L_{m}^{\left(s\right)}\left(\alpha\right)$\footnote{Similar to the Love symmetry example, there exists a particular local boost on the tetrad vectors with the same boost parameter $\lambda=e^{\left(t-\phi/\Omega\right)/\beta}$ whose implementation in the spin weighted Lie derivative correctly reproduces the appropriate scalar pieces in $L_{m}^{\left(s\right)}\left(\alpha\right)$, but is not globally defined and does have a smooth $\Omega\rightarrow0$ limit. Interestingly, this boost parameter is independent of $\alpha$, but the Starobinsky near zone algebra is in fact invariant under such transformations involving the co-rotating azimuthal angle, which allows to always gauge away such pathological factors.}. In fact, the $s\ne0$ corrections predicted from the spin weighted Lie derivative are independent of $\alpha$ and they are always equal to the $s\ne0$ pieces of the Starobinsky near zone $\SL$ algebra generators. However, in contrast to the Love symmetry example, the $s\ne0$ pieces of $L_{m}^{\left(s\right)}\left(\alpha\right)$ cannot be gauged away by any globally defined local boost or rotation.

\section{Further generalizations}
\label{sec:furt}

We have seen 
that many properties of static
Love numbers of non-extremal black holes
can be explained by means of
the Love symmetry
and its extensions.
We also discussed the relation of the Love symmetry to the near horizon isometries in the extremal case.
However, the near zone approximation breaks down for the extremal black holes.
In this section we show that the vanishing 
of Love numbers of extremal black holes
can be proven using a novel so-called ``middle zone''
Love symmetry.

In the second part of this section we present the most general 
infinite-dimensional and globally defined extension of the Love symmetry and of the
Starobinsky near zone algebra.
Finally, we discuss some implications 
of the Love symmetry for modified
gravity theories.

\subsection{Extremal middle zones}

Let us see that in the extremal case there is another
set of generators that allows us to derive
the vanishing of Love numbers from the group theory 
arguments. Let us first focus on the Reissner-Nordstr{\"o}m case.
Consider the following set of vector fields
regular at the future horizon,
\be
	\begin{split}
		& L_0 = -2M\d_t +s \,, \\ 
		&L_{\pm 1}= e^{\pm \left(\frac{t}{2M} -\frac{M}{2\left(r-M\right)}\right)} \left[\mp \left(r-M\right)\d_r  + 2M\left(1
		\pm \frac{M}{2\left(r-M\right)}\right)\d_t -s\left(1\pm 1\right)\right] \,.
	\end{split} 
\ee
The corresponding  Casimir is given by 
\be
	\begin{split}
		\mathcal{C}_2=&\left(r-M\right)^{-2s}\d_r\left(r-M\right)^{2\left(s+1\right)}\d_r 
		+\frac{M^4}{\left(r-M\right)^2}\,\d_t^2 \\
		&-2M^2\,\d_t \d_r - 2s\frac{M^2}{r-M}\,\d_t + s\left(s+1\right)
		\,.
	\end{split} 
\ee
Strictly speaking, it matches the
Teukolsky equation 
only in the static limit:
the $\d_t^2$ term has a wrong sign, and hence 
the Casimir provides a good approximation 
to the physical Teukolsky operator only in the regime
\be
\label{eq:mz}
	\omega M^2 \ll r-M \ll 1/\omega \,.
\ee
We call this region the ``middle zone''
in what follows.
The appearance of the middle zone
is natural because the extremal near horizon patch 
obviously decouples from the asymptotically flat 
region.
Previously, we had reproduced the 
static solution and derived the properties of Love numbers
from the near horizon symmetry.
However, the Love numbers themselves
are defined through the matching in the asymptotically 
flat space patch,
and the fact that they can be fully 
extracted from 
the near horizon 
approximation looks 
like a miracle. This might simply be a result 
of some accidental degeneracy that takes place 
in the Reissner-Nordstr{\"o}m case.

In contrast to the near horizon region, 
the middle zone interpolates between the asymptotic infinity
and the near horizon region. 
Hence, it is natural to expect 
that it is this symmetry that should be important 
for Love numbers of extremal black holes. 
Indeed, we 
can use this symmetry 
for an alternative derivation of the vanishing of Love 
numbers of extremal four dimensional black holes.
However, the near horizon symmetry
is a symmetry of the background geometry itself,
while right now we do not have a similar interpretation 
for the ``middle zone'' symmetry. It remains to be seen
if there is a deeper reason behind the appearance of this symmetry.

Importantly, 
there is an analog of the middle zone 
for the extremal rotating black holes, which captures
solutions that are not even present in the AdS$_2$
near horizon throat.
The static part of the extreme Kerr-Newman black hole near zone Teukolsky equation can be  
reproduced with the following $\SL$ vector fields,
\be
\label{eq:algmid}
	\begin{split}
		L_0 &= -2M\d_t+s\,, \\ 
		L_{\pm 1} &= e^{\pm \left(\frac{t}{2M} -\frac{M^2+a^2}{2M\left(r-M\right)}\right)} \bigg[\mp \left(r-M\right)\d_r  + 2M \left(1\pm \frac{M^2+a^2}{2M\left(r-M\right)}\right)\d_t  \\
		&\qquad\qquad\qquad\qquad\,\,\,\, +\frac{a}{r-M}\,\d_\phi -s\left(1\pm 1\right)\bigg] \,,
	\end{split} 
\ee
which are regular in the advanced Kerr coordinates.
The Casimir of this algebra 
\be
	\begin{split}
		&\mathcal{C}_2= \left(r-M\right)^{-2s}\d_r \left(r-M\right)^{2\left(s+1\right)} \d_r +\frac{\left(M^2+a^2\right)^2}{\left(r-M\right)^2} \left(\d_{t}^2 - \Omega^2\,\d_\phi^2\right) \\
		&-\frac{4M a}{r-M}\,\d_t\d_\phi - 2\left(M^2+a^2\right) \d_t \d_r - 2s\frac{M^2+a^2}{r-M}\left(\d_t - \Omega\,\d_\phi\right) + s\left(s+1\right) \,,
	\end{split}
\ee
is again somewhat different from the full extremal Kerr-Newman low-frequency 
Teukolsky operator~\eqref{eq:NZextr},
but it is still accurate 
provided that Eq.~\eqref{eq:mz} holds true.
Importantly, the middle zone symmetry also 
captures states 
with mode frequencies different from the locking one,
and hence it is
suitable to describe dynamics outside
the throat.

We can use the middle zone symmetry for an algebraic 
derivation of the vanishing of Love numbers
for general static non-axisymmetric perturbations.
Unlike the Reissner-Nordsr{\"o}m case,
this statement cannot be made within the non-extremal
Love symmetry or the near horizon symmetry. 
The proof goes essentially the same way as
in the non-extremal Kerr-Newman case. The crucial observation
is that the static solution belongs to the highest weight
$\SL$ representation with $h=-\ell$, which dictates its 
polynomial form 
in $r$.

\subsection{Infinite zones of Love from local translations of time}
Interestingly, by carefully solving the constraints that need to be satisfied for a near zone $\SL$ symmetry to exist as is sketched in Appendix \ref{ApGenerators}, the most general forms of the generators come into two classes. These are generalizations of the Love and Starobinsky near zones and are controlled by an arbitrary radial function $g\left(r\right)$ that is regular at the horizon, $g\left(r_{+}\right)=\text{finite}$, according to,
\be\label{eq:SL2RExtension}
	\begin{gathered}
		L_0\left[g\left(r\right)\right] = L_0\left[0\right] \\
		L_{\pm1}\left[g\left(r\right)\right] = e^{\pm g\left(r\right)/\beta}L_{\pm1}\left[0\right] \pm e^{\pm\left(t+g\left(r\right)\right)/\beta}\sqrt{\Delta}g^{\prime}\left(r\right)\partial_{t}
	\end{gathered}
\ee
where primes denote radial derivatives and $L_0\left[0\right]$ and $L_{\pm1}\left[0\right]$ are the already found expressions for the Love and Starobinsky near zone $\SL$ generators in \eqref{eq:SL2RsKerrNewman} and \eqref{eq:SL2RStarobinsky} respectively. For higher-dimensional spherically symmetric black holes, the same extension holds but this time with primes denoting derivatives with respect to the variable $\rho=r^{d-3}$. These generalizations are not simple rescalings of $L_{\pm1}$, which trivially leave the algebra and the Casimir unchanged. They are \textit{local} rescalings accompanied with an additional $t$-component which are realized as local, $r$-dependent, translations of the temporal coordinate,
\be
	\tilde{t} \equiv t + g\left(r\right) \,.
\ee
This is seen more transparently by comparing \eqref{eq:SL2RsKerrNewman} and \eqref{eq:SL2RStarobinsky} in $\left(t,r,\phi\right)$ coordinates with the explicit expressions for the generalized generators in $\left(\tilde{t},r,\phi\right)$ coordinates,
\begin{itemize}
	\item \underline{Generalized Love near zone}:
	\be
		\begin{gathered}
			L_0\left[g\right] = -\beta\partial_{\tilde{t}} + s \\
			L_{\pm1}\left[g\right] = e^{\pm\tilde{t}/\beta} \left[ \mp\sqrt{\Delta}\,\partial_{r} + \partial_{r}\left(\sqrt{\Delta}\right)\beta\,\partial_{\tilde{t}} + \frac{a}{\sqrt{\Delta}}\,\partial_{\phi} - s\left(1\pm1\right)\partial_{r}\left(\sqrt{\Delta}\right) \right]
		\end{gathered}
	\ee
	
	\item \underline{Generalized Starobinsky near zone}:
	\be
		\begin{gathered}
			L_0^{\text{Star}}\left[g\right] = -\beta\left(\partial_{\tilde{t}}+\Omega\partial_{\phi}\right) \\
			L_{\pm1}^{\text{Star}}\left[g\right] = e^{\pm\tilde{t}/\beta} \left[ \mp\sqrt{\Delta}\,\partial_{r} + \partial_{r}\left(\sqrt{\Delta}\right)\beta\left(\partial_{\tilde{t}}+\Omega\,\partial_{\phi}\right) \mp s\frac{r-r_{\mp}}{\sqrt{\Delta}} \right]
		\end{gathered}
	\ee
\end{itemize}
The Love symmetry argument implying vanishing Love numbers can still be applied in the same way as before but with this translated temporal coordinate $\tilde{t}$ appearing in place of $t$ in the elements of highest weight representations. This does not alter the conclusion that static Love numbers vanish due to the polynomial form, up to overall irrelevant form-factors, of the static solution.

The corresponding equations of motion arising from the Casimirs of the above two generalized near zone $\SL$'s are the same as the ``fundamental'' ($g=0$) ones but with $t$ replaced by $\tilde{t}$.

\subsection{Love symmetry beyond general relativity}

In this section we investigate general conditions for the existence of the Love symmetry in general relativity modifications. Let us focus on a simple example of a massless scalar field in the background of a generalized spherically symmetric black hole geometry which can always be brought to the form,
\be
	ds^2 = -f_{t}\left(r\right)dt^2 + \frac{dr^2}{f_{r}\left(r\right)} + r^2d\Omega_{d-2}^2\,.
\ee
The functions $f_{t}\left(r\right)$ and $f_{r}\left(r\right)$ are arbitrary at this point. In vacuum general relativity, $f_{r}\left(r\right)=f_{t}\left(r\right)$ but in general modified gravity this is not true anymore. The preliminary assumptions we impose on these radial functions are that there exists a horizon at $r=r_{h}$ where $f_{t}\left(r_{h}\right)=f_{r}\left(r_{h}\right)=0$ with multiplicity one,
\be
	f_{t}\left(r_{h}\right)=f_{r}\left(r_{h}\right)=0\,,\quad f_{t}^{\prime}\left(r_{h}\right)\ne0\,,\quad f_{r}^{\prime}\left(r_{h}\right)\ne0\,,
\ee
that is, the geometry describes a non-extremal spherically symmetric black hole. The full radial Klein-Gordon equation after expanding over monochromatic spherical harmonic modes of orbital number $\ell$ reads,
\be
	\begin{gathered}
		\mathbb{O}_{\text{full}}\Phi_{\omega\ell m} = \hat{\ell}(\hat{\ell}+1)\Phi_{\omega\ell m} \\
		\mathbb{O}_{\text{full}} = \partial_{\rho}\Delta_{r}\partial_{\rho} + \frac{\Delta_{r}^2}{2\Delta_{t}}\left(\frac{\Delta_{t}}{\Delta_{r}}\right)^{\prime}\partial_{\rho} - \frac{r^{2\left(d-2\right)}}{\left(d-3\right)^2\Delta_{t}}\partial_{t}^2
	\end{gathered}
\ee
where $\rho=r^{d-3}$ and $\hat{\ell}=\ell/\left(d-3\right)$ as before, $\Delta_{t}\equiv\rho^2f_{t}$, $\Delta_{r}\equiv\rho^2f_{r}$ and primes denote derivatives with respect to $\rho$.

We now explore whether there exist near zone truncations that are equipped with an $\SL$ structure. Of particular interest is the following near zone approximation of the radial Klein-Gordon operator,
\be\label{eq:NZModGR}
	\mathbb{O}_{\text{NZ}} = \partial_{\rho}\Delta_{r}\partial_{\rho} + \frac{\Delta_{r}^2}{2\Delta_{t}}\left(\frac{\Delta_{t}}{\Delta_{r}}\right)^{\prime}\partial_{\rho} - \frac{r_{h}^{2\left(d-2\right)}}{\left(d-3\right)^2\Delta_{t}}\partial_{t}^2 \,.
\ee
This differential operator can be represented as a Casimir of a certain regular and globally defined $\SL$ algebra if and only if the following condition is satisfied,
\be\label{eq:SL2RModGRConstraint}
	\frac{\Delta_{r}}{\Delta_{t}}\Delta_{t}^{\prime\prime} + \frac{1}{2}\Delta_{t}^{\prime}\left(\frac{\Delta_{r}}{\Delta_{t}}\right)^{\prime} = 2 \,.
\ee
The generators of the $\SL$ near zone symmetry are given by,
\be\label{eq:SL2RModGR}
	\begin{gathered}
		L_0 = -\beta\,\partial_{t} \,,\quad
		L_{\pm1} = e^{\pm t/\beta}\left[\mp\sqrt{\Delta_{r}}\,\partial_{\rho} + \sqrt{\frac{\Delta_{r}}{\Delta_{t}}}\partial_{\rho}\left(\sqrt{\Delta_{t}}\right)\beta\,\partial_{t}\right]\,,
	\end{gathered}
\ee
where $\beta$ is the inverse Hawking temperature,
\be
	\beta = \frac{2}{\sqrt{f_{t}^{\prime}\left(r_{h}\right)f_{r}^{\prime}\left(r_{h}\right)}}\,,
\ee
and are regular at both the future and past event horizons. The geometric constraint \eqref{eq:SL2RModGRConstraint} can be solved explicitly,
\be
	\Delta_{r}\left(\rho\right) = \Delta_{t}\left(\rho\right)\frac{4\Delta_{t}\left(\rho\right)+\left(\frac{\beta_{s}}{\beta}\rho_{h}\right)^2}{\Delta_{t}^{\prime2}\left(\rho\right)} \,,
\ee
with $\beta_{s}=\frac{2r_{\text{h}}}{d-3}$ the inverse Hawking temperature for the Schwarzschild black hole. At the level of the functions $f_{t}\left(r\right)$ and $f_{r}\left(r\right)$ themselves, the above condition reads,
\be
	f_{r}\left(r\right) = f_{t}\left(r\right) \frac{\left(d-3\right)^2r^{2\left(d-4\right)}}{\left(r^{2\left(d-3\right)}f_{t}\left(r\right)\right)^{\prime2}} \left[4r^{2\left(d-3\right)}f_{t}\left(r\right) + \left(\frac{\beta_{s}}{\beta}\right)^2r_{\text{h}}^{2\left(d-3\right)}\right] \,.
\ee
As can be checked explicitly, the asymptotic flatness condition is automatically imposed by the above condition if either $f_{t}$ or $f_{r}$ is asymptotically flat. For the case where $f_{r}\left(r\right)=f_{t}\left(r\right)$, we get that the most general such geometry is the higher-dimensional Reissner-Nordstr\"{o}m black hole. Following the procedure outlined in Appendix~\ref{ApGenerators}, one can in fact show that the near zone truncation \eqref{eq:NZModGR} employed here is the only one that is a candidate of being equipped with an $\SL$ structure. Indeed, the above results set absolute geometric constraints on the existence of Love symmetry beyond General Relativity as long as we ignore possible scalar field redefinitions and only consider fields minimally coupled to gravity.

One can also check that the above near zone $\SL$ implies the vanishing of static Love numbers when $\hat{\ell}\in\mathbb{N}$. Using the same symmetry argument of the regular static solution being an element of a highest weight representation of this $\SL$, we obtain $\left(L_{+1}\right)^{\ell+1}\upsilon_{-\ell,\ell} = 0$. We now get a modified ``polynomial'' requirement. In particular, noticing that,
\be
	\left(L_{+1}\right)^{n}F\left(\rho\right) = \left(-e^{t/\beta}\sqrt{\Delta_{t}}\right)^{n}\left[\sqrt{\frac{\Delta_{r}}{\Delta_{t}}}\frac{d}{d\rho}\right]^{n}F\left(\rho\right)
\ee
for an arbitrary purely radial function $F\left(\rho\right)$, we again get that the static radial wavefunction $R_{\omega=0,\ell m}$ belongs to a highest weight representation if and only if $\hat\ell\in\mathbb{N}$, in which case it is again a polynomial, but this time not in the radial variable $\rho=r^{d-3}$, but in the variable $\tilde \rho$, defined as
\be
	d\tilde{\rho} \equiv \sqrt{\frac{\Delta_{t}}{\Delta_{r}}} \,d\rho \Rightarrow \tilde{\rho} = \sqrt{\Delta_{t}+\left(\frac{\beta_{s}}{2\beta}\rho_{\text{h}}\right)^2} +\tilde{\rho}_{h} - \frac{\beta_{s}}{2\beta}\rho_{\text{h}} \,,
\ee
where $\tilde{\rho}_{h}$ is an integration constant indicating the location of the event horizon in this new radial coordinate,
\be
	R_{\omega=0,\ell m}\left(r\right) = \sum_{n=0}^{\hat{\ell}}c_{n}^{\left(m\right)}\tilde{\rho}^{n}\left(r\right)\,\,\,\,\text{if $\hat{\ell}\in\mathbb{N}$} \,.
\ee
We note that, asymptotically, $\tilde{\rho}\rightarrow\rho$ due to the asymptotic flatness of $f_{t}$. Expanding this polynomial in $\tilde{\rho}$ at large distance in the initial radial variable $\rho$, one observes the appearance of an $\rho^{-\hat{\ell}-1}=r^{-\ell-d+3}$ term. However, this term is a relativistic correction in the profile of the ``source'' part of the solution, rather than a response effect from induced multipole moments. Indeed, if the geometric condition \eqref{eq:SL2RModGRConstraint} for the existence of a near zone $\SL$ symmetry is satisfied, we arrive at a situation practically identical to the case of Schwarzschild black hole, Eq.~\eqref{eq:SchwarzschildD_KG}, when working with the variable $\tilde{\rho}$. More explicitly, the full radial Klein-Gordon operator reads,
\be
	\mathbb{O}_{\text{full}} = \partial_{\tilde{\rho}}\Delta_{t}\partial_{\tilde{\rho}} - \frac{r^{2\left(d-2\right)}}{\left(d-3\right)^2\Delta_{t}}\,\partial_{t}^2 \,,
\ee
and $\Delta_{t}$ is a quadratic polynomial in $\tilde{\rho}$,
\be
	\Delta_{t} = \left(\tilde{\rho}-\tilde{\rho}_{\text{h}}\right)\left(\tilde{\rho}-\tilde{\rho}_{\text{h}}+\frac{\beta_{s}}{\beta}\rho_{h}\right) \,.
\ee
Matching onto the worldline EFT is equivalent to solving the equations motion after analytically continuing the orbital number to perform the source/response split of the scalar field, and only in the end sending $\ell$ to take its physical integer values. Doing this, we see that the ``response'' part of the static scalar field is singular at the horizon when $\hat{\ell}\in\mathbb{N}$ and is therefore absent, while the ``source'' part becomes a polynomial of degree $\hat{\ell}$ in $\tilde{\rho}$. Consequently, the corresponding static Love numbers vanish identically and we see again how a polynomial form of the solution is indicative of this vanishing. For generic $\hat{\ell}$, the procedure just described gives the following static scalar Love numbers,
\be
	k_{\ell}^{\left(0\right)} = \frac{1}{2^{4\hat{\ell}+2}}\frac{\Gamma^2\left(\hat{\ell}+1\right)}{\Gamma\left(\hat{\ell}+\frac{1}{2}\right)\Gamma\left(\hat{\ell}+\frac{3}{2}\right)}\tan\pi\hat{\ell}\,\left(\frac{\beta_{s}}{\beta}\frac{\rho_{\text{h}}}{\rho_{s}}\right)^{2\hat{\ell}+1} \,,
\ee
which are exactly analogous to the ones for the higher-dimensional Schwarzschild black hole obtained in \cite{Kol:2011vg}.

All in all, we observe that the Love symmetry can be present beyond general relativity. We have derived a generic class of geometries enjoying the near zone $\SL$ symmetries. All these geometries must have exactly zero static Love numbers when $\hat{\ell}\in\mathbb{N}$ as a result of the highest weight property.

An instructive application of our construction is the non-vanishing of Love numbers in Riemann-cubed gravity considered in Sec.~\ref{sec:Rcubed}. In this case, the Klein-Gordon equation does not posses an $\SL$ symmetry, which can be seen
from the fact that the geometric constraint \eqref{eq:SL2RModGRConstraint} is violated for any non-zero $\alpha$. Thus, Riemann-cubed gravity gives an explicit example where the absence of the Love symmetry is accompanied by running Love numbers.

\section{Discussion}
\label{sec:disc}

We have presented the $\SL$ Love symmetry of 
black hole perturbations that 
allows for a group theory 
description of Love numbers' properties
and, in particular, 
addresses 
the vanishing of black hole Love numbers
in four dimensions.

On the one hand, it is satisfactory that the ``Love hierarchy problem''~\cite{Porto:2016zng} has lead us to the identification of a novel (approximate) black hole symmetry. Static solutions of the near zone
Teukolsky equation belong to highest weight $\SL$
representations, which dictate a polynomial
form of these solutions (in appropriate coordinates)
and consequently forces the Love numbers to vanish.
At first glance, this picture is consistent with the 't Hooft naturalness dogma \cite{tHooft:1979rat}. 

On the other hand, the Love symmetry has an unconventional property that it mixes UV and IR modes. This happens
because of the $e^{\pm t/\beta}$ factors in the $L_{\pm 1}$ generators. As a result, 
$\SL$ multiplets 
contain both the static solution
and modes with large (imaginary) frequencies.
However, only in the near extreme limit $M/\beta\ll1$ the action of the Love symmetry is compatible with the near zone validity conditions \eqref{eq:NZ}.
This does not, of course, invalidate our arguments. 
Our logic is first to 
work in the near zone limit $\epsilon=0$,
solve the resulting theory exactly,
and then perturb around this exact solution. 
Formally, in the general Kerr-Newman case, 
the near zone limit is valid only for 
small frequency modes, which include, of course, 
the static solution. 
For them the Love symmetry argument 
allows us to obtain exact results
despite the UV/IR mixing.

All in all, it is somewhat 
unclear whether 
the discovery of the Love symmetry should be considered 
as a triumph of naturalness 
in the sense of 't Hooft, 
or rather an 
example of 
a ``UV miracle.'' At the moment all evidence suggests 
that it is
a UV symmetry whose full action cannot be 
understood from the EFT point of view.
It remains to 
be seen whether 
this UV/IR mixing example may 
be useful for other known hierarchy problems.

Related to the UV/IR mixing it is
worth commenting
on the recently presented ``ladder symmetry''
of black hole perturbations~\cite{Hui2021}.
This construction appears to be very different from ours. 
Unlike the Love 
symmetry, the ``ladder symmetry''
does not have a conventional spacetime algebra. Instead, the ladder
operators act directly on the multipole moments of the fields, mixing modes with different orbital number $\ell$. 
Using the ladder operators one can 
obtain that the profile of a 
static perturbing filed, regular at 
the horizon, has a polynomial 
form. This form thus is dictated by the 
structure of the ladder operators
and the regularity requirement. 
In contrast, the Love symmetry gives a group theory explanation of vanishing Love numbers as a consequence of the highest weight property of the corresponding representation.
Comparing the two, we see that
the ``ladder symmetry'' does not have
a similar group theoretical 
interpretation of Love numbers' vanishing.

Another peculiar property of the ``ladder symmetry'' proposal is that it does not, in general, transform physical solutions into other physical solutions. In particular, the ladder
``multiplets'' of Kerr black hole
perturbations contain unphysical states with $|m|\geq \ell$.
In this sense the ladder symmetry 
is somewhat different from conventional 
symmetries that transform a
physical solution into another physical 
solution. 

It should also be pointed out that the ladder construction operates at the level of static
perturbations. Thus, it has a potential 
to address the vanishing of Love numbers
in a more conventional way, i.e. by providing
symmetry selection rules at the IR level.
It remains to be seen if this symmetry can 
successfully 
address all peculiar properties of Love numbers,
including their
fine-tuning for higher-dimensional black holes.

As a continuation of that work, \cite{Hui:2022vbh} appeared while our paper 
was being prepared. There, the ladder symmetry structure arises from a larger conformal group of an effective conformally flat near zone metric. As mentioned above, such effective black hole metrics are known 
as 
``subtracted geometries''
in the literature, see e.g.~\cite{Cvetic:2011dn,Cvetic:2011hp,Kim:2012mh}.
The effective Kerr near zone geometry 
presented in \cite{Hui:2022vbh} actually
corresponds to our Starobinsky near zone approximation \eqref{eq:starCas}, i.e. the $s=0$ Starobinsky near zone $\SL$ generators \eqref{eq:SL2RStarobinsky} are Killing vectors of that effective near zone geometry. 
This construction offers 
an interpretation of the ladder
generators for Schwarzschild
black holes as boost-like
conformal Killing vectors of the 
subtracted geometry. 

A different geometric interpretation 
of the ladder operators 
similar to \cite{Hui2021} is given
in an earlier work~\cite{Cardoso:2017qmj}.
This work has shown that if a spacetime possesses a closed conformal 
Killing vector, 
this vector can be used
to construct ladder
operators that change the 
effective mass of a scalar
field in this spacetime. 
In this context, 
the Schwarzschild 
ladder operators 
of \cite{Hui2021}
match the 
ladder operators 
generated by an
effective conformal
Killing vector
of the subtracted 
geometry.

It is useful to mention that 
the Love symmetry 
parallels the spherical symmetry of the hydrogen atom. 
The Love symmetry generators $L_{\pm}$ raise and lower 
the imaginary frequency ($\SL$ weight $h$) 
of black hole
perturbations just like the $SO(3)$
operators raise and lower the magnetic 
quantum number $m$. The orbital number 
$\ell$ stays the same because the
symmetry operators only transform 
vectors within the same representation. 
The spherical symmetry, however, 
also implies 
the existence of a ladder structure that changes the orbital number $\ell$~\cite{Cardoso:2017qmj}. This 
ladder structure parallels that of \cite{Hui2021}.

All in all, we stress that 
we do not see a 
direct connection between the ladder construction \cite{Hui2021}
and our Love symmetry generators \eqref{eq:SL2RKerr}.
While some link exists in the context 
of the effective 
subtracted geometry~\cite{Hui:2022vbh}, 
its physical meaning is
unclear to us.
Our above results
suggest that a proper
interpretation of the Love 
symmetry should be in terms 
of the broken near horizon
isometries of extremal black holes. 




\section{Future directions}
\label{sec:future}

Our work suggests a number of future research directions.

\textbf{Complete symmetry structure.}
First of all, it would be important to understand
the full symmetry structure of black hole perturbations. 
We have already seen that the Love symmetry can be 
extended to cover other near zones 
as well as perturbations with real 
frequencies. It is natural to expect 
that the $\SL$ algebra of the Love 
symmetry can be extended into the full 
Virasoro algebra by including
the asymptotic Bondi--Metzner--Sachs symmetries~\cite{Bondi1962}.

It should be pointed out that 
even if such an extension exists, it can only  
be approximate.
This is because the full solution of the Teukolsky 
equation is known only in terms of infinite 
series~\cite{Mano:1996vt,Mano:1996mf,Mano:1996gn}, which cannot be interpreted as 
a basis function in
some particular Lie algebra representation.

\textbf{Kerr/CFT conjecture.} The Love symmetry can be 
contrasted with the non-critical 
Kerr/CFT proposal~\cite{Castro:2010fd}, see also~\cite{Guica:2008mu}.
It is based on the fact that 
for a different choice of the near zone split
the wave equation in the Kerr black hole background
enjoys a local ``hidden'' 
$\SL_R\times \SL_L$
conformal symmetry.
However, its generators are not globally well-defined because they do not respect the $\phi\to \phi+2\pi$ periodicity, and therefore regular solutions of the Teukolsky equation 
do not form $\SL_R\times \SL_L$ representations.
Furthermore, the Kerr/CFT $\SL_R\times \SL_L$ generators
do not have a smooth Schwarzschild limit.

In contrast, our Love symmetry generators are well defined 
globally and have a smooth Schwarzschild limit.  
This suggest 
that the Love symmetry may be a better starting point for 
a hypothetical holographic description of Kerr black holes. 
This expectation is further supported by 
the observation that the Love
symmetry is a cousin of the extremal Kerr-Newman
near horizon isometry~\cite{Bardeen:1999px,Amsel:2009et}:
they are subalgebras of the same common algebra.


\textbf{Reorganization of black hole perturbation theory.}
An immediate practical application of the Love symmetry
is that it can be used as a guiding principle
in organizing black hole perturbation theory calculations. 
The standard approach is to first find a formal 
full solution 
to the Teukolsky equation in 
terms of a series of hypergeometric functions,
and then Taylor expand it 
at small frequencies~\cite{Mano:1996vt,Mano:1996mf,Mano:1996gn}. However, this way 
the information that the Teukolsky equation is exactly solvable
in the near zone approximation is lost and can be recovered
only \textit{a posteriori}. A better way to build perturbation
theory is to start with an expansion around the near zone, where
perturbations obey the Love symmetry. In this approach the Love 
symmetry can play a role similar 
to the chiral symmetry in pion perturbation theory.  
Treating symmetry breaking parameters in Eq.~\eqref{eq:V0V1}
as spurions under the Love symmetry, one should be able
to derive analogues of the Gell-Mann–Okubo relations between 
finite frequency black hole responses and quasinormal modes.

\textbf{Quasinormal modes.}
Another natural question is
to what extent the Love symmetry can be useful 
for the study of quasinormal modes. We have argued that 
states of Love highest weight multiplets with negative imaginary 
frequencies can be interpreted as highly-damped 
quasinormal modes in the Schwarzschild case
and as total transmission modes in the Kerr and Reissner-N\"{o}rdstrom case.
It is quite unexpected that the leading 
imaginary frequency of the highly
damped quasinormal modes
easily follows from the Love 
algebra. This can be compared
with the celebrated simple
derivation from the poles
of the scattering amplitude in the Born approximation~\cite{Padmanabhan:2003fx}.
In fact, the Born approximation gives 
a universal result of the $2\pi T_H$
spacing of highly-damped QNMs 
for all black hole backgrounds, 
which agrees with the 
spacing in Love multiplets. 
This suggests a connection between the two
approaches and it would be interesting to investigate
it in the future. 

In addition, it would be 
important  
to study the relationship 
between the Love multiplet 
and QNMs in more detail 
and understand how this relationship transforms
beyond the near zone expansion 
and the highly-damped limit.

\textbf{Higher spin fields in higher dimensions.}
We have presented a generalization of the Love symmetry that 
describes the properties of scalar Love numbers
of higher dimensional Schwarzschild black holes. 
In order to complete the Love symmetry arguments
in higher dimensions, it would be important 
to generalize our results to the case of spin-1 and spin-2
perturbations. In particular, it would be interesting if the Love symmetry can address the intricate structure
of electric and magnetic type Love numbers that takes place
in higher dimensions as well~\cite{Hui:2020xxx}. Related to this, it would be interesting to also investigate the existence of Love symmetry in non-spherically symmetric black hole geometries in higher dimensions, such as Myers-Perry black holes~\cite{Myers:1986un}, or even black objects with non-compact horizons, i.e. black $p$-branes~\cite{Duff:1993ye}, where the associated near zone symmetry structure is expected to follow from the $SO\left(p+1,2\right)$ subgroup of the enhanced isometry of the near horizon geometry of extremal black $p$-branes.

\textbf{Love symmetry beyond General Relativity.}
Another interesting outcome of the analysis done in this paper is the possibility of Love symmetry existing in theories beyond General Relativity. We have in fact derived a geometric constraint and theories of gravity supporting such black hole solutions will have vanishing static scalar Love numbers due to the highest weight property of the near zone $\SL$. It is natural to study better what theories of gravity do support such solutions. It should be noted here that our current analysis has been quite restrictive on the form of the generators, namely, we have ignored possible field redefinitions and global rescalings that are not a priori forbidden, and some relaxations on this front will allow for a more proper investigation. An immediate interest that arises here is to apply such analyses to string theories, e.g. the spherically symmetric Callan-Myers-Perry black hole of bosonic/heterotic string theory \cite{Myers:1998gt} or the type-II supertstring theory $\alpha^{\prime3}$-corrections to the Schwarzschild black hole \cite{Myers:1987qx} which do not satisfy the geometric constraint \eqref{eq:SL2RModGRConstraint} and do not appear to posses a Love symmetry at first sight.


\section*{Acknowledgments}
This work is supported in part by the NSF award PHY 2210349 and by the BSF grant 2018068.

\appendix

\section{Teukolsky equation for extremal black holes}
\label{app:extr}

The behavior of extremal Kerr-Newman black hole 
perturbations is somewhat different from 
the non-extremal case because of the presence 
of additional singularities as $r_-\to r_+$
in the Teukolsky equation.
We treat this case in this Appendix.


The purely incoming boundary condition has the following form 
in the tortoise coordinate $r_{\ast}$,
\be 
R\sim \Delta^{-s}\e^{-i(\omega - m\Omega)r_{\ast}}\,,\quad 
\text{as}\quad r_{\ast}\to -\infty~(r\to r_+). 
\ee
where 
\be
\frac{dr_{\ast}}{dr}=\frac{r^2+a^2}{\Delta}\,,\quad \Rightarrow 
\quad 
r_{\ast}=r+2M\ln\frac{r-M}{M}-\frac{M^2+a^2}{r-M}\,.
\ee
In the Boyer-Lindquist coordinates it reads
\be 
\label{eq:regBL0}
R = \text{const}\cdot \left(r-M\right)^{-2s-i2M\left(\omega-m\Omega\right)}\e^{i\frac{M^2+a^2}{r-M}\left(\omega-m\Omega\right)}\,.
\ee
In what follows it is useful to introduce the variables
\be
\alpha \equiv 1-\frac{\omega}{m\Omega}\,,
\quad x\equiv \frac{r-M}{M}\,,\quad \varepsilon \equiv 2M\omega \,,
\ee
in which the boundary condition \eqref{eq:regBL0} reads
\be
\label{eq:regBL1}
R = 
\text{const}\cdot x^{-2s+2i\alpha m\Omega M}\e^{-im\frac{a\alpha}{M x}}
\quad \text{as}\quad x\to 0\,. 
\ee
The radial Teukolsky equation can be written as
\be
\begin{split}
	\Bigg[ &\left(r-M\right)^{-2s}\partial_{r}\left(r-M\right)^{2\left(s+1\right)}\partial_{r} + \frac{\left(M^2+a^2\right)^2}{\left(r-M\right)^2}\left(\omega-m\Omega\right)^2 - \frac{4Ma}{r-M}m\omega \\
	&+ \frac{r+M}{r-M}\left(r^2+M^2+2a^2\right)\omega^2 -2is\frac{M^2+a^2}{r-M}(\omega-m\Omega) \\
	&+2is\left(r-M\right)\omega + s(s+1) -\ell(\ell+1) \Bigg]R=0\,.
\end{split} 
\ee
In the new variables it reads
\be
	\begin{split}
		\Bigg[ &x^{-2s}\partial_{x}\,x^{2\left(s+1\right)}\,\partial_{x} + \frac{m^2\alpha^{\prime2}}{x^2} - 2\left(\varepsilon-is\right)\frac{m\alpha^{\prime}}{x} + is\varepsilon x \\
		& + \frac{\varepsilon^2}{2}\left(\frac{x^2}{2}+2x+3+\frac{a^2}{M^2}\right) + s\left(s+1\right) - \ell\left(\ell+1\right) \Bigg]R = 0\,,
	\end{split} 
\ee
where we have introduced,
\be
    \alpha^{\prime} \equiv \frac{a}{M}\alpha \,.
\ee
Let us solve this equation in some approximations.

\subsection{Low frequency limit and Love numbers}

Let us neglect the terms which are suppressed provided that 
\be
	\varepsilon \ll 1\,,\quad \varepsilon x \ll 1\,.
\ee
This is 
identical to the near zone approximation.
It is convenient to make a new transformation $z=1/x$, with which 
the approximate Teukolsky equation
reads:
\be
\label{eq:NZextr}
	\begin{split}
		\bigg[ z^{2\left(s+1\right)}\partial_{z}\,z^{-2s}\,\partial_{z} + m^2\alpha^{\prime2}z^2 - 2\left(\varepsilon-is\right)m\alpha^{\prime}z + s\left(s+1\right) - \ell\left(\ell+1\right) \bigg]R = 0\,.
	\end{split} 
\ee
This equation can be brought to the Whittaker form 
after transforming to the new variable
\be
 	y =2im \alpha^{\prime} z \,.
\ee
and the field redefinition $R=y^{s}\varphi$,
\be
	\Bigg[\d_y^2 - \frac{1}{4}+\frac{i\varepsilon+s}{y}-\frac{\ell\left(\ell+1\right)}{y^2} \Bigg]\varphi = 0 \,.
\ee
The general solution has the form:
\be 
	\begin{split}
		R = e^{-im\alpha^{\prime} z}(2im\alpha^{\prime} z)^{\ell+1+s} \bigg(& \Phi\left(-i\varepsilon +\ell -s+1,2\ell+2;2im\alpha^{\prime}z\right) \\
		& + \Psi\left(-i\varepsilon +\ell -s +1,2\ell+2;2im\alpha^{\prime} z\right) \bigg) \,,
	\end{split}
\ee
where $\Phi$ and $\Psi$ are the confluent hypergeometric functions of the first and second kind. 
Their expansions at the horizon $z\to \infty$ read~\cite{Bateman:100233}
\be
	\begin{split}
		&\Psi\left(a,b,y\right)=\sum_{n=0}\left(-1\right)^n\frac{\left(a\right)_n\left(a-b+1\right)_n}{n!}y^{-a-n} \,,\\
		&\Phi\left(a,b,y\right)= \frac{\Gamma\left(b\right)}{\Gamma\left(b-a\right)}\left(-\frac{1}{y}\right)^a\left(1+...\right) + \frac{\Gamma\left(b\right)}{\Gamma\left(a\right)}e^y y^{a-b}\left(1+...\right)\,,
	\end{split} 
\ee
from which we see that $\Psi$ is the solution regular 
at the horizon. We also used
\be 
	\left(a-b+1\right)_n = \frac{\Gamma\left(a-b+2\right)}{\Gamma\left(a-b+2-n\right)} \,.
\ee
To find the Love numbers we expand $\Psi\left(a,n+1,x\right)$ at spatial infinity $z\to 0$ and find logs,
\be
	\begin{split}
		&\Psi\left(a,n+1;y\right)=\frac{\left(-1\right)^{n-1}}{n!\Gamma\left(a-n\right)} \Bigg( \Phi\left(a,n+1;x\right)\log\left(y\right) \\
		&+\sum_{r=0}^{\infty}\frac{\left(a\right)_r}{\left(n+1\right)_r} \left[\psi\left(a+r\right)-\psi\left(1+r\right)-\psi\left(1+n+r\right)\right]\frac{y^r}{r!} \Bigg) \\
		&+\frac{\left(n-1\right)!}{\Gamma\left(a\right)}\sum_{r=0}^{n-1}\frac{\left(a-n\right)_r}{\left(1-n\right)_r}\frac{y^{r-n}}{r!} \,.
	\end{split}
\ee
At the leading order we have the following source and response contributions,
\be 
\begin{split}
	& \Psi\left(-i\varepsilon +\ell -s +1,2\ell+2;2im\alpha^{\prime}z\right)\\
	&=\frac{\left(2\ell\right)!}{\Gamma\left(-i\varepsilon -s +\ell +1\right)}\left(2im\alpha^{\prime}z\right)^{-2\ell-1} +\frac{\log\left(2im\alpha^{\prime}z\right)}{\left(2\ell+1\right)!\Gamma\left(-i\varepsilon -s -\ell\right)} \,,
	\end{split}
\ee
which gives the following expression for the response 
coefficient
\be 
	\begin{split}
		k^{\left(s\right)}_{\ell m}& =\left(\log\left(2m\alpha^{\prime}z\right)+i\frac{\pi}{2}\right) \frac{\left(im\alpha^{\prime}\right)^{2\ell+1}\Gamma\left(-i\varepsilon -s +\ell +1\right)}{\left(2\ell\right)!\left(2\ell+1\right)!\Gamma\left(-i\varepsilon -s -\ell\right)} \\
		& =\left(-1\right)^{s} \varepsilon {\left(m\alpha^{\prime}\right)^{2\ell+1}} \left(\log\frac{2m\alpha^{\prime}M}{r-M}+i\frac{\pi}{2}\right) \frac{\left(\ell-s\right)!\left(\ell+s\right)!}{\left(2\ell\right)!\left(2\ell+1\right)!} \,,
	\end{split}
\ee
where 
in the last line we kept 
terms linear in $\varepsilon$ and
used formulas from Appendix D of Ref.~\cite{Charalambous:2021mea}. 
We see that frequency-dependent response coefficients
are not zero and run with the distance.
However, the exact static ($\varepsilon=0$) Love numbers 
vanish.

\subsection{Near horizon at the locking frequency}

The Teukolsky equation has a different behavior
in the regime 
\be
	mx \ll 1\quad \text{and}\quad m^2\alpha\ll 1\,,
\ee
which corresponds to a near horizon approximation and modes
with frequencies closed to the locking one. In this case we have,
\be
	\begin{split}
		\Bigg[ &x^{-2s}\partial_{x}\,x^{2\left(s+1\right)}\partial_{x} + \frac{m^2\alpha^{\prime2}}{x^2} -2\left(2Mm\Omega-is\right)\frac{m\alpha^{\prime}}{x} \\
		&+ 2 \left(3M^2+a^2\right)m^2\Omega^2 + s\left(s+1\right) - \ell\left(\ell+1\right)\Bigg]R = 0\,.
\end{split} 
\ee
We proceed by defining~\cite{Starobinski1973,Starobinski1974}
\be
	\ell(\ell+1)-2 \left(3M^2+a^2\right)m^2\Omega^2 \equiv \frac{1}{4} \left(\delta ^2-1\right)\,,
\ee
$z=x^{-1}$, $\varphi =z^{-s}  R$ and $m^{\prime}=2M\Omega m$ such that our equation takes the form
\be
	\left[z^2\partial_{z}^2 + m^2\alpha^{\prime2}z^2 - 2m\alpha^{\prime}\left(m^{\prime}-is\right)z + \frac{1-\delta ^2}{4}\right]\varphi = 0
\ee
Introducing 
$y=2im\alpha^{\prime}z$,
we arrive at the Whittaker equation
\be
	\Bigg[ \d_y^2 - \frac{1}{4} + \frac{im^{\prime}+s}{y}-\frac{\delta^2-1}{4y^2} \Bigg] \varphi= 0 \,.
\ee
whose general solution is expressed in terms of the Whittaker functions,
\be
	\varphi = c_1 M_{i m^{\prime}+s,\frac{\delta }{2}}\left(2 i m\alpha^{\prime}z \right)+c_2 W_{i m^{\prime}+s,\frac{\delta }{2}}\left(2 i m \alpha^{\prime}z \right) \,.
\ee

\section{Love symmetry vectors in the advanced/retarded coordinates}
\label{app:adv}

In this section we present the Love symmetry 
generators in the advanced and retarded coordinates ~\cite{frolov1998black}.
From these expressions, it will be clear that 
the Love vectors are regular at the horizon $r= r_+$, both the future and the past one.

Investigation of regularity at the future and the past event horizon is achieved by transitioning to advanced and retarded null coordinates $\left(\upsilon,r,\theta,\varphi\right)$ and $\left(u,r,\theta,\tilde{\varphi}\right)$ respectively. These are related to the Boyer-Lindquist coordinates $\left(t,r,\theta,\phi\right)$ according to\footnote{We are writing here the expressions for a non-extremal Kerr-Newman black hole. However, we have fixed the integration constants such that, besides the asymptotic behaviors $\upsilon\rightarrow t + r$, $u\rightarrow t - r$ and $\varphi\rightarrow \phi$, $\tilde{\varphi}\rightarrow\phi$ as $r\rightarrow\infty$, we also get a smooth extremal limit. Namely, taking the extremal limit $a^2+Q^2\rightarrow M^2$ ($r_{+}-r_{-}\rightarrow 0$), we retrieve the correct expressions,
\be
	a^2+Q^2 \rightarrow M^2 \Rightarrow
	\begin{cases}
	    \upsilon \rightarrow t + r + 2M\ln\frac{r-M}{M} - \frac{M^2+a^2}{r-M} \,,\quad \varphi \rightarrow \phi - \frac{a}{r-M} \\
	    u \rightarrow t - r - 2M\ln\frac{r-M}{M} + \frac{M^2+a^2}{r-M} \,,\quad \tilde{\varphi} \rightarrow \phi + \frac{a}{r-M}
	\end{cases} \,,
\ee
},
\be\label{eq:AdvancedCoords}
\begin{split}
	& d\upsilon = dt + \frac{r^2+a^2}{\Delta}\,dr \Rightarrow \upsilon = t + r + \frac{r_{+}^2+a^2}{r_{+}-r_{-}}\ln\frac{r-r_{+}}{r_{+}} - \frac{r_{-}^2+a^2}{r_{+}-r_{-}}\ln\frac{r-r_{-}}{r_{+}} \,, \\
	& d\varphi = d\phi + \frac{a}{\Delta}\,dr \quad\,\,\,\,\, \Rightarrow \varphi = \phi + \frac{a}{r_{+}-r_{-}}\ln\frac{r-r_{+}}{r-r_{-}} \,.
\end{split}
\ee
and,
\be\label{eq:RetardedCoords}
\begin{split}
	& du = dt - \frac{r^2+a^2}{\Delta}\,dr \Rightarrow u = t - r - \frac{r_{+}^2+a^2}{r_{+}-r_{-}}\ln\frac{r-r_{+}}{r_{+}} + \frac{r_{-}^2+a^2}{r_{+}-r_{-}}\ln\frac{r-r_{-}}{r_{+}} \,, \\
	& d\tilde{\varphi} = d\phi - \frac{a}{\Delta}\,dr \quad\,\,\,\,\, \Rightarrow \tilde{\varphi} = \phi - \frac{a}{r_{+}-r_{-}}\ln\frac{r-r_{+}}{r-r_{-}} \,.
\end{split}
\ee

In advanced null coordinates $\left(\upsilon,r,\theta,\varphi\right)$, the Love vector fields~\eqref{eq:SL2RKerr} read,
\be\label{eq:LoveAdvanced}
\begin{split}
	& L_0 = -\beta\,\partial_{\upsilon} \,,\\
	& L_{\pm1}=e^{\pm\left(\upsilon-r\right)/\beta} \left(\frac{r-r_{-}}{r_{+}}\right)^{\mp 2M/\beta} \\
	&\left[\mp\left(r-r_{\mp}\right)\partial_{r}+\frac{r-r_{\mp}}{r-r_{-}}\left(\beta\mp\left(r+r_{+}\right)\right)\partial_{\upsilon} + \left(1\mp1\right)\frac{r_{+}^2+a^2}{r-r_{-}}\left(\partial_{\upsilon}+\Omega\,\partial_{\varphi}\right)\right] \,,
\end{split}
\ee
and regularity at the future event horizon becomes clear, at least for non-extremal black holes.

Similarly, in retarded null coordinates $\left(u,r,\theta,\tilde{\varphi}\right)$, the corresponding components are,
\be\label{eq:LoveRetarded}
\begin{split}
	& L_0 = -\beta\,\partial_{u} \,,\\
	& L_{\pm1}=e^{\pm\left(u+r\right)/\beta} \left(\frac{r-r_{-}}{r_{+}}\right)^{\pm 2M/\beta} \\
	&\left[\mp\left(r-r_{\pm}\right)\partial_{r}+\frac{r-r_{\pm}}{r-r_{-}}\left(\beta\pm\left(r+r_{+}\right)\right)\partial_{u} + \left(1\pm1\right)\frac{r_{+}^2+a^2}{r-r_{-}}\left(\partial_{u}+\Omega\,\partial_{\tilde{\varphi}}\right)\right] \,,
\end{split}
\ee
revealing regularity also at the past event horizon in the non-extremal case.

Last, the extreme middle zone algebra vectors~\eqref{eq:algmid} have 
the following form in advanced null coordinates,
\be 
\begin{split}
	& L_0 = -2M\d_v\,,\\
	& L_{+1}=-M e^{\frac{v-r}{2M}}\left(
	\d_v + \d_r
	\right)\,,\\
	& L_{-1}=M^{-1}e^{-\frac{v-r}{2M}}\left(
	(r-M)(r+3M) \d_v + (M-r)^2\d_r 
	+2a \d_{\varphi}
	\right)\,.
\end{split} 
\ee

\section{Some properties of $\SL$ representations}
\label{app:sl2r}

The lowest and highest weight representations are
two standard reducible representations 
of $\SL$. Here we present some basic 
properties 
of these representations (see~\cite{Howe1992}
for more detail).\footnote{In the notation of that book we have 
$e^{\pm}=\pm L_{\mp1}$, $h=2L_0$, $\mathcal{C}_2^{\rm here}=\mathcal{C}_2^{\rm there}/4$, $\l=2h$. }
The highest weight module $V_{h}$ 
has a basis of $L_0-$ eigenvectors $v_j$ ($j={0,1,2,...}$) that satisfy
\be
\begin{split}
& L_0 v_j=\left(h + j\right)v_j\,,\quad L_{-1} v_j = v_{j+1}\,,\\
& L_{+1} v_j = j\left(2h+j -1\right)v_{j-1}\,,\quad L_{+1} v_0 = 0\,,\\
& \mathcal{C}v_j=h\left(h-1\right)v_j\,.
\end{split} 
\ee
The lowest weight module $\bar V_{h}$ 
has a basis of $L_0-$ eigenvectors $\bar v_j$ ($j={0,1,2,...}$)
that satisfy
\be
\begin{split}
& L_0 \bar{v}_j=\left(h - j\right)\bar{v}_j\,,\quad L_{+1} \bar{v}_j = -\bar{v}_{j+1}\,,\\
& L_{-1} \bar{v}_j = j\left(2h -j + 1\right)\bar{v}_{j-1}\,,\quad L_{-1} \bar{v}_0 = 0\,,\\
& \mathcal{C}\bar{v}_j=h\left(h+1\right)\bar{v}_j\,.
\end{split} 
\ee

Note that Ref.~\cite{doi:10.1137/0501037} offers 
a somewhat different classification of reducible 
representations of $\SL$.
In particular, the representation relevant for 
the $d=4$ Schwarzschild case is the finite-dimensional
representation $D(2\ell)$,
whereas the generic highest/lowest weight Verma modules
correspond to the representations $D^-(2\ell)/D^+(2\ell)$,
respectively.

For completeness, we also 
present here the three additional standard modules $W\left(\mu,\l\right)$, $\bar{W}\left(\mu,\lambda\right)$ and $U\left(\nu^+,\nu^-\right)$, spanned by infinite sets of vectors 
$v_j$ ($j\in \mathbb{Z}$), which satisfy
\be
\begin{split}
W\left(\mu,\lambda\right):\quad 
& L_0 v_j=\left(\frac{\lambda}{2} + j\right)v_j\,,\quad L_{-1} v_j = v_{j+1}
\,,\quad \mathcal{C}v_j=\left(\mu/4\right) v_j
\,,\\
& L_{+1} v_j = -\frac{1}{4}
\left[\mu-\left(\lambda+2j-1\right)^2 +1 \right]
v_{j-1}\,,
\\
\bar W\left(\mu,\lambda\right):\quad 
& L_0 \bar{v}_j=\left(\frac{\lambda}{2} + j\right)\bar{v}_j\,,
\quad L_{+1} \bar{v}_j = -\bar{v}_{j-1}
\,,\quad \mathcal{C}\bar{v}_j=\left(\mu/4\right) \bar{v}_j
\,,\\
& L_{-1} \bar{v}_j = \frac{1}{4}
\left[\mu-\left(\lambda+2j+1\right)^2 +1 \right]
\bar{v}_{j+1}\,,\\
U\left(\nu^{+},\nu^{-}\right):\quad 
& L_0 v_j=\left(\frac{\nu^{+}-\nu^{-}}{2} + j\right)v_j\,,\quad 
L_{\mp1} v_j = \left(j\pm\nu^{\pm}\right)v_{j\pm1}
\,,\\ 
&
\mathcal{C}v_j=\frac{\nu^{+} + \nu^{-}}{2}\left(\frac{\nu^{+} + \nu^{-}}{2} - 1\right) v_j
\,.
\end{split} 
\ee

\section{Derivation of $\SL$ generators}
\label{ApGenerators}

In this appendix we will derive the form of the generators of the $\SL$ symmetry of the near zone equations of motion. Our method will be to first find those truncations of the Teukolsky equation that enjoy an enhanced $\SL$ structure and preserve the characteristic exponents of the NP scalars in the vicinity of the event horizon. These truncations of the Teukolsky equation are realized as wave operators built from effective background black hole geometries, which have the property of preserving the internal structure of the black hole, namely, its thermodynamic properties. Such geometries were originally studied in \cite{Cvetic:2011hp,Cvetic:2011dn} and were coined the term ``subtracted geometries''\footnote{Strictly speaking, subtracted geometries as introduced in \cite{Cvetic:2011hp,Cvetic:2011dn} preserve more structure than just the thermodynamic variables. For example, they also preserve the location of the ergosphere which is a property of the black hole geometry that extends beyond the behavior of observables in the vicinity of the event horizon.}. We will adopt the same terminology here for these effective black hole geometries.

After finding those subtracted geometry truncations of the Teukolsky operator equipped with an $\SL$ structure, we then identify the ones that are true near zone approximations, i.e. the ones that, besides preserving the characteristic exponents of the NP scalars in the vicinity of the event horizon, also preserve the entire static Teukolsky operator. These will in general \textit{not} be globally defined, spontaneously broken by the periodicity of the azimuthal angle similarly to \cite{Castro:2010fd}. However, we will see that there exist two particular families of near zone truncations with an $\SL$ symmetry which are globally defined; these are the Love symmetry and the Starobinsky near zone $\SL$ symmetry examined in the main text, up to local $r$-dependent translations in time.

The analysis in this Appendix has the upshot of generalizing some already known results in the literature, such as the local $\SL\times\SL$ originally found in \cite{Castro:2010fd}. We focus on the non-extremal case but, as we discuss in the main text, there is a particular subset of the subtracted geometry truncations of the Teukolsky operator equipped with $\SL$ symmetries for non-extremal black holes that can be utilized to get the correct spin weighted $\SL$ Killing vectors of the near horizon extremal black hole configuration as well.

\subsection{Subtracted geometry truncations}
The full master Teukolsky operator \eqref{eq:teuk}-\eqref{eq:teukExplicit} for the spin-$s$ NP scalar has the form,
\be\ba
	\mathbb{T}^{\left(s\right)}_{\text{full}}\Psi_{s} &= \frac{1}{\Sigma}\left[G^{\mu\nu}_{\text{full}}\left(r,\theta\right)\nabla_{\mu}\nabla_{\nu} + s\left(\gamma^{\mu}_{\text{full}}\nabla_{\mu}\left(r,\theta\right) + F_{\text{full},0}\left(r,\theta\right)\right)\right]\Psi_{s} \\
	&= \frac{1}{\Sigma}\left[\mathbb{O}^{\left(s\right)}_{\text{full}} - \mathbb{P}^{\left(s\right)}_{\text{full}}\right]\Psi_{s}\,,
\ea\ee
where $G^{\mu\nu}_{\text{full}}\equiv\Sigma g^{\mu\nu}_{\text{full}}$ is the rescaled Kerr-Newman metric and $\gamma^{\mu}_{\text{full}}$ and $F_{\text{full},0}$ are tetrad-dependent modifications of the Klein-Gordon operator due a non-zero spin weight. All of these functions depend only on the radial and polar coordinates by virtue of the $\mathbb{R}_{t}\times U_{\phi}\left(1\right)$ isometry group of the background geometry. In the second equality above, we have dumbed as $\mathbb{O}^{\left(s\right)}_{\text{full}}$ and $\mathbb{P}^{\left(s\right)}_{\text{full}}$ the full radial and angular Teukolsky operators respectively to stress out the separability of the Teukolsky equation. The important point here is that the second derivatives terms in the equations of motion always come from the inverse metric.

Let us momentarily focus on the $s=0$ Teukolsky operator. As a preliminary approximation on the geometry, we wish to truncate the Klein-Gordon operator such that the characteristic exponents of the scalar field in the vicinity of the event horizon are preserved for any frequency $\omega$ of the perturbation\footnote{For the separated scalar field, $\Phi = e^{-i\omega t}e^{im\phi}R\left(r\right)S\left(\theta\right)$, this would mean,
\be
	R\left(r\right) \sim A_{\text{in}}\left(r-r_{+}\right)^{-iZ\left(\omega\right)} + A_{\text{out}}\left(r-r_{+}\right)^{+iZ\left(\omega\right)}\,\,\,,\,\,\,Z\left(\omega\right) = \frac{\beta}{2}\left(\omega-m\Omega\right)\,,
\ee
with the first and second terms describing ingoing and outgoing waves respectively.}. For the second derivative terms, this means that,
\be
	G^{\mu\nu}\left(r,\theta\right)\partial_{\mu}\partial_{\nu} = \left[\Delta\,\partial_{r}^2 -\frac{\left(r_{+}^2+a^2\right)^2}{\Delta}\left(\partial_{t}+\Omega\,\partial_{\phi}\right)^2\right] + \left[\partial_{\theta}^2 + \frac{1}{\sin^2\theta}\partial_{\phi}^2\right] + \delta G^{\mu\nu}\left(r,\theta\right)\partial_{\mu}\partial_{\nu} \,,
\ee
with $\delta G^{\mu\nu}$ being ``far-horizon'' corrections, while we have enclosed into square brackets the separated contributions to the radial and angular wave operators. The most general such subtracted geometry metric that makes the second derivative terms separable in the Boyer-Lindquist coordinates has the following form,
\be
	\begin{gathered}
		\delta G^{rr}\left(r,\theta\right) = \Delta^2f_{rr}\left(r\right) \,, \\
		\delta G^{tt}\left(r,\theta\right) = f_{tt}\left(r\right) + \tilde{f}_{tt}\left(\theta\right) \,, \\
		\delta G^{t\phi}\left(r,\theta\right) = \Omega\left[ f_{t\phi}\left(r\right) + \tilde{f}_{t\phi}\left(\theta\right) \right] \,, \\
		\delta G^{\phi\phi}\left(r,\theta\right) = \Omega^2\left[ f_{\phi\phi}\left(r\right) + \tilde{f}_{\phi\phi}\left(\theta\right) \right] \,, \\
		\delta G^{tr}\left(r,\theta\right) = \Delta\,f_{tr}\left(r\right) \,\,\,,\,\,\, \delta G^{r\phi}\left(r,\theta\right) = \Omega\,\Delta\,f_{r\phi}\left(r\right) \,, \\
		\delta G^{\mu\theta}\left(r,\theta\right) = 0 \,,
	\end{gathered}
\ee
where all $f_{\mu\nu}$ functions are regular at $r=r_{+}$ and the $\Omega$ insertions are to remind us that these functions are absent for non-rotating black holes. The fact that $\delta G^{\mu\theta}=0$ is just a consequence of the separability condition. We remark here that we are also allowing the generation of the non-diagonal terms $\delta G^{tr}$ and $\delta G^{r\phi}$, as well as the offset diagonal term $\delta G^{rr}$. These are zero in the full background geometry but are in general allowed in the current subtracted geometry treatment, as long as they vanish on the horizon\footnote{The need for $\delta G^{tr}$ and $\delta G^{r\phi}$ to vanish on the horizon can also be physically motivated by requiring the angular velocity of the black hole to be $\Omega = \frac{a}{r_{+}^2+a^2}$, i.e. that the Killing vector $\partial_{t} + \Omega\,\partial_{\phi}$ becomes null at the horizon for this particular value of $\Omega$, along with a finite and non-zero horizon surface area, practically meaning that the induced spatial metric at $r=r_{+}$ has finite and non-zero determinant.}. All these terms, however, can be removed by performing appropriate $r$-dependent ``far-horizon'' transformations, namely, we can introduce coordinates $(\tilde{t},\tilde{r},\tilde{\phi})$, related to $\left(t,r,\phi\right)$, according to,
\be
	\frac{d\tilde{r}}{\sqrt{\Delta\left(\tilde{r}\right)}} = \frac{dr}{\sqrt{G^{rr}\left(r\right)}} \,\,\,,\,\,\, d\tilde{t} = dt - \frac{G^{tr}\left(r\right)}{G^{rr}\left(r\right)}\,dr \,\,\,,\,\,\, d\tilde{\phi} = d\phi - \frac{G^{r\phi}\left(r\right)}{G^{rr}\left(r\right)}\,dr \,.
\ee
Then,
\be
	G^{\tilde{\mu}\tilde{\nu}}\partial_{\tilde{\mu}}\partial_{\tilde{\nu}} = \left[\Delta\left(\tilde{r}\right)\,\partial_{\tilde{r}}^2 -\frac{\left(r_{+}^2+a^2\right)^2}{\Delta\left(\tilde{r}\right)}\left(\partial_{\tilde{t}}+\Omega\,\partial_{\tilde{\phi}}\right)^2\right] + \left[\partial_{\theta}^2 + \frac{1}{\sin^2\theta}\partial_{\tilde{\phi}}^2\right] + \delta G^{\tilde{\mu}\tilde{\nu}}\left(\tilde{r},\theta\right)\partial_{\tilde{\mu}}\partial_{\tilde{\nu}} \,,
\ee
with,
\be
	\begin{gathered}
		\delta G^{\tilde{t}\tilde{t}}\left(\tilde{r},\theta\right) = f_{\tilde{t}\tilde{t}}\left(\tilde{r}\right) + \tilde{f}_{tt}\left(\theta\right) \,, \\
		\delta G^{\tilde{t}\tilde{\phi}}\left(\tilde{r},\theta\right) = \Omega\left[ f_{\tilde{t}\tilde{\phi}}\left(\tilde{r}\right) + \tilde{f}_{t\phi}\left(\theta\right) \right] \,, \\
		\delta G^{\tilde{\phi}\tilde{\phi}}\left(\tilde{r},\theta\right) = \Omega^2\left[ f_{\tilde{\phi}\tilde{\phi}}\left(\tilde{r}\right) + \tilde{f}_{\phi\phi}\left(\theta\right) \right] \,, \\
		\delta G^{\tilde{r}\tilde{r}}\left(\tilde{r},\theta\right) = 0 \,\,\,,\,\,\, \delta G^{\tilde{t}\tilde{r}}\left(\tilde{r},\theta\right) = 0 \,\,\,,\,\,\, \delta G^{\tilde{r}\tilde{\phi}}\left(\tilde{r},\theta\right) = 0 \,, \\
		\delta G^{\tilde{\mu}\theta}\left(\tilde{r},\theta\right) = 0 \,,
	\end{gathered}
\ee
and $f_{\tilde{\mu}\tilde{\nu}}\left(\tilde{r}\right)$ are related to $f_{\mu\nu}\left(r\right)$ according to,
\be\ba
	f_{\tilde{t}\tilde{t}}\left(\tilde{r}\right) &= f_{tt}\left(r\right) - \frac{\left(G^{tr}\left(r\right)\right)^2}{G^{rr}\left(r\right)} \,, \\
	f_{\tilde{t}\tilde{\phi}}\left(\tilde{r}\right) &= f_{t\phi}\left(r\right) - \frac{G^{tr}\left(r\right)G^{r\phi}\left(r\right)}{G^{rr}\left(r\right)} \,, \\
	f_{\tilde{\phi}\tilde{\phi}}\left(\tilde{r}\right) &= f_{\phi\phi}\left(r\right) - \frac{\left(G^{r\phi}\left(r\right)\right)^2}{G^{rr}\left(r\right)} \,.
\ea\ee
Interestingly, these subtracted geometries give rise to a separable full wave operator, that is, the following conditions on the Christoffel symbols,
\be\label{eq:SeperabilityConditions}
\begin{gathered}
	\partial_{\theta}\left(G^{\mu\nu}\Gamma^{r}_{\mu\nu}\right)=0\,\,\,,\,\,\, \partial_{r}\left(G^{\mu\nu}\Gamma^{\theta}_{\mu\nu}\right)=0 \,, \\
	\partial_{r}\partial_{\theta}\left(G^{\mu\nu}\Gamma^{t}_{\mu\nu}\right)=0\,\,\,,\,\,\,\partial_{r}\partial_{\theta}\left(G^{\mu\nu}\Gamma^{\phi}_{\mu\nu}\right)=0 \,,
\end{gathered}
\ee
are automatically satisfied.

In the rest of this appendix, we will be working in the $(\tilde{t},\tilde{r},\tilde{\phi},\theta)$ coordinates but we will drop the tildes to ease our notation. Nevertheless, in the final expressions, we will always have in mind to supplement with the more general results corresponding to the replacements,
\be\label{eq:t_r_phi_Rep}
	r \rightarrow r + \Delta^2 g_{r}\left(r\right) \,\,\,,\,\,\, t\rightarrow t + g_{t}\left(r\right) \,\,\,,\,\,\, \phi\rightarrow \phi + g_{\phi}\left(r\right) \,,
\ee
with $g_{r}$, $g_{t}$ and $g_{\phi}$ radial functions that are regular near the horizon, giving rise to non-zero $\delta G^{rr}$, $\delta G^{tr}$ and $\delta G^{r\phi}$ respectively.

Last, let us analyze the $s\ne0$ tetrad-dependent part of the Teukolsky operator. Requiring the preservation of the near horizon characteristic exponents, we write,
\be\ba
	s\left(\gamma^{\mu}\left(r,\theta\right)\partial_{\mu} + F_0\left(r,\theta\right)\right) &= s\left[\Delta^{\prime}\,\partial_{r} + \frac{\left(r_{+}^2+a^2\right)\Delta^{\prime}}{\Delta}\left(\partial_{t}+\Omega\,\partial_{\phi}\right) + s+1\right]  \\
	&+ s\left[ -\frac{s}{\sin^2\theta}\right] + s\left(\delta\gamma^{\mu}\left(r,\theta\right)\partial_{\mu} + \delta F_0\left(r,\theta\right)\right) \,,
\ea\ee
with the ``far-horizon'' terms having the form,
\be
	\begin{gathered}
		\delta\gamma^{r}\left(r,\theta\right) = \Delta\,f_{r}\left(r\right) \,\,\,,\,\,\, \delta\gamma^{t}\left(r,\theta\right) = f_{t}\left(r\right) + \tilde{f}_{t}\left(\theta\right) \,, \\
		\delta\gamma^{\phi}\left(r,\theta\right) = \Omega\left[f_{\phi}\left(r\right) + \tilde{f}_{\phi}\left(\theta\right)\right] \,\,\,,\,\,\, \delta\gamma^{\theta}\left(r,\theta\right) = 0 \,, \\
		\delta F_0\left(r,\theta\right) = \Delta\,f_0\left(r\right) \,.
	\end{gathered}
\ee
We can now utilize the additional gauge actions, associated with local Lorentz transformations acting on the tetrad vectors, to further simplify the above operator in a general manner. Given that the Teukolsky operator acts on the perturbed NP scalars $\Psi_{s}$ which have boost weights equal to their spin weights, $b=s$, and since the Teukolsky operator itself is built from background quantities, that is, it is invariant under gauge transformations of the perturbations, we can always perform a ``far-horizon'' local boost transformation,
\be\label{eq:s_Rep}
	\Psi_{s} \xrightarrow{\text{boost}} e^{s\,\Delta^{\varepsilon}\,\eta\left(r\right)} \Psi_{s} \,,
\ee
with $\eta\left(r\right)$ a purely radial function that is regular at the horizon, chosen such that we remove the $s\,\delta\gamma^{r}\partial_{r}$ term, while the exponent $\varepsilon$ must be such that the near horizon characteristic exponents of the equations of motion are simultaneously preserved. Indeed, choosing, $\eta\left(r\right)$ to satisfy,
\be
	\left[\Delta^{\varepsilon}\,\eta\left(r\right)\right]^{\prime} = -2f_{r} \,,
\ee
removes the term that goes like $s\,\partial_{r}$, while the modifications in the scalar part are always subleading in the vicinity of the horizon as long as $\varepsilon>1$. We will adopt this particular gauge where $\delta\gamma^{r}=0$ but always have in mind to supplement with the more general results corresponding to the above local boost transformation.

\subsection{Construction of $\SL$'s}
With this background effective geometry in hand, we now investigate the existence of the three generators $L_0$, $L_{+1}$ and $L_{-1}$ satisfying the $\SL$ algebra,
\be
	\left[L_{\pm1},L_0\right]=\pm L_{\pm1}\,\,\,,\,\,\,\left[L_{\pm1},L_{\mp1}\right]=\pm2L_0 \,,
\ee
and whose Casimir is given by the truncated radial Teukolsky operator,
\be\ba
	\mathcal{C}_2 &= L_0^2 - \frac{1}{2}\left(L_{\pm1}L_{\mp1}+L_{\mp1}L_{\pm1}\right) \\
	&\equiv \mathcal{C}_2^{\text{Star}} + f_{tt}\left(r\right)\,\partial_{t}^2 + 2 f_{t\phi}\left(r\right)\Omega\,\partial_{t}\partial_{\phi} + f_{\phi\phi}\left(r\right)\Omega^2\,\partial_{\phi}^2 \\
	&\,\,\,\,\,\,\,\,\,\,\,\,\,\,\,\,\,\,\,+ \Delta\,f_{r}\left(r\right)\,\partial_{r} + s\,\left[f_{t}\left(r\right)\,\partial_{t} + f_{\phi}\left(r\right)\Omega\,\partial_{\phi} + \Delta f_0\left(r\right)\right] \,,
\ea\ee
where $\mathcal{C}_2^{\text{Star}}$ is the Starobinsky near zone approximation,
\be\ba
    \mathcal{C}_2^{\text{Star}} &=  \Delta^{-s}\,\partial_{r}\Delta^{s+1}\partial_{r} -\frac{\left(r_{+}^2+a^2\right)^2}{\Delta}\left(\partial_{t}+\Omega\,\partial_{\phi}\right)^2 \\
	&\,\,\,\,\,\,\,+ s\,\frac{\left(r_{+}^2+a^2\right)\Delta^{\prime}}{\Delta}\left(\partial_{t}+\Omega\,\partial_{\phi}\right) + s\left(s+1\right) \,.
\ea\ee
In the above radial operator, we used the facts that $G^{\mu\nu}\Gamma^{t}_{\mu\nu}=0$ and $G^{\mu\nu}\Gamma^{\phi}_{\mu\nu}=0$ according to our generic subtracted geometry treatment above, i.e. we identified any $\partial_{t}$ and $\partial_{\phi}$ terms as arising due to a non-zero spin weight, while $f_{r}$ is independent of $s$ as per our last discussion regarding local boost transformations of the perturbed NP scalars.

The goal here is to be able write solutions of the subtracted equations charged under the action of this $\SL$, i.e. labeled by $L_0$ and $\mathcal{C}_2$. Given that these solutions are also charged under the action of the $\mathbb{R}_{t}\times U_{\phi}\left(1\right)$ isometry group, we make the following general ansatz for $L_0$,
\be
	L_0 = -\left(\beta_{t}\partial_{t} + \beta_{\phi}\Omega\,\partial_{\phi}\right) + s\,\beta_0 \,,
\ee
with $\beta_{t}$, $\beta_{\phi}$ and $\beta_0$ constants. This is supplemented with a completely general form of the remaining generators,
\be
	L_{\pm 1} = \tilde{g}_{\pm}\left(t,r,\phi\right)\,\partial_{r} + \tilde{k}_{\pm}\left(t,r,\phi\right)\beta_{t}\,\partial_{t} + \tilde{h}_{\pm}\left(t,r,\phi\right)\Omega\,\partial_{\phi} + s\,\tilde{\lambda}_{\pm}\left(t,r,\phi\right) \,,
\ee
where all scalar parts of the generators are assigned to a non-zero spin weight. However, we are not explicitly assuming that the vector part functions do not depend on $s$, even though this will turn out to be the case as we will see. We note here that we are explicitly assuming that $\beta_{t}\ne0$, indicated also by the explicit appearance of $\beta_{t}$ in the $t$-component of $L_{\pm1}$. This should be regarded as an additional requirement, rather as an assumption, associated to the fact that we are also looking for generators that have a smooth non-rotating limit, where the equations of motion become spherically symmetric and no $\phi$-derivatives appear in the radial operator.

With this starting ansatz for the generators, we begin applying the $\SL$ algebra constraints. First of all, $\left[L_{\pm 1},L_0\right]=\pm L_{\pm 1}$ implies that,
\be
	\left(\beta_{t}\partial_{t} + \beta_{\phi}\Omega\,\partial_{\phi}\right)\tilde{\chi}_{\pm} = \pm \tilde{\chi}_{\pm}\,\,\,,\,\,\,\tilde{\chi}_{\pm}=\tilde{g}_{\pm},\tilde{k}_{\pm},\tilde{h}_{\pm},\tilde{\lambda}_{\pm} \,,
\ee
which can be solved to eliminate the explicit $t$-dependence,
\be
	\tilde{\chi}_{\pm}\left(t,r,\phi\right) = e^{\pm t/\beta_{t}}\chi_{\pm}(r,\hat{\phi})\,\,\,,\,\,\,\hat{\phi}\equiv\phi-\frac{\beta_{\phi}}{\beta_{t}}\,\Omega t\,\,\,,\,\,\,\chi_{\pm}=g_{\pm},k_{\pm},h_{\pm},\lambda_{\pm} \,.
\ee
From the above form of the functions $\tilde{\chi}_{\pm}$, it is instructive to reformulate the problem in the $(t,r,\hat{\phi})$ coordinates, instead of $\left(t,r,\phi\right)$. In these coordinates, the generators read,
\be
	\begin{gathered}
		L_0 = -\beta_{t}\,\partial_{t} + s\,\beta_0 \,, \\
		L_{\pm 1} = e^{\pm t/\beta_{t}}\left[g_{\pm}(r,\hat{\phi})\,\partial_{r} + k_{\pm}(r,\hat{\phi})\beta_{t}\,\partial_{t} + \hat{h}_{\pm}(r,\hat{\phi})\Omega\,\partial_{\hat{\phi}} + s\,\lambda_{\pm}(r,\hat{\phi})\right] \,,
	\end{gathered}
\ee
with $\hat{h}_{\pm} = h_{\pm} - \beta_{\phi}k_{\pm}$. The remaining algebra constraints $\left[L_{\pm1},L_{\mp1}\right]=\pm2L_0$ then become,
\begin{subequations}
	\begin{gather}
	\label{eq:AC1}
		g_{[\pm}\partial_{r}g_{\mp]} + \hat{h}_{[\pm}\Omega\,\partial_{\hat{\phi}}g_{\mp]} = \pm k_{(\pm}g_{\mp)} \,, \\
	\label{eq:AC2}
		g_{[\pm}\partial_{r}\hat{h}_{\mp]} + \hat{h}_{[\pm}\Omega\,\partial_{\hat{\phi}}\hat{h}_{\mp]} = \pm k_{(\pm}\hat{h}_{\mp)} \,, \\
	\label{eq:AC3}
		g_{[\pm}\partial_{r}k_{\mp]} + \hat{h}_{[\pm}\Omega\,\partial_{\hat{\phi}}k_{\mp]} = \pm \left(k_{\pm}k_{\mp}-1\right) \,, \\
	\label{eq:AC4}
		g_{[\pm}\partial_{r}\lambda_{\mp]} + \hat{h}_{[\pm}\Omega\,\partial_{\hat{\phi}}\lambda_{\mp]} = \pm \left(k_{(\pm}\lambda_{\mp)} + \beta_0\right) \,,
	\end{gather}
\end{subequations}
where we have defined the symmetric and antisymmetric operations with respect to the ``$\pm$'' indices $A_{(\pm}B_{\mp)}\equiv \left(A_{\pm}B_{\mp}+A_{\mp}B_{\pm}\right)/2$ and $A_{[\pm}B_{\mp]}\equiv \left(A_{\pm}B_{\mp}-A_{\mp}B_{\pm}\right)/2$. These are supplemented with the Casimir constraints associated with the near horizon characteristic exponents of the equations of motion,
\begin{subequations}
	\begin{gather}
	\label{eq:CC1}
		g_{\pm}g_{\mp} = -\Delta \,\,\,,\,\,\, g_{(\pm}k_{\mp)} = 0 \,\,\,,\,\,\, g_{(\pm}\hat{h}_{\mp)} = 0 \,, \\
	\label{eq:CC2}
		k_{\pm}k_{\mp}-1 = -\beta_{t}^{-2}\left[-\frac{\left(r_{+}^2+a^2\right)^2}{\Delta} + f_{tt}\left(r\right)\right] \,, \\
	\label{eq:CC3}
		k_{(\pm}\hat{h}_{\mp)} = -\beta_{t}^{-1}\left[-\frac{\left(r_{+}^2+a^2\right)^2}{\Delta}\left(1-\frac{\beta_{\phi}}{\beta_{t}}\right) + f_{t\hat{\phi}}\left(r\right)\right] \,, \\
	\label{eq:CC4}
		\hat{h}_{\pm}\hat{h}_{\mp} = -\left[-\frac{\left(r_{+}^2+a^2\right)^2}{\Delta}\left(1-\frac{\beta_{\phi}}{\beta_{t}}\right)^2 + f_{\hat{\phi}\hat{\phi}}\left(r\right)\right] \,,
	\end{gather}
\end{subequations}
\begin{subequations}
	\begin{gather}
	\label{eq:CC5}
		\frac{1}{2}\partial_{r}\left(g_{\pm}g_{\mp}\right) + \hat{h}_{(\pm}\Omega\,\partial_{\hat{\phi}}g_{\mp)} \mp k_{[\pm}g_{\mp]} + 2s\, \lambda_{(\pm}g_{\mp)} = -\left[\left(s+1\right)\Delta^{\prime} + \Delta\,f_{r}\left(r\right)\right] \,, \\
	\label{eq:CC6}
		g_{(\pm}\partial_{r}k_{\mp)} + \hat{h}_{(\pm}\Omega\,\partial_{\hat{\phi}}k_{\mp)} + 2s \left(\lambda_{(\pm}k_{\mp)} + \beta_0\right) = -\beta_{t}^{-1}s\left[\frac{\left(r_{+}^2+a^2\right)\Delta^{\prime}}{\Delta} + f_{t}\left(r\right)\right] \,, \\
	\label{eq:CC7}
		g_{(\pm}\partial_{r}\hat{h}_{\mp)} + \frac{1}{2}\Omega\,\partial_{\hat{\phi}}(\hat{h}_{\pm}\hat{h}_{\mp}) \mp k_{[\pm}\hat{h}_{\mp]} + 2s\, \lambda_{(\pm}\hat{h}_{\mp)} = -s\left[\frac{\left(r_{+}^2+a^2\right)\Delta^{\prime}}{\Delta}\left(1-\frac{\beta_{\phi}}{\beta_{t}}\right) + f_{\hat{\phi}}\left(r\right)\right] \,, \\
	\label{eq:CC8}
		g_{(\pm}\partial_{r}\lambda_{\mp)} + \hat{h}_{(\pm}\Omega\,\partial_{\hat{\phi}}\lambda_{\mp)} \mp k_{[\pm}\lambda_{\mp]} + s\,\left(\lambda_{\pm}\lambda_{\mp} - \beta_0^2\right) = -\left[s+1 + \Delta f_0\left(r\right)\right] \,.
	\end{gather}
\end{subequations}
We can now proceed to solve the above algebra and Casimir constraints. The first thing to notice is that the Casimir constraints imply that products of any component of $\chi_{\pm}$ and any component of $\chi_{\mp}$ are $\hat{\phi}$-independent and $s$-independent. As such, any $\hat{\phi}$-dependence must come in the form,
\be
	\chi_{\pm}(r,\hat{\phi}) = e^{\pm A(r,\hat{\phi})}X_{\pm}\left(r\right)\,\,\,,\,\,\,X_{\pm} = G_{\pm}, K_{\pm}, \hat{H}_{\pm}, \Lambda_{\pm}
\ee
with $X_{\pm}\left(r\right)$ being independent of $s$ as well.

For the radial components $g_{\pm}(r,\hat{\phi})$, \eqref{eq:CC1} and the differential equation \eqref{eq:AC1} allow to completely fix the radial dependence. In particular, the second Casimir constraint in \eqref{eq:CC1} makes the RHS of \eqref{eq:AC1} zero, while the second term in the algebra constraint \eqref{eq:AC1} also vanishes by virtue of the third Casimir constraint in \eqref{eq:CC1} since, due to the first of \eqref{eq:CC1}, $\hat{h}_{[\pm}\Omega\,\partial_{\hat{\phi}}g_{\mp]} = g_{(\pm}\hat{h}_{\mp)}\Omega\,\partial_{\hat{\phi}}\ln g_{\pm} = 0$. As such, all the $r$-dependence in $A(r,\hat{\phi})$ can be separated with the end result,
\be
	g_{\pm}(r,\hat{\phi}) = e^{\pm A(\hat{\phi})}\,\left[\mp\sqrt{\Delta}\right] \,.
\ee
Here, we note that we also have the freedom of an overall integration constant $c$ that comes in the form $e^{\pm c}$. Such overall constant reciprocal rescalings of $L_{\pm1}$, however, are automorphisms and are, therefore, algebraically trivial. With this extracted function form of $G_{\pm} = \mp\sqrt{\Delta}$, the second and third of the Casimir constraints \eqref{eq:CC1} imply that the radial functions in $t$- and $\hat{\phi}$-components of the generators $L_{\pm 1}$ are equal,
\be
	K_{\pm}\left(r\right) = K\left(r\right) \,\,\,,\,\,\, \hat{H}_{\pm}\left(r\right) = \hat{H}\left(r\right) \,.
\ee
It is then straightforward to conclude that the purely angular function $A(\hat{\phi})$ is linear in $\hat{\phi}$. For example, taking the $\hat{\phi}$-derivative of the algebra constraint \eqref{eq:AC2} or \eqref{eq:AC3}, we get,
\be
	A^{\prime\prime}(\hat{\phi}) = 0 \Rightarrow A(\hat{\phi}) = \tau\hat{\phi} \,,
\ee
with $\tau$ an arbitrary integration constant, while we have again ignored algebraically trivial reciprocal rescalings of $L_{\pm 1}$.

Moving on, multiplying \eqref{eq:AC2} with $\tau\Omega$ and adding with it \eqref{eq:AC3}, we end up with a first order non-linear differential equation for the function $K\left(r\right)+\tau\Omega\hat{H}\left(r\right)$ which can be solved to find,
\be\label{eq:KmH}
	K\left(r\right)+\tau\Omega\hat{H}\left(r\right) = \frac{\left(\sqrt{r-r_{+}}+\sqrt{r-r_{-}}\right)^4 + c_1}{\left(\sqrt{r-r_{+}}+\sqrt{r-r_{-}}\right)^4 - c_1} \,,
\ee
with $c_1$ an integration constant. In the mean time, the Casimir constraints \eqref{eq:CC2} and \eqref{eq:CC4} tell us that, up to an overall sign, which falls into the category of automorphisms of the $\SL$ algebra,
\be
	K\left(r\right) \sim \beta_{t}^{-1}\frac{r_{+}^2+a^2}{\sqrt{\Delta}} \,\,\,\text{and}\,\,\, \hat{H}\left(r\right) \sim \left(1-\frac{\beta_{\phi}}{\beta_{t}}\right)\frac{r_{+}^2+a^2}{\sqrt{\Delta}}\,\,\,\text{as }\Delta\rightarrow0 \,,
\ee
where the sign of $\hat{H}\left(r\right)$ relative to $K\left(r\right)$ was also fixed by using the near horizon behavior of \eqref{eq:CC3}. From these, we infer the near horizon behavior of \eqref{eq:KmH} which fixes $c_1=\left(r_{+}-r_{-}\right)^2$ regardless of the values of $\beta_{t}$, $\beta_{\phi}$ and $\tau$. This greatly simplifies \eqref{eq:KmH} to,
\be
	K\left(r\right)+\tau\Omega \hat{H}\left(r\right) = \partial_{r}\left(\sqrt{\Delta}\right) \,,
\ee
and results in the following quite pleasing exact expressions of $K\left(r\right)$ and $\hat{H}\left(r\right)$ after solving \eqref{eq:AC2} and matching the near horizon behavior of $\hat{H}\left(r\right)$,
\be
	K\left(r\right) = \partial_{r}\left(\sqrt{\Delta}\right) - \tau\Omega\left(1-\frac{\beta_{\phi}}{\beta_{t}}\right)\frac{r_{+}^2+a^2}{\sqrt{\Delta}}\,\,\,,\,\,\,\hat{H}\left(r\right) = \left(1-\frac{\beta_{\phi}}{\beta_{t}}\right)\frac{r_{+}^2+a^2}{\sqrt{\Delta}}
\ee
The near horizon behavior of $K\left(r\right)$ also eliminates $\beta_{t}$ or $\beta_{\phi}$, depending on the value of $\tau$. More explicitly,
\be
	\begin{gathered}
		\beta_{t} = \beta\,\frac{1-\tau\beta_{\phi}\Omega}{1-\tau\beta\Omega} \equiv \beta^{\left(\tau\right)} \,\,\,\text{and}\,\,\,\beta_{\phi}\in\mathbb{R}\,\,\,\,\,\text{if }\,\,\tau\ne\frac{1}{\beta\Omega} \,, \\
		\text{ OR } \\
		\beta_{\phi} = \beta \,\,\,\text{and}\,\,\,\beta_{t}\in\mathbb{R} \,\,\,\,\,\text{if }\,\,\tau=\frac{1}{\beta\Omega} \,.
	\end{gathered}
\ee

We now move on to the remaining constraints \eqref{eq:AC4} and \eqref{eq:CC5}-\eqref{eq:CC8}. \eqref{eq:AC4} results in a linear first-order inhomogeneous differential equation for $\Lambda_{+}\left(r\right)+\Lambda_{-}\left(r\right)$ which is easily solved to give,
\be
	\Lambda_{+}\left(r\right) + \Lambda_{-}\left(r\right) = -\left[2\beta_0\sqrt{\frac{r-r_{+}}{r-r_{-}}} + \frac{r_{+}-r_{-}}{2\sqrt{\Delta}}\right] \,,
\ee
where the integration constant was fixed by the near horizon behavior of \eqref{eq:CC6}. Next, \eqref{eq:CC5} simplifies to,
\be
	s\left[\sqrt{\Delta}\left(\Lambda_{+}\left(r\right)-\Lambda_{-}\left(r\right)\right) + \Delta^{\prime}\right] = -\Delta f_{r}\left(r\right)
\ee
and, since $f_{r}\left(r\right)$ is independent of the spin weight, $f_{r}\left(r\right)=0$, leaving us with,
\be
	\Lambda_{+}\left(r\right)-\Lambda_{-}\left(r\right) = -2\partial_{r}\left(\sqrt{\Delta}\right)
\ee
These two finally tells us that,
\be
	\Lambda_{\pm}\left(r\right) = -\left(1\pm1\right)\partial_{r}\left(\sqrt{\Delta}\right) + \left(1-\beta_0\right)\sqrt{\frac{r-r_{+}}{r-r_{-}}} \,,
\ee
and the remaining constraints fix the ``far-horizon'' corrections $f_{t}\left(r\right)$, $f_{\hat{\phi}}\left(r\right)$ and $f_0\left(r\right)$.

In summary, after some rearrangement, the most general subtracted geometry $\SL$ is generated by,
\be\label{eq:SL2RGeneral}
	\begin{gathered}
		L_0 = -\left(\beta_{t}\,\partial_{t} + \beta_{\phi}\Omega\,\partial_{\phi}\right) + s\,\beta_0 \,,
		\\
		\ba
			L_{\pm1} = e^{\pm \left[t/\beta + \tau\left(\phi-\Omega t\right)\right]}&\bigg[\mp\sqrt{\Delta}\,\partial_{r} + \left(\frac{\beta_{t}}{\beta}\sqrt{\frac{r-r_{+}}{r-r_{-}}} + \frac{r_{+}-r_{-}}{2\sqrt{\Delta}}\right)\beta\,\partial_{t} \\
			&+ \left(\frac{\beta_{\phi}}{\beta}\sqrt{\frac{r-r_{+}}{r-r_{-}}} + \frac{r_{+}-r_{-}}{2\sqrt{\Delta}}\right)\beta\Omega\,\partial_{\phi} \\
			&- s\left(\left(\beta_0\pm1\right)\sqrt{\frac{r-r_{+}}{r-r_{-}}} + \left(1\pm1\right)\frac{r_{+}-r_{-}}{2\sqrt{\Delta}}\right) \bigg] \,,
		\ea
	\end{gathered}
\ee
with $\beta_{t}=\beta^{\left(\tau\right)}$ and generic $\beta_{\phi}$, $\beta_0$ and $\tau$ if $\tau\beta\Omega\ne1$ or $\beta_{\phi}=\beta$ and generic $\beta_{t}$ and $\beta_0$ if $\tau\beta\Omega=1$. The associated Casimir is given by,
\be
	\mathcal{C}_2 = \mathcal{C}_2^{\text{Star}} + \frac{r_{+}-r_{-}}{r-r_{-}}\big[\beta_{t}\,\partial_{t}+\beta_{\phi}\Omega\,\partial_{\phi}+s\left(1-\beta_0\right)\big]\left[\left(\beta_{t}-\beta\right)\partial_{t} + \left(\beta_{\phi}-\beta\right)\Omega\,\partial_{\phi}-s\,\beta_0\right] \,.
\ee

From all of the above subtracted Casimirs, only the subset with $(\beta_{\phi},\beta_0)=(0,1)$ or $(\beta_{\phi},\beta_0)=(\beta,0)$ give rise to valid near zone approximations, for which all the static terms must be kept explicitly, and precisely correspond to $\tau$-generalizations of the Love and Starobinsky near zones respectively.

Interestingly, all the generators written above are automatically regular at both the future and the past event horizons, as can be seen by working in advanced and retarded null coordinates respectively. This statement of course only holds for the vector, $s=0$, parts, since the scalar $s\ne0$ parts are tetrad-dependent and we are working here in the singular at the horizon Kinnerslay tetrad. More explicitly,
\be\label{eq:SL2RGeneralAdvanced}
\begin{split}
	& L_0\big|_{s=0} = -\left(\beta_{t}\,\partial_{\upsilon} + \beta_{\phi}\Omega\,\partial_{\varphi}\right) \,,\\
	& L_{\pm1}\big|_{s=0} = \exp\left\{\pm\left[\frac{\upsilon-r}{\beta}-\tau\left(\varphi-\Omega\left(\upsilon-r\right)\right)\right]\right\} \left(\frac{r-r_{-}}{r_{+}}\right)^{\mp 2M\left(1+\tau\beta\Omega\right)/\beta} \\
	&\left[\mp\left(r-r_{\mp}\right)\partial_{r}+\frac{r-r_{\mp}}{r-r_{-}}\left(\left(\beta_{t}\mp\left(r+r_{+}\right)\right)\partial_{\upsilon} + \beta_{\phi}\Omega\,\partial_{\varphi}\right) + \frac{1\mp1}{2}\frac{r_{+}-r_{-}}{r-r_{-}}\beta\left(\partial_{\upsilon}+\Omega\,\partial_{\varphi}\right)\right] \,,
\end{split}
\ee
in advanced coordinates \eqref{eq:AdvancedCoords} and,
\be\label{eq:SL2RGeneralRetarded}
\begin{split}
	& L_0\big|_{s=0} = -\left(\beta_{t}\,\partial_{u} + \beta_{\phi}\Omega\,\partial_{\tilde{\varphi}}\right) \,,\\
	& L_{\pm1}\big|_{s=0} = \exp\left\{\pm\left[\frac{u+r}{\beta}-\tau\left(\tilde{\varphi}-\Omega\left(u+r\right)\right)\right]\right\} \left(\frac{r-r_{-}}{r_{+}}\right)^{\pm 2M\left(1+\tau\beta\Omega\right)/\beta} \\
	&\left[\mp\left(r-r_{\pm}\right)\partial_{r}+\frac{r-r_{\pm}}{r-r_{-}}\left(\left(\beta_{t}\pm\left(r+r_{+}\right)\right)\partial_{u} + \beta_{\phi}\Omega\,\partial_{\tilde{\varphi}}\right) + \frac{1\pm1}{2}\frac{r_{+}-r_{-}}{r-r_{-}}\beta\left(\partial_{u}+\Omega\,\partial_{\tilde{\varphi}}\right)\right] \,,
\end{split}
\ee
in retarded coordinates \eqref{eq:RetardedCoords}.

\subsection{Globally defined, time-reversal symmetric subtracted $\SL$'s}
In the main text, we are mostly interested in those subtracted geometry approximations which preserve all the symmetries of the full geometry, including time-reversal invariance\footnote{Time-reversal invariance here refers to the simultaneous time-reversal transformation $t\rightarrow-t$ and the flip of the direction of rotation of the black hole $a\rightarrow-a$.}, and are globally defined. These correspond to $\tau=0$ and make up a $2$-parameter family of subtracted geometry approximations enjoying an $\SL$ symmetry, labeled by the $\beta_{\phi}$ and $\beta_0$ parameters. The generators read,
\be
	\begin{gathered}
		L_0 = -\beta\,\partial_{t} + s + \left(-1\right)\big(\beta_{\phi}\Omega\,\partial_{\phi} + s\left(1-\beta_0\right)\big) \,,
		\\
		\ba
			L_{\pm1} = e^{\pm t/\beta}&\left[\mp\sqrt{\Delta}\,\partial_{r} + \partial_{r}\left(\sqrt{\Delta}\right)\beta\,\partial_{t} + \frac{a}{\sqrt{\Delta}}\,\partial_{\phi} - s\,\left(1\pm1\right)\partial_{r}\left(\sqrt{\Delta}\right)\right] \\
			&+\left(e^{\pm t/\beta}\sqrt{\frac{r-r_{+}}{r-r_{-}}}\right)\big(\beta_{\phi}\Omega\,\partial_{\phi} + s\left(1-\beta_0\right)\big) \,,
		\ea
	\end{gathered}
\ee
and the associated Casimir operator is given by,
\be
	\mathcal{C}_2 = \mathcal{C}_2^{\text{Star}} + \frac{r_{+}-r_{-}}{r-r_{-}}\big[\beta\,\partial_{t}+\beta_{\phi}\Omega\,\partial_{\phi}+s\left(1-\beta_0\right)\big]\big[\left(\beta_{\phi}-\beta\right)\Omega\,\partial_{\phi} - s\,\beta_0\big] \,.
\ee
As already mentioned, only the subset with $(\beta_{\phi},\beta_0)=(0,1)$ (Love near zone) or $(\beta_{\phi},\beta_0)=(\beta,0)$ (Starobinsky near zone) give rise to valid near zone approximations. We also remind here that we still have the freedom of performing the coordinate transformations \eqref{eq:t_r_phi_Rep} and the local boost transformation \eqref{eq:s_Rep}. From these, only the temporal translations $t\rightarrow t + g_{t}\left(r\right)$ preserve the near zone behavior which, however, break the time-reflection symmetry.

A nice remark here is that the $\tau$-generalized generators of the spontaneously broken near zone $\SL$ can be obtained by the $\tau=0$ globally defined ones after performing a particular $\phi$-dependent temporal translation,
\be
	L_{m}^{\left(\tau=0\right)} \xrightarrow[\tau\beta\Omega\ne1]{t \rightarrow t + \tau\beta\left(\phi-\Omega t\right)} L_{m}^{\left(\tau\ne0\right)}
\ee

\subsection{Subtracted $\SL\times\SL$'s}
As a last piece of analysis, let us investigate under what conditions it is possible to enhance the $\SL$ symmetry of the subtracted region to $\SL\times\SL$. Let $\SL^{\left(1\right)}$ and $\SL^{\left(2\right)}$ be two generic subtracted geometry $\SL$'s with parameters $(\beta_{t}^{\left(1\right)},\beta_{\phi}^{\left(1\right)},\beta_0^{\left(1\right)},\tau^{\left(1\right)})$ and $(\beta_{t}^{\left(2\right)},\beta_{\phi}^{\left(2\right)},\beta_0^{\left(2\right)},\tau^{\left(2\right)})$ respectively. If $\tau^{\left(i\right)}\beta\Omega\ne1$, then $\beta_{t}^{\left(i\right)}=\beta^{\tau^{\left(i\right)}}$. Otherwise, if $\tau^{\left(i\right)}\beta\Omega=1$, then $\beta_{\phi}^{\left(i\right)}=\beta$. We wish to work out the requirements,
\be
	\left[L_{m}^{\left(1\right)},L_{n}^{\left(2\right)}\right] = 0 \,\,\,,\,\,\,m,n=0,\pm1 \,,
\ee
for the various cases of possibles values of the parameters.

First, for the case where $\tau^{\left(1\right)}\beta\Omega\ne1$ and $\tau^{\left(2\right)}\beta\Omega\ne1$, where $\beta_{t}^{\left(1\right)}=\beta^{\tau^{\left(1\right)}}$ and $\beta_{t}^{\left(2\right)}=\beta^{\tau^{\left(2\right)}}$, one finds the unique solution,
\be
	\beta_{\phi}^{\left(2\right)} = \beta - \beta_{\phi}^{\left(1\right)} \,,\quad \beta_0^{\left(2\right)} = 1 -\beta_0^{\left(1\right)}\,,\quad \tau^{\left(2\right)} = \frac{1-\tau^{\left(1\right)}\beta_{\phi}^{\left(1\right)}\Omega}{\left(\beta-\beta_{\phi}^{\left(1\right)}\right)\Omega} \,,
\ee
with generic $\tau^{\left(1\right)}\ne 1/\left(\beta\Omega\right)$, $\beta_{\phi}^{\left(1\right)}\ne\beta$ and $\beta_0^{\left(1\right)}$.

For the case where $\tau^{\left(1\right)}\beta\Omega=1$ and $\tau^{\left(2\right)}\beta\Omega=1$, no solutions exists for which the two $\SL$'s commute.

Last, for the case of $\tau^{\left(1\right)}\beta\Omega\ne1$ and $\tau^{\left(2\right)}\beta\Omega=1$, for which $\beta_{t}^{\left(1\right)}=\beta^{\tau^{\left(1\right)}}$ and $\beta_{\phi}^{\left(2\right)}=\beta$, we find the solution,
\be
	\beta_{\phi}^{\left(1\right)}=0 \,,\quad \beta_{t}^{\left(2\right)}=\beta\frac{-\tau^{\left(1\right)}\beta\Omega}{1-\tau^{\left(1\right)}\beta\Omega} \,,
\ee
with generic $\tau^{\left(1\right)}\ne 1/\left(\beta\Omega\right)$, $\beta_0^{\left(1\right)}$ and $\beta_0^{\left(2\right)}$.

For all of these enhancements the two Casimir operators are automatically equal, $\mathcal{C}_2^{\left(2\right)} = \mathcal{C}_2^{\left(1\right)}$. Interestingly, all of the above requirements for the subtracted geometry truncations to be equipped with an $\SL\times\SL$ symmetry can be elegantly reproduced from the following absolute condition,
\be
	L_0^{\left(1\right)} + L_0^{\left(2\right)} = -\beta\left(\partial_{t}+\Omega\,\partial_{\phi}\right)+s \,.
\ee
This allows to write the $L_{\pm1}$ generators and the Casimir very compactly in terms of the $L_0$ generators as,
\be
	\begin{gathered}
		L_{\pm1}^{\left(i\right)} = e^{\pm\left[t/\beta+\tau^{\left(i\right)}\left(\phi-\Omega t\right)\right]}\left[\mp\sqrt{\Delta}\,\partial_{r}-\partial_{r}\left(\sqrt{\Delta}\right)\left(L_0^{\left(i\right)}\pm s\right) + \frac{r_{+}-r_{-}}{2\sqrt{\Delta}}L_0^{\left(j\right)}\right] \,, \\
		\mathcal{C}_2^{\left(i\right)} = \mathcal{C}_2^{\left(j\right)} = \mathcal{C}_2^{\text{Star}} - \frac{r_{+}-r_{-}}{r-r_{-}}\left(L_0^{\left(i\right)}-s\right)\left(L_0^{\left(j\right)}-s\right) \,,
	\end{gathered}
\ee
with $i=1,2$ and $j\ne i$. Even though there are infinitely many such enhancements, none can have simultaneously $\tau^{\left(1\right)}=\tau^{\left(2\right)}=0$, that is, these subtracted geometry conformal symmetries are unavoidably spontaneously broken by the periodicity of the azimuthal angle.

\subsection{Near zone $\SL\times\SL$'s}
Finally, it is possible to construct enhancements of the near zone $\SL$'s, not just the subtracted $\SL$'s. These now correspond to the third case above, supplemented with the near zone requirement $\beta_0^{\left(1\right)} = 1$ and $\beta_0^{\left(2\right)} = 0$, but for generic $\tau^{\left(1\right)}\equiv\tau$. The new element here compared to \cite{Castro:2010fd} is that we find a $1$-parameter family of near zone $\SL\times\SL$'s controlled by generic\footnote{In \cite{Castro:2010fd}, $\tau=-\frac{r_{s}}{2a}$ and $\beta_{t}=r_{s}\frac{r_{s}}{r_{+}-r_{-}}$.} $\tau$, with generators,
\be
	\begin{gathered}
		L_0^{\left(1\right)} = -\frac{1}{1-\tau\beta\Omega}\beta\,\partial_{t} + s \,, \\
		\ba
			L_{\pm1}^{\left(1\right)} = e^{\pm \left[t/\beta + \tau\left(\phi-\Omega t\right)\right]}&\bigg[\mp\sqrt{\Delta}\,\partial_{r} + \left(\partial_{r}\left(\sqrt{\Delta}\right) + \frac{\tau\beta\Omega}{1-\tau\beta\Omega}\sqrt{\frac{r-r_{+}}{r-r_{-}}}\right)\beta\,\partial_{t} \\
			&+ \frac{a}{\sqrt{\Delta}}\,\partial_{\phi} - s\,\left(1\pm1\right)\partial_{r}\left(\sqrt{\Delta}\right)\bigg] \,,
		\ea \\
		\\
		L_0^{\left(2\right)} = -\beta\left(\frac{-\tau\beta\Omega}{1-\tau\beta\Omega}\,\partial_{t} + \Omega\,\partial_{\phi}\right) \,, \\
		\ba
			L_{\pm1}^{\left(2\right)} = e^{\pm \phi/\beta\Omega}&\bigg[\mp\sqrt{\Delta}\,\partial_{r} + \left(\partial_{r}\left(\sqrt{\Delta}\right) -\frac{1}{1-\tau\beta\Omega}\sqrt{\frac{r-r_{+}}{r-r_{-}}}\right)\beta\,\partial_{t} \\
			&+\partial_{r}\left(\sqrt{\Delta}\right)\beta\Omega\,\partial_{\phi} \mp s\frac{r-r_{\mp}}{\sqrt{\Delta}}\bigg] \,,
		\ea
	\end{gathered}
\ee
and Casimir,
\be
	\mathcal{C}_2^{\left(1\right)} = \mathcal{C}_2^{\left(2\right)} = \mathcal{C}_2^{\text{Star}} - \frac{r_{+}-r_{-}}{r-r_{-}}\left[\beta\left(\frac{-\tau\beta\Omega}{1-\tau\beta\Omega}\,\partial_{t} + \Omega\,\partial_{\phi}\right) + s\right]\frac{\beta\,\partial_{t}}{1-\tau\beta\Omega} \,.
\ee
The same generalization was provided in \cite{Lowe2012} but only for $s=0$. To our best knowledge, no such extensions have been previously introduced beyond a specific case of particular values of $\tau$ for electromagnetic fields in \cite{Shi2018}, even though these were not performed in the framework of the NP formalism employed in the Teukolsky equation. We also remark here how the globally defined Love symmetry, corresponding to $\SL^{\left(1\right)}$ with $\tau=0$, can be extended to $\SL\times\SL$, in contrast to the other globally defined near zone $\SL$ symmetry, the Starobinsky near zone.

\section{Generalized Lie derivative on GHP tensors}
\label{ApLieDerivative}

In this appendix, we present the construction of the generalized ``spin weighted'' Lie derivative. In particular, we will construct the most general generalized Lie derivative acting on objects that are spacetime tensors of some particular rank and also carry definite GHP weights. We will refer to such object as GHP tensors, with the ``tensors'' and ``GHP'' parts referring to their homogeneous transformation rules under diffeomorphisms and local boosts and rotations respectively.

There are some minimal conditions that such a Lie derivative must obey for it to be well defined. In the following conditions, we denote a general GHP tensor of GHP weights $\left\{p,q\right\}$ and spacetime tensorial structure $\left(k,l\right)$, with $k$ the contravariant rank and $l$ the covariant rank, as $A_{\left\{q,p\right\}}^{\left(k,l\right)}$.
\begin{itemize}
	\item It reduces to the usual unique Lie derivative $\mathcal{L}_{\xi}$ in the case the GHP tensor it acts on has zero GHP weights, i.e. when it acts on pure spacetime tensors,
	\be
		\Lstr_{\xi}A_{\left\{0,0\right\}}^{\left(k,l\right)} = \mathcal{L}_{\xi}A_{\left\{0,0\right\}}^{\left(k,l\right)} \,.
	\ee
	
	\item It obeys the Leibniz rule,
	\be
		\Lstr_{\xi}\left(A_{\left\{p,q\right\}}^{\left(k,l\right)}B_{\left\{p^{\prime},q^{\prime}\right\}}^{\left(k^{\prime},l^{\prime}\right)}\right) = \left(\Lstr_{\xi}A_{\left\{p,q\right\}}^{\left(k,l\right)}\right)B_{\left\{p^{\prime},q^{\prime}\right\}}^{\left(k^{\prime},l^{\prime}\right)} + A_{\left\{p,q\right\}}^{\left(k,l\right)}\left(\Lstr_{\xi}B_{\left\{p^{\prime},q^{\prime}\right\}}^{\left(k^{\prime},l^{\prime}\right)}\right) \,.
	\ee
	
	\item It acts covariantly, leaving the GHP weights and spacetime rank unaltered, i.e. when it acts on a GHP tensors, it outputs a GHP tensor of the same type,
	\be
		\Lstr_{\xi}A_{\left\{p,q\right\}}^{\left(k,l\right)} \xrightarrow{\lambda,\chi}\lambda^{b}e^{is\chi}\Lstr_{\xi}A_{\left\{p,q\right\}}^{\left(k,l\right)} \,,
	\ee
	with the spin and boost weights related to the GHP weights according to $b=\frac{p+q}{2}$ and $s=\frac{p-q}{2}$.
	
	\item It is linear in the vector $\xi^{\mu}$ with respect to which we are Lie dragging.
\end{itemize}

Then, in order to satisfy the first two conditions, one starts with an ansatz that contains at most first spacetime derivatives, such that the Leibniz rule is satisfied,
\be
	\Lstr_{\xi} = \mathcal{L}_{\xi} + \lambda_{0,\xi}\left(p,q;k,l\right) + \lambda_{1,\xi}^{\mu}\left(p,q;k,l\right)\partial_{\mu} \,,
\ee
where $\left(k,l\right)$ and $\left\{p,q\right\}$ are the spacetime rank and GHP weights of the GHP tensor the generalized Lie derivative acts on. The Leibniz rule then further imposes,
\be\ba
	\lambda_{0,\xi}\left(p+p^{\prime},q+q^{\prime};k+k^{\prime},l+l^{\prime}\right) &= \lambda_{0,\xi}\left(p,q;k,l\right) + \lambda_{0,\xi}\left(p^{\prime},q^{\prime};k^{\prime},l^{\prime}\right) \,, \\
	\lambda_{1,\xi}^{\mu}\left(p+p^{\prime},q+q^{\prime};k+k^{\prime},l+l^{\prime}\right) &= \lambda_{1,\xi}^{\mu}\left(p,q;k,l\right) = \lambda_{1,\xi}^{\mu}\left(p^{\prime},q^{\prime};k^{\prime},l^{\prime}\right) \,.
\ea\ee
The first constraint is the usual multi-variable Cauchy functional equation implying that $\lambda_{0,\xi}\left(p,q;k,l\right)$ must be linear in $k$, $l$, $p$ and $q$, while the second constraint tells us that that $\lambda_{1,\xi}^{\mu}\left(p,q;k,l\right)$ is independent of the spacetime rank and GHP weights. Combining with the first condition of getting the usual Lie derivative $\mathcal{L}_{\xi}$ when acting on pure spacetime tensors with zero GHP weights,
\be
	\lambda_{0,\xi}\left(0,0;k,l\right) = 0\,\,\,\,,\,\,\,\,\lambda_{1,\xi}^{\mu}\left(0,0;k,l\right) = 0 \,,
\ee
we see that so far we have,
\be\label{eq:LieIntermediate}
	\Lstr_{\xi} = \mathcal{L}_{\xi} + p\,\alpha_{\xi} + q\,\beta_{\xi} \,,
\ee
with $\alpha_{\xi}$ and $\beta_{\xi}$ some spacetime scalar functions, linear in $\xi$, that are independent of $k$, $l$, $p$ and $q$.

We next impose the third condition of this generalized Lie derivative to act covariantly on GHP tensors. Since the additive modifications on the usual Lie derivative are independent of the spacetime tensorial structure of the GHP tensor it acts on, it is sufficient to consider the case of spacetime scalars of general GHP weights for which $\mathcal{L}_{\xi}=\xi^{\mu}\nabla_{\mu}$. The spacetime covariant derivative is written in terms of the NP directional derivatives,
\be
	D \equiv \ell^{\mu}\nabla_{\mu}\,\,\,\,,\,\,\,\,\triangle \equiv n^{\mu}\nabla_{\mu}\,\,\,\,,\,\,\,\,\delta \equiv m^{\mu}\nabla_{\mu}\,\,\,\,,\,\,\,\,\bar{\delta} \equiv \bar{m}^{\mu}\nabla_{\mu} \,,
\ee
as,
\be
	\nabla_{\mu} =-\ell_{\mu}\triangle - n_{\mu}D + m_{\mu}\bar{\delta} + \bar{m}_{\mu}\delta \,,
\ee
so,
\be
	\xi^{\mu}\nabla_{\mu} = -\xi_{\ell}\triangle - \xi_{n}D + \xi_{m}\bar{\delta} + \xi_{\bar{m}}\delta \,.
\ee
It is well known that the NP derivatives do not transform covariantly under local boosts and rotations when acting on NP scalars. Their ``bad'' transformation rules read,
\be
	\begin{pmatrix}
		D \\
		\triangle \\
		\delta \\
		\bar{\delta}
	\end{pmatrix} \xrightarrow{\lambda,\chi}
	\begin{pmatrix}
		\lambda\left[D + \frac{1}{2}\left(D\ln\left(\mathscr{B}^{p}\bar{\mathscr{B}}^{q}\right)\right)\right] \\
		\lambda^{-1}\left[\triangle + \frac{1}{2}\left(\triangle\ln\left(\mathscr{B}^{p}\bar{\mathscr{B}}^{q}\right)\right)\right] \\
		e^{i\chi}\left[\delta + \frac{1}{2}\left(\delta\ln\left(\mathscr{B}^{p}\bar{\mathscr{B}}^{q}\right)\right)\right] \\
		e^{-i\chi}\left[\bar{\delta} + \frac{1}{2}\left(\bar{\delta}\ln\left(\mathscr{B}^{p}\bar{\mathscr{B}}^{q}\right)\right)\right]
	\end{pmatrix} \,,
\ee
where we have defined,
\be
	\mathscr{B}\equiv \lambda e^{i\chi} \,.
\ee
As a result, the usual Lie derivative transforms as,
\be
	\mathcal{L}_{\xi} \xrightarrow{\lambda,\chi}\mathcal{L}_{\xi} + \frac{1}{2}\left(\xi^{\mu}\nabla_{\mu}\ln\left(\mathscr{B}^{p}\bar{\mathscr{B}}^{q}\right)\right) \,,
\ee
and the covariance of the generalized Lie derivative implies the following transformation rules for the unknown scalar functions $\alpha_{\xi}$ and $\beta_{\xi}$,
\be\ba
	\alpha_{\xi} &\xrightarrow{\lambda,\chi} \alpha_{\xi} - \frac{1}{2}\xi^{\mu}\nabla_{\mu}\ln\mathscr{B} \,, \\
	\beta_{\xi} &\xrightarrow{\lambda,\chi} \beta_{\xi} - \frac{1}{2}\xi^{\mu}\nabla_{\mu}\ln\bar{\mathscr{B}} \,.
\ea\ee
Comparing with the transformation laws for the ``bad'' spin coefficients \eqref{eq:NPSpinCoefficientsBad},
\be
	\begin{pmatrix}
		\varepsilon \\
		\gamma \\
		\beta \\
		\alpha
	\end{pmatrix} \xrightarrow{\lambda,\chi}
	\begin{pmatrix}
		\lambda\left(\varepsilon + \frac{1}{2}D\ln\mathscr{B}\right) \\
		\lambda^{-1}\left(\gamma + \frac{1}{2}\triangle\ln\mathscr{B}\right) \\
		e^{i\chi}\left(\beta + \frac{1}{2}\delta\ln\mathscr{B}\right) \\
		e^{-i\chi}\left(\alpha + \frac{1}{2}\bar{\delta}\ln\mathscr{B}\right)
	\end{pmatrix} \,,
\ee
we see that the non-homogeneous part of the transformation laws for the scalar functions $\alpha_{\xi}$ and $\beta_{\xi}$ can be reproduced from,
\be
	\alpha_{\xi}^{\text{bad}} = -\xi^{\mu}\zeta_{\mu} \,\,\,\,,\,\,\,\,\beta_{\xi}^{\text{bad}} = -\xi^{\mu}\bar{\zeta}_{\mu} \,,
\ee
where,
\be
	\zeta_{\mu} = -\ell_{\mu}\gamma - n_{\mu}\varepsilon + m_{\mu}\alpha + \bar{m}_{\mu}\beta = -\frac{1}{2}\left(n^{\nu}\nabla_{\mu}\ell_{\nu} - \bar{m}^{\nu}\nabla_{\mu}m_{\nu}\right) \,.
\ee
In conclusion, the most general generalized Lie derivative that acts on GHP tensors of GHP weights $\left\{p,q\right\}$ reads,
\be
	\Lstr_{\xi} = \mathcal{L}_{\xi} -\xi^{\mu}\left(p\,\zeta_{\mu} + q\,\bar{\zeta}_{\mu}\right) + p\,\eta_{\xi} + q\,\vartheta_{\xi} \,,
\ee
with $\eta_{\xi}$ and $\vartheta_{\xi}$ two scalar functions independent of $p$ and $q$ that transform covariantly with zero GHP weight,
\be
	\eta_{\xi}\xrightarrow{\lambda,\chi}\eta_{\xi}\,\,\,\,,\,\,\,\,\vartheta_{\xi}\xrightarrow{\lambda,\chi}\vartheta_{\xi} \,,
\ee
that are also linear in the vector field $\xi^{\mu}$. For the minimal choice $\eta_{\xi}=\vartheta_{\xi}=0$, we retrieve the usual GHP derivative (\cite{Geroch1973}). These scalar functions that appear above are arbitrary but part of them can be fixed by uniquely constructing a generalized Lie derivative when Lie dragging along a Killing vector. This was first proposed by Ludwig et al (\cite{Ludwig2000,Ludwig2002}) and is obtained from our above construction by requiring the existence of a Killing vector, satisfying $\mathcal{L}_{\xi}g_{\mu\nu}=0$, to be apparent directly at the level of the tetrad vectors. It is sufficient to impose,
\be
	n_{\mu}\Lstr_{\xi}\ell^{\mu} = \bar{m}_{\mu}\mathcal{L}_{\xi}m^{\mu} = 0 \,\,\,\,\text{when $\mathcal{L}_{\xi}g_{\mu\nu}=0$} \,,
\ee
uniquely fixing $\eta_{\xi}$ and $\vartheta_{\xi}$ with the end result for the generalized Lie derivative,
\be\label{eq:LieKilling}
	\Lstr_{\xi} = \mathcal{L}_{\xi} + b\,n_{\mu}\mathcal{L}_{\xi}\ell^{\mu} - s\,\bar{m}_{\mu}\mathcal{L}_{\xi}m^{\mu}\,\,\,,\,\,\,\text{when $\mathcal{L}_{\xi}g_{\mu\nu}=0$} \,.
\ee

\subsection{Preserving the algebra}
A further feature we would like the generalized Lie derivative to have is for it to preserve the algebra already satisfied by the usual Lie derivative with respect to some vector generators of the algebra. In particular, we would like to preserve the identity,
\be
	\left[\mathcal{L}_{\xi_1},\mathcal{L}_{\xi_2}\right] = \mathcal{L}_{\left[\xi_1,\xi_2\right]_{\text{LB}}} \,,
\ee
where $\left[\xi_1,\xi_2\right]_{\text{LB}} = \mathcal{L}_{\xi_1}\xi_2$ is the Lie bracket of the two vector fields. At the level of the scalar functions $\alpha_{\xi}$ and $\beta_{\xi}$ in \eqref{eq:LieIntermediate}, prior to requiring the generalized Lie derivative to act covariantly on a general GHP tensor, the above algebra-preserving identity implies,
\be\label{eq:LieAlgebraPreserving}
	\left[\Lstr_{\xi_1},\Lstr_{\xi_2}\right] = \Lstr_{\left[\xi_1,\xi_2\right]_{\text{LB}}} \Rightarrow
	\begin{cases}
		\mathcal{L}_{\xi_1}\alpha_{\xi_2} - \mathcal{L}_{\xi_2}\alpha_{\xi_1} = \alpha_{\left[\xi_1,\xi_2\right]_{\text{LB}}} \\
		\mathcal{L}_{\xi_1}\beta_{\xi_2} - \mathcal{L}_{\xi_2}\beta_{\xi_1} = \beta_{\left[\xi_1,\xi_2\right]_{\text{LB}}}
	\end{cases} \,.
\ee
Then, after writing $\alpha_{\xi} = -\xi^{\mu}\zeta_{\mu} + \eta_{\xi}$ and $\beta_{\xi} = -\xi^{\mu}\bar{\zeta}_{\mu} + \vartheta_{\xi}$ to ensure the covariance of the generalized Lie derivative and since the tensorial structure associated with the geometry itself is carried only by the tetrad vectors, the most general form of the scalar function $\eta_{\xi}$ is some linear operator containing at least one spacetime covariant derivative acting on the vector field $\xi^{\mu}$,
\be\ba
	\eta_{\xi} &= \left[H^{\left(\mu\nu\right)} + \frac{1}{2}\left(\ell^{[\mu}n^{\nu]} - m^{[\mu}\bar{m}^{\nu]}\right)\right]\nabla_{\nu}\xi_{\mu} + H^{\left(\mu\nu\right)\rho}\nabla_{\rho}\nabla_{\nu}\xi_{\mu} + \dots \,, \\
	\vartheta_{\xi} &= \left[\Theta^{\left(\mu\nu\right)} + \frac{1}{2}\left(\ell^{[\mu}n^{\nu]} + m^{[\mu}\bar{m}^{\nu]}\right)\right]\nabla_{\nu}\xi_{\mu} + \Theta^{\left(\mu\nu\right)\rho}\nabla_{\rho}\nabla_{\nu}\xi_{\mu} + \dots \,,
\ea\ee
where $H^{\left(\mu\nu\right)}$, $\Theta^{\left(\mu\nu\right)}$, $H^{\left(\mu\nu\right)\rho}$, $\Theta^{\left(\mu\nu\right)\rho}$, $\dots$ are pure spacetime tensors carrying zero GHP weights that are symmetric in their first two indices, while the antisymmetric parts of the rank-$2$ tensors were fixed such that we reproduce the generalized Lie derivative \eqref{eq:LieKilling} when Lie dragging along a Killing vector. We want the algebra preserving identity to be independent of the involved vector fields $\xi_1^{\mu}$ and $\xi_2^{\mu}$. Unfortunately, it can be shown directly from the algebra preserving constraints \eqref{eq:LieAlgebraPreserving} that this is never possible, unless $\xi^{\mu}$ is a Killing vector.


\bibliographystyle{JHEP}
\bibliography{short_new_v3}

\providecommand{\href}[2]{#2}\begingroup\raggedright\begin{thebibliography}{100}

\bibitem{LIGOScientific:2016aoc}
{\scshape LIGO Scientific, Virgo} collaboration, B.~P. Abbott et~al.,
  \emph{{Observation of Gravitational Waves from a Binary Black Hole Merger}},
  \href{https://doi.org/10.1103/PhysRevLett.116.061102}{\emph{Phys. Rev. Lett.}
  {\bfseries 116} (2016) 061102}
  [\href{https://arxiv.org/abs/1602.03837}{{\ttfamily 1602.03837}}].

\bibitem{Flanagan2008}
E.~E. Flanagan and T.~Hinderer, \emph{Constraining neutron-star tidal love
  numbers with gravitational-wave detectors},
  \href{https://doi.org/10.1103/PhysRevD.77.021502}{\emph{Phys. Rev. D}
  {\bfseries 77} (2008) 021502}.

\bibitem{Binnington:2009bb}
T.~Binnington and E.~Poisson, \emph{{Relativistic theory of tidal Love
  numbers}}, \href{https://doi.org/10.1103/PhysRevD.80.084018}{\emph{Phys. Rev.
  D} {\bfseries 80} (2009) 084018}
  [\href{https://arxiv.org/abs/0906.1366}{{\ttfamily 0906.1366}}].

\bibitem{Yagi:2013bca}
K.~Yagi and N.~Yunes, \emph{{I-Love-Q}},
  \href{https://doi.org/10.1126/science.1236462}{\emph{Science} {\bfseries 341}
  (2013) 365} [\href{https://arxiv.org/abs/1302.4499}{{\ttfamily 1302.4499}}].

\bibitem{Chatziioannou2020}
K.~Chatziioannou, \emph{Neutron-star tidal deformability and equation-of-state
  constraints}, \href{https://doi.org/10.1007/s10714-020-02754-3}{\emph{General
  Relativity and Gravitation} {\bfseries 52} (2020) 109}.

\bibitem{LIGOScientific:2017vwq}
{\scshape LIGO Scientific, Virgo} collaboration, B.~P. Abbott et~al.,
  \emph{{GW170817: Observation of Gravitational Waves from a Binary Neutron
  Star Inspiral}},
  \href{https://doi.org/10.1103/PhysRevLett.119.161101}{\emph{Phys. Rev. Lett.}
  {\bfseries 119} (2017) 161101}
  [\href{https://arxiv.org/abs/1710.05832}{{\ttfamily 1710.05832}}].

\bibitem{Fang2005}
H.~Fang and G.~Lovelace, \emph{Tidal coupling of a schwarzschild black hole and
  circularly orbiting moon},
  \href{https://doi.org/10.1103/PhysRevD.72.124016}{\emph{Phys. Rev. D}
  {\bfseries 72} (2005) 124016}.

\bibitem{PhysRev.55.374}
J.~R. Oppenheimer and G.~M. Volkoff, \emph{On massive neutron cores},
  \href{https://doi.org/10.1103/PhysRev.55.374}{\emph{Phys. Rev.} {\bfseries
  55} (1939) 374}.

\bibitem{Porto:2016zng}
R.~A. Porto, \emph{{The Tune of Love and the Nature(ness) of Spacetime}},
  \href{https://doi.org/10.1002/prop.201600064}{\emph{Fortsch. Phys.}
  {\bfseries 64} (2016) 723}
  [\href{https://arxiv.org/abs/1606.08895}{{\ttfamily 1606.08895}}].

\bibitem{Goldberger:2004jt}
W.~D. Goldberger and I.~Z. Rothstein, \emph{{An Effective field theory of
  gravity for extended objects}},
  \href{https://doi.org/10.1103/PhysRevD.73.104029}{\emph{Phys. Rev. D}
  {\bfseries 73} (2006) 104029}
  [\href{https://arxiv.org/abs/hep-th/0409156}{{\ttfamily hep-th/0409156}}].

\bibitem{Goldberger:2005cd}
W.~D. Goldberger and I.~Z. Rothstein, \emph{{Dissipative effects in the
  worldline approach to black hole dynamics}},
  \href{https://doi.org/10.1103/PhysRevD.73.104030}{\emph{Phys. Rev. D}
  {\bfseries 73} (2006) 104030}
  [\href{https://arxiv.org/abs/hep-th/0511133}{{\ttfamily hep-th/0511133}}].

\bibitem{Goldberger:2006bd}
W.~D. Goldberger and I.~Z. Rothstein, \emph{{Towers of Gravitational
  Theories}}, \href{https://doi.org/10.1142/S0218271806009698}{\emph{Gen. Rel.
  Grav.} {\bfseries 38} (2006) 1537}
  [\href{https://arxiv.org/abs/hep-th/0605238}{{\ttfamily hep-th/0605238}}].

\bibitem{Porto:2005ac}
R.~A. Porto, \emph{{Post-Newtonian corrections to the motion of spinning bodies
  in NRGR}}, \href{https://doi.org/10.1103/PhysRevD.73.104031}{\emph{Phys. Rev.
  D} {\bfseries 73} (2006) 104031}
  [\href{https://arxiv.org/abs/gr-qc/0511061}{{\ttfamily gr-qc/0511061}}].

\bibitem{Rothstein2014}
I.~Z. Rothstein, \emph{Progress in effective field theory approach to the
  binary inspiral problem},
  \href{https://doi.org/10.1007/s10714-014-1726-y}{\emph{General Relativity and
  Gravitation} {\bfseries 46} (2014) 1726}.

\bibitem{Porto:2016pyg}
R.~A. Porto, \emph{{The effective field theorist\textquoteright{}s approach to
  gravitational dynamics}},
  \href{https://doi.org/10.1016/j.physrep.2016.04.003}{\emph{Phys. Rept.}
  {\bfseries 633} (2016) 1} [\href{https://arxiv.org/abs/1601.04914}{{\ttfamily
  1601.04914}}].

\bibitem{Levi:2018nxp}
M.~Levi, \emph{{Effective Field Theories of Post-Newtonian Gravity: A
  comprehensive review}},
  \href{https://doi.org/10.1088/1361-6633/ab12bc}{\emph{Rept. Prog. Phys.}
  {\bfseries 83} (2020) 075901}
  [\href{https://arxiv.org/abs/1807.01699}{{\ttfamily 1807.01699}}].

\bibitem{Kol:2011vg}
B.~Kol and M.~Smolkin, \emph{{Black hole stereotyping: Induced gravito-static
  polarization}}, \href{https://doi.org/10.1007/JHEP02(2012)010}{\emph{JHEP}
  {\bfseries 02} (2012) 010} [\href{https://arxiv.org/abs/1110.3764}{{\ttfamily
  1110.3764}}].

\bibitem{Hui:2020xxx}
L.~Hui, A.~Joyce, R.~Penco, L.~Santoni and A.~R. Solomon, \emph{{Static
  response and Love numbers of Schwarzschild black holes}},
  \href{https://doi.org/10.1088/1475-7516/2021/04/052}{\emph{JCAP} {\bfseries
  04} (2021) 052} [\href{https://arxiv.org/abs/2010.00593}{{\ttfamily
  2010.00593}}].

\bibitem{Charalambous:2021mea}
P.~Charalambous, S.~Dubovsky and M.~M. Ivanov, \emph{{On the Vanishing of Love
  Numbers for Kerr Black Holes}},
  \href{https://doi.org/10.1007/JHEP05(2021)038}{\emph{JHEP} {\bfseries 05}
  (2021) 038} [\href{https://arxiv.org/abs/2102.08917}{{\ttfamily
  2102.08917}}].

\bibitem{Ivanov:2022hlo}
M.~M. Ivanov and Z.~Zhou, \emph{{Black Hole Tidal Love Numbers and Dissipation
  Numbers in Worldline Effective Field Theory}},
  \href{https://arxiv.org/abs/2208.08459}{{\ttfamily 2208.08459}}.

\bibitem{tHooft:1979rat}
G.~'t~Hooft, \emph{{Naturalness, chiral symmetry, and spontaneous chiral
  symmetry breaking}},
  \href{https://doi.org/10.1007/978-1-4684-7571-5_9}{\emph{NATO Sci. Ser. B}
  {\bfseries 59} (1980) 135}.

\bibitem{Chia:2020yla}
H.~S. Chia, \emph{{Tidal deformation and dissipation of rotating black holes}},
  \href{https://doi.org/10.1103/PhysRevD.104.024013}{\emph{Phys. Rev. D}
  {\bfseries 104} (2021) 024013}
  [\href{https://arxiv.org/abs/2010.07300}{{\ttfamily 2010.07300}}].

\bibitem{Goldberger2020c}
W.~D. Goldberger, J.~Li and I.~Z. Rothstein, \emph{{Non-conservative effects on
  Spinning Black Holes from World-Line Effective Field Theory}},
  \href{https://arxiv.org/abs/2012.14869}{{\ttfamily 2012.14869}}.

\bibitem{LeTiec2020}
A.~Le~Tiec and M.~Casals, \emph{Spinning black holes fall in love},
  {\emph{arXiv e-prints} (2020) arXiv:2007.00214}.

\bibitem{LeTiec2020a}
A.~Le~Tiec, M.~Casals and E.~Franzin, \emph{Tidal love numbers of kerr black
  holes}, {\emph{arXiv e-prints} (2020) arXiv:2010.15795}.

\bibitem{Poisson:2020mdi}
E.~Poisson, \emph{{Gravitomagnetic Love tensor of a slowly rotating body:
  post-Newtonian theory}},
  \href{https://doi.org/10.1103/PhysRevD.102.064059}{\emph{Phys. Rev. D}
  {\bfseries 102} (2020) 064059}
  [\href{https://arxiv.org/abs/2007.01678}{{\ttfamily 2007.01678}}].

\bibitem{Charalambous:2021kcz}
P.~Charalambous, S.~Dubovsky and M.~M. Ivanov, \emph{{Hidden Symmetry of
  Vanishing Love Numbers}},
  \href{https://doi.org/10.1103/PhysRevLett.127.101101}{\emph{Phys. Rev. Lett.}
  {\bfseries 127} (2021) 101101}
  [\href{https://arxiv.org/abs/2103.01234}{{\ttfamily 2103.01234}}].

\bibitem{Teukolsky1972}
S.~A. Teukolsky, \emph{Rotating black holes: Separable wave equations for
  gravitational and electromagnetic perturbations},
  \href{https://doi.org/10.1103/PhysRevLett.29.1114}{\emph{Physical Review
  Letters} {\bfseries 29} (1972) 1114}.

\bibitem{Teukolsky1973}
S.~A. Teukolsky, \emph{Perturbations of a rotating black hole. i. fundamental
  equations for gravitational, electromagnetic, and neutrino-field
  perturbations}, \href{https://doi.org/10.1086/152444}{\emph{The Astrophysical
  Journal} {\bfseries 185} (1973) 635}.

\bibitem{Starobinski1973}
A.~A. {Starobinski{\v{i}}}, \emph{{Amplification of waves during reflection
  from a rotating ``black hole''}}, {\emph{Soviet Journal of Experimental and
  Theoretical Physics} {\bfseries 37} (1973) 28}.

\bibitem{Starobinski1974}
A.~A. {Starobinski{\v{i}}} and S.~M. {Churilov}, \emph{{Amplification of
  electromagnetic and gravitational waves scattered by a rotating ``black
  hole''}}, {\emph{Soviet Journal of Experimental and Theoretical Physics}
  {\bfseries 38} (1974) 1}.

\bibitem{Bardeen:1999px}
J.~M. Bardeen and G.~T. Horowitz, \emph{{The Extreme Kerr throat geometry: A
  Vacuum analog of
  ${\mathrm{AdS}}_{2}{\ifmmode\times\else\texttimes\fi{}\mathrm{S}}^{2}$}},
  \href{https://doi.org/10.1103/PhysRevD.60.104030}{\emph{Phys. Rev. D}
  {\bfseries 60} (1999) 104030}
  [\href{https://arxiv.org/abs/hep-th/9905099}{{\ttfamily hep-th/9905099}}].

\bibitem{Amsel:2009et}
A.~J. Amsel, G.~T. Horowitz, D.~Marolf and M.~M. Roberts, \emph{{Uniqueness of
  Extremal Kerr and Kerr-Newman Black Holes}},
  \href{https://doi.org/10.1103/PhysRevD.81.024033}{\emph{Phys. Rev. D}
  {\bfseries 81} (2010) 024033}
  [\href{https://arxiv.org/abs/0906.2367}{{\ttfamily 0906.2367}}].

\bibitem{Cvetic:2011hp}
M.~Cvetic and F.~Larsen, \emph{{Conformal Symmetry for General Black Holes}},
  \href{https://doi.org/10.1007/JHEP02(2012)122}{\emph{JHEP} {\bfseries 02}
  (2012) 122} [\href{https://arxiv.org/abs/1106.3341}{{\ttfamily 1106.3341}}].

\bibitem{Cvetic:2011dn}
M.~Cvetic and F.~Larsen, \emph{{Conformal Symmetry for Black Holes in Four
  Dimensions}}, \href{https://doi.org/10.1007/JHEP09(2012)076}{\emph{JHEP}
  {\bfseries 09} (2012) 076} [\href{https://arxiv.org/abs/1112.4846}{{\ttfamily
  1112.4846}}].

\bibitem{Edgar2000}
S.~B. Edgar and G.~Ludwig, \emph{Integration in the ghp formalism iv: A new lie
  derivative operator leading to an efficient treatment of killing vectors},
  \href{https://doi.org/10.1023/A:1001915118339}{\emph{General Relativity and
  Gravitation} {\bfseries 32} (2000) 637}.

\bibitem{Ludwig2000}
G.~Ludwig and S.~B. Edgar, \emph{A generalized lie derivative and homothetic or
  killing vectors in the geroch-held-penrose formalism},
  \href{https://doi.org/10.1088/0264-9381/17/7/308}{\emph{Classical and Quantum
  Gravity} {\bfseries 17} (2000) 1683}.

\bibitem{Ludwig2002}
G.~Ludwig and S.~B. Edgar, \emph{(conformal) killing vectors in the
  newman-penrose formalism},
  \href{https://doi.org/10.1023/A:1016361729933}{\emph{General Relativity and
  Gravitation} {\bfseries 34} (2002) 807}.

\bibitem{Love1912}
H.~Love, \emph{Some problems of geodynamics},
  \href{https://doi.org/10.1038/089471a0}{\emph{Nature} {\bfseries 89} (1912)
  471}.

\bibitem{PoissonWill2014}
E.~Poisson and C.~M. Will, \emph{Gravity: Newtonian, Post-Newtonian,
  Relativistic}. Cambridge University Press, 2014,
  \href{https://doi.org/10.1017/CBO9781139507486}{10.1017/CBO9781139507486}.

\bibitem{Newman1962}
E.~Newman and R.~Penrose, \emph{An approach to gravitational radiation by a
  method of spin coefficients},
  \href{https://doi.org/10.1063/1.1724257}{\emph{Journal of Mathematical
  Physics} {\bfseries 3} (1962) 566}
  [\href{https://arxiv.org/abs/https://doi.org/10.1063/1.1724257}{{\ttfamily
  https://doi.org/10.1063/1.1724257}}].

\bibitem{Geroch1973}
R.~Geroch, A.~Held and R.~Penrose, \emph{A space‐time calculus based on pairs
  of null directions}, \href{https://doi.org/10.1063/1.1666410}{\emph{Journal
  of Mathematical Physics} {\bfseries 14} (1973) 874}
  [\href{https://arxiv.org/abs/https://doi.org/10.1063/1.1666410}{{\ttfamily
  https://doi.org/10.1063/1.1666410}}].

\bibitem{Levi:2015msa}
M.~Levi and J.~Steinhoff, \emph{{Spinning gravitating objects in the effective
  field theory in the post-Newtonian scheme}},
  \href{https://doi.org/10.1007/JHEP09(2015)219}{\emph{JHEP} {\bfseries 09}
  (2015) 219} [\href{https://arxiv.org/abs/1501.04956}{{\ttfamily
  1501.04956}}].

\bibitem{Page:1976df}
D.~N. Page, \emph{{Particle Emission Rates from a Black Hole: Massless
  Particles from an Uncharged, Nonrotating Hole}},
  \href{https://doi.org/10.1103/PhysRevD.13.198}{\emph{Phys. Rev. D} {\bfseries
  13} (1976) 198}.

\bibitem{Goldberger:2020wbx}
W.~D. Goldberger and I.~Z. Rothstein, \emph{{Horizon radiation reaction
  forces}}, \href{https://doi.org/10.1007/JHEP10(2020)026}{\emph{JHEP}
  {\bfseries 10} (2020) 026}
  [\href{https://arxiv.org/abs/2007.00731}{{\ttfamily 2007.00731}}].

\bibitem{Goldberger:2007hy}
W.~D. Goldberger, \emph{{Les Houches lectures on effective field theories and
  gravitational radiation}},  in \emph{{Les Houches Summer School - Session 86:
  Particle Physics and Cosmology: The Fabric of Spacetime}}, 1, 2007,
  \href{https://arxiv.org/abs/hep-ph/0701129}{{\ttfamily hep-ph/0701129}}.

\bibitem{Cheung:2020sdj}
C.~Cheung and M.~P. Solon, \emph{{Tidal Effects in the Post-Minkowskian
  Expansion}},
  \href{https://doi.org/10.1103/PhysRevLett.125.191601}{\emph{Phys. Rev. Lett.}
  {\bfseries 125} (2020) 191601}
  [\href{https://arxiv.org/abs/2006.06665}{{\ttfamily 2006.06665}}].

\bibitem{Bern:2020uwk}
Z.~Bern, J.~Parra-Martinez, R.~Roiban, E.~Sawyer and C.-H. Shen, \emph{{Leading
  Nonlinear Tidal Effects and Scattering Amplitudes}},
  \href{https://doi.org/10.1007/JHEP05(2021)188}{\emph{JHEP} {\bfseries 05}
  (2021) 188} [\href{https://arxiv.org/abs/2010.08559}{{\ttfamily
  2010.08559}}].

\bibitem{Creci:2021rkz}
G.~Creci, T.~Hinderer and J.~Steinhoff, \emph{{Tidal response from scattering
  and the role of analytic continuation}},
  \href{https://doi.org/10.1103/PhysRevD.104.124061}{\emph{Phys. Rev. D}
  {\bfseries 104} (2021) 124061}
  [\href{https://arxiv.org/abs/2108.03385}{{\ttfamily 2108.03385}}].

\bibitem{Goldberger:2019sya}
W.~D. Goldberger and I.~Z. Rothstein, \emph{{An Effective Field Theory of
  Quantum Mechanical Black Hole Horizons}},
  \href{https://doi.org/10.1007/JHEP04(2020)056}{\emph{JHEP} {\bfseries 04}
  (2020) 056} [\href{https://arxiv.org/abs/1912.13435}{{\ttfamily
  1912.13435}}].

\bibitem{Goldberger:2020fot}
W.~D. Goldberger, J.~Li and I.~Z. Rothstein, \emph{{Non-conservative effects on
  spinning black holes from world-line effective field theory}},
  \href{https://doi.org/10.1007/JHEP06(2021)053}{\emph{JHEP} {\bfseries 06}
  (2021) 053} [\href{https://arxiv.org/abs/2012.14869}{{\ttfamily
  2012.14869}}].

\bibitem{Aharony:1999ti}
O.~Aharony, S.~S. Gubser, J.~M. Maldacena, H.~Ooguri and Y.~Oz, \emph{{Large N
  field theories, string theory and gravity}},
  \href{https://doi.org/10.1016/S0370-1573(99)00083-6}{\emph{Phys. Rept.}
  {\bfseries 323} (2000) 183}
  [\href{https://arxiv.org/abs/hep-th/9905111}{{\ttfamily hep-th/9905111}}].

\bibitem{Maldacena:1997re}
J.~M. Maldacena, \emph{{The Large N limit of superconformal field theories and
  supergravity}}, \href{https://doi.org/10.1023/A:1026654312961}{\emph{Adv.
  Theor. Math. Phys.} {\bfseries 2} (1998) 231}
  [\href{https://arxiv.org/abs/hep-th/9711200}{{\ttfamily hep-th/9711200}}].

\bibitem{Klebanov:1997kc}
I.~R. Klebanov, \emph{{World volume approach to absorption by nondilatonic
  branes}}, \href{https://doi.org/10.1016/S0550-3213(97)00235-6}{\emph{Nucl.
  Phys. B} {\bfseries 496} (1997) 231}
  [\href{https://arxiv.org/abs/hep-th/9702076}{{\ttfamily hep-th/9702076}}].

\bibitem{Gubser:1997yh}
S.~S. Gubser, I.~R. Klebanov and A.~A. Tseytlin, \emph{{String theory and
  classical absorption by three-branes}},
  \href{https://doi.org/10.1016/S0550-3213(97)00325-8}{\emph{Nucl. Phys. B}
  {\bfseries 499} (1997) 217}
  [\href{https://arxiv.org/abs/hep-th/9703040}{{\ttfamily hep-th/9703040}}].

\bibitem{Guica:2008mu}
M.~Guica, T.~Hartman, W.~Song and A.~Strominger, \emph{{The Kerr/CFT
  Correspondence}},
  \href{https://doi.org/10.1103/PhysRevD.80.124008}{\emph{Phys. Rev. D}
  {\bfseries 80} (2009) 124008}
  [\href{https://arxiv.org/abs/0809.4266}{{\ttfamily 0809.4266}}].

\bibitem{Castro:2010fd}
A.~Castro, A.~Maloney and A.~Strominger, \emph{{Hidden Conformal Symmetry of
  the Kerr Black Hole}},
  \href{https://doi.org/10.1103/PhysRevD.82.024008}{\emph{Phys. Rev. D}
  {\bfseries 82} (2010) 024008}
  [\href{https://arxiv.org/abs/1004.0996}{{\ttfamily 1004.0996}}].

\bibitem{Dudley1977}
A.~L. Dudley and J.~D. Finley, \emph{Separation of wave equations for
  perturbations of general type-$d$ space-times}, {\emph{PRL} {\bfseries 38}
  (1977) 1505}.

\bibitem{Kinnersley1969}
W.~Kinnersley, \emph{Type d vacuum metrics},
  \href{https://doi.org/10.1063/1.1664958}{\emph{Journal of Mathematical
  Physics} {\bfseries 10} (1969) 1195}
  [\href{https://arxiv.org/abs/https://doi.org/10.1063/1.1664958}{{\ttfamily
  https://doi.org/10.1063/1.1664958}}].

\bibitem{Maldacena1997}
J.~Maldacena and A.~Strominger, \emph{Universal low-energy dynamics for
  rotating black holes},
  \href{https://doi.org/10.1103/PhysRevD.56.4975}{\emph{Phys. Rev. D}
  {\bfseries 56} (1997) 4975}.

\bibitem{Berti2005}
E.~Berti and K.~D. Kokkotas, \emph{Quasinormal modes of kerr-newman black
  holes: Coupling of electromagnetic and gravitational perturbations},
  \href{https://doi.org/10.1103/PhysRevD.71.124008}{\emph{Phys. Rev. D}
  {\bfseries 71} (2005) 124008}.

\bibitem{Porto:2007qi}
R.~A. Porto, \emph{{Absorption effects due to spin in the worldline approach to
  black hole dynamics}},
  \href{https://doi.org/10.1103/PhysRevD.77.064026}{\emph{Phys. Rev. D}
  {\bfseries 77} (2008) 064026}
  [\href{https://arxiv.org/abs/0710.5150}{{\ttfamily 0710.5150}}].

\bibitem{Donoghue:2017pgk}
J.~F. Donoghue, M.~M. Ivanov and A.~Shkerin, \emph{{EPFL Lectures on General
  Relativity as a Quantum Field Theory}},
  \href{https://arxiv.org/abs/1702.00319}{{\ttfamily 1702.00319}}.

\bibitem{Goroff:1985th}
M.~H. Goroff and A.~Sagnotti, \emph{{The Ultraviolet Behavior of Einstein
  Gravity}}, \href{https://doi.org/10.1016/0550-3213(86)90193-8}{\emph{Nucl.
  Phys. B} {\bfseries 266} (1986) 709}.

\bibitem{Cai2019}
S.~Cai and K.-D. Wang, \emph{Non-vanishing of tidal love numbers}, {\emph{arXiv
  e-prints} (2019) arXiv:1906.06850}.

\bibitem{Cardoso2018}
V.~Cardoso, M.~Kimura, A.~Maselli and L.~Senatore, \emph{Black holes in an
  effective field theory extension of general relativity}, {\emph{PRL}
  {\bfseries 121} (2018) 251105}.

\bibitem{Bekenstein:1972ky}
J.~D. Bekenstein, \emph{{Nonexistence of baryon number for black holes. ii}},
  \href{https://doi.org/10.1103/PhysRevD.5.2403}{\emph{Phys. Rev. D} {\bfseries
  5} (1972) 2403}.

\bibitem{Cardoso:2016ryw}
V.~Cardoso and L.~Gualtieri, \emph{{Testing the black hole
  \textquoteleft{}no-hair\textquoteright{} hypothesis}},
  \href{https://doi.org/10.1088/0264-9381/33/17/174001}{\emph{Class. Quant.
  Grav.} {\bfseries 33} (2016) 174001}
  [\href{https://arxiv.org/abs/1607.03133}{{\ttfamily 1607.03133}}].

\bibitem{Dubovsky:2007zi}
S.~Dubovsky, P.~Tinyakov and M.~Zaldarriaga, \emph{{Bumpy black holes from
  spontaneous Lorentz violation}},
  \href{https://doi.org/10.1088/1126-6708/2007/11/083}{\emph{JHEP} {\bfseries
  11} (2007) 083} [\href{https://arxiv.org/abs/0706.0288}{{\ttfamily
  0706.0288}}].

\bibitem{Berti:2015itd}
E.~Berti et~al., \emph{{Testing General Relativity with Present and Future
  Astrophysical Observations}},
  \href{https://doi.org/10.1088/0264-9381/32/24/243001}{\emph{Class. Quant.
  Grav.} {\bfseries 32} (2015) 243001}
  [\href{https://arxiv.org/abs/1501.07274}{{\ttfamily 1501.07274}}].

\bibitem{Herdeiro:2015waa}
C.~A.~R. Herdeiro and E.~Radu, \emph{{Asymptotically flat black holes with
  scalar hair: a review}},
  \href{https://doi.org/10.1142/S0218271815420146}{\emph{Int. J. Mod. Phys. D}
  {\bfseries 24} (2015) 1542014}
  [\href{https://arxiv.org/abs/1504.08209}{{\ttfamily 1504.08209}}].

\bibitem{Hollands:2012xy}
S.~Hollands and A.~Ishibashi, \emph{{Black hole uniqueness theorems in higher
  dimensional spacetimes}},
  \href{https://doi.org/10.1088/0264-9381/29/16/163001}{\emph{Class. Quant.
  Grav.} {\bfseries 29} (2012) 163001}
  [\href{https://arxiv.org/abs/1206.1164}{{\ttfamily 1206.1164}}].

\bibitem{Bertini2012}
S.~Bertini, S.~L. Cacciatori and D.~Klemm, \emph{Conformal structure of the
  schwarzschild black hole},
  \href{https://doi.org/10.1103/PhysRevD.85.064018}{\emph{Phys. Rev. D}
  {\bfseries 85} (2012) 064018}.

\bibitem{Nollert:1993zz}
H.-P. Nollert, \emph{{Quasinormal modes of Schwarzschild black holes: The
  determination of quasinormal frequencies with very large imaginary parts}},
  \href{https://doi.org/10.1103/PhysRevD.47.5253}{\emph{Phys. Rev. D}
  {\bfseries 47} (1993) 5253}.

\bibitem{Berti:2004um}
E.~Berti, V.~Cardoso and S.~Yoshida, \emph{{Highly damped quasinormal modes of
  Kerr black holes: A Complete numerical investigation}},
  \href{https://doi.org/10.1103/PhysRevD.69.124018}{\emph{Phys. Rev. D}
  {\bfseries 69} (2004) 124018}
  [\href{https://arxiv.org/abs/gr-qc/0401052}{{\ttfamily gr-qc/0401052}}].

\bibitem{Berti:2009kk}
E.~Berti, V.~Cardoso and A.~O. Starinets, \emph{{Quasinormal modes of black
  holes and black branes}},
  \href{https://doi.org/10.1088/0264-9381/26/16/163001}{\emph{Class. Quant.
  Grav.} {\bfseries 26} (2009) 163001}
  [\href{https://arxiv.org/abs/0905.2975}{{\ttfamily 0905.2975}}].

\bibitem{Hod:2003jn}
S.~Hod, \emph{{Kerr black hole quasinormal frequencies}},
  \href{https://doi.org/10.1103/PhysRevD.67.081501}{\emph{Phys. Rev. D}
  {\bfseries 67} (2003) 081501}
  [\href{https://arxiv.org/abs/gr-qc/0301122}{{\ttfamily gr-qc/0301122}}].

\bibitem{Berti:2003jh}
E.~Berti, V.~Cardoso, K.~D. Kokkotas and H.~Onozawa, \emph{{Highly damped
  quasinormal modes of Kerr black holes}},
  \href{https://doi.org/10.1103/PhysRevD.68.124018}{\emph{Phys. Rev. D}
  {\bfseries 68} (2003) 124018}
  [\href{https://arxiv.org/abs/hep-th/0307013}{{\ttfamily hep-th/0307013}}].

\bibitem{Cook:2016fge}
G.~B. Cook and M.~Zalutskiy, \emph{{Purely imaginary quasinormal modes of the
  Kerr geometry}},
  \href{https://doi.org/10.1088/0264-9381/33/24/245008}{\emph{Class. Quant.
  Grav.} {\bfseries 33} (2016) 245008}
  [\href{https://arxiv.org/abs/1603.09710}{{\ttfamily 1603.09710}}].

\bibitem{Cook:2016ngj}
G.~B. Cook and M.~Zalutskiy, \emph{{Modes of the Kerr geometry with purely
  imaginary frequencies}},
  \href{https://doi.org/10.1103/PhysRevD.94.104074}{\emph{Phys. Rev. D}
  {\bfseries 94} (2016) 104074}
  [\href{https://arxiv.org/abs/1607.07406}{{\ttfamily 1607.07406}}].

\bibitem{nla.cat-vn2139291}
W.~Miller, \emph{Lie theory and special functions / Willard Miller}. Academic
  Press New York, 1968.

\bibitem{doi:10.1137/0501037}
W.~Miller, Jr., \emph{Lie theory and some special solutions of the
  hypergeometric equations}, \href{https://doi.org/10.1137/0501037}{\emph{SIAM
  Journal on Mathematical Analysis} {\bfseries 1} (1970) 405}
  [\href{https://arxiv.org/abs/https://doi.org/10.1137/0501037}{{\ttfamily
  https://doi.org/10.1137/0501037}}].

\bibitem{Howe1992}
R.~Howe and E.~Chye~Tan, \emph{{Non-abelian harmonic analysis : applications of
  $\mathrm{SL}(2,\mathbb{R})$}}, Universitext. Springer-Verlag, New York, 1992.

\bibitem{zamolodchikov1989integrable}
A.~B. Zamolodchikov, \emph{Integrable field theory from conformal field
  theory},  in \emph{Integrable Sys Quantum Field Theory}, pp.~641--674.
\newblock Elsevier, 1989.

\bibitem{Konechny:2003yy}
A.~Konechny, \emph{{g function in perturbation theory}},
  \href{https://doi.org/10.1142/S0217751X04019469}{\emph{Int. J. Mod. Phys. A}
  {\bfseries 19} (2004) 2545}
  [\href{https://arxiv.org/abs/hep-th/0310258}{{\ttfamily hep-th/0310258}}].

\bibitem{Opdam:2001wrq}
E.~M. Opdam, \emph{{Multivariable Hypergeometric Functions}},  in \emph{{3rd
  European Congress of Mathematics: Shaping the 21st Century}}, vol.~201 of
  \emph{Progress in Mathematics}, 2001,
  \href{https://doi.org/10.1007/978-3-0348-8268-2_29}{DOI}.

\bibitem{Barut1965}
A.~O. Barut, C.~Fronsdal and M.~A. Salam, \emph{On non-compact groups. ii.
  representations of the 2+1 lorentz group},
  \href{https://doi.org/10.1098/rspa.1965.0195}{\emph{Proceedings of the Royal
  Society of London. Series A. Mathematical and Physical Sciences} {\bfseries
  287} (1965) 532}
  [\href{https://arxiv.org/abs/https://royalsocietypublishing.org/doi/pdf/10.1098/rspa.1965.0195}{{\ttfamily
  https://royalsocietypublishing.org/doi/pdf/10.1098/rspa.1965.0195}}].

\bibitem{Hui:2022vbh}
L.~Hui, A.~Joyce, R.~Penco, L.~Santoni and A.~R. Solomon, \emph{{Near-Zone
  Symmetries of Kerr Black Holes}},
  \href{https://arxiv.org/abs/2203.08832}{{\ttfamily 2203.08832}}.

\bibitem{Hui2021}
L.~Hui, A.~Joyce, R.~Penco, L.~Santoni and A.~R. Solomon, \emph{{Ladder
  Symmetries of Black Holes: Implications for Love Numbers and No-Hair
  Theorems}},  \href{https://arxiv.org/abs/2105.01069}{{\ttfamily 2105.01069}}.

\bibitem{Kim:2012mh}
Y.-W. Kim, Y.~S. Myung and Y.-J. Park, \emph{{Quasinormal modes and hidden
  conformal symmetry in the Reissner-Nordstr\"om black hole}},
  \href{https://doi.org/10.1140/epjc/s10052-013-2440-8}{\emph{Eur. Phys. J. C}
  {\bfseries 73} (2013) 2440}
  [\href{https://arxiv.org/abs/1205.3701}{{\ttfamily 1205.3701}}].

\bibitem{Cardoso:2017qmj}
V.~Cardoso, T.~Houri and M.~Kimura, \emph{{Mass Ladder Operators from Spacetime
  Conformal Symmetry}},
  \href{https://doi.org/10.1103/PhysRevD.96.024044}{\emph{Phys. Rev. D}
  {\bfseries 96} (2017) 024044}
  [\href{https://arxiv.org/abs/1706.07339}{{\ttfamily 1706.07339}}].

\bibitem{Bondi1962}
H.~Bondi, M.~G.~J. Van~der Burg and A.~W.~K. Metzner, \emph{Gravitational waves
  in general relativity, vii. waves from axi-symmetric isolated system},
  \href{https://doi.org/10.1098/rspa.1962.0161}{\emph{Proceedings of the Royal
  Society of London. Series A. Mathematical and Physical Sciences} {\bfseries
  269} (1962) 21}
  [\href{https://arxiv.org/abs/https://royalsocietypublishing.org/doi/pdf/10.1098/rspa.1962.0161}{{\ttfamily
  https://royalsocietypublishing.org/doi/pdf/10.1098/rspa.1962.0161}}].

\bibitem{Mano:1996vt}
S.~Mano, H.~Suzuki and E.~Takasugi, \emph{{Analytic solutions of the Teukolsky
  equation and their low frequency expansions}},
  \href{https://doi.org/10.1143/PTP.95.1079}{\emph{Prog. Theor. Phys.}
  {\bfseries 95} (1996) 1079}
  [\href{https://arxiv.org/abs/gr-qc/9603020}{{\ttfamily gr-qc/9603020}}].

\bibitem{Mano:1996mf}
S.~Mano, H.~Suzuki and E.~Takasugi, \emph{{Analytic solutions of the
  Regge-Wheeler equation and the postMinkowskian expansion}},
  \href{https://doi.org/10.1143/PTP.96.549}{\emph{Prog. Theor. Phys.}
  {\bfseries 96} (1996) 549}
  [\href{https://arxiv.org/abs/gr-qc/9605057}{{\ttfamily gr-qc/9605057}}].

\bibitem{Mano:1996gn}
S.~Mano and E.~Takasugi, \emph{{Analytic solutions of the Teukolsky equation
  and their properties}}, \href{https://doi.org/10.1143/PTP.97.213}{\emph{Prog.
  Theor. Phys.} {\bfseries 97} (1997) 213}
  [\href{https://arxiv.org/abs/gr-qc/9611014}{{\ttfamily gr-qc/9611014}}].

\bibitem{Padmanabhan:2003fx}
T.~Padmanabhan, \emph{{Quasinormal modes: A Simple derivation of the level
  spacing of the frequencies}},
  \href{https://doi.org/10.1088/0264-9381/21/1/L01}{\emph{Class. Quant. Grav.}
  {\bfseries 21} (2004) L1}
  [\href{https://arxiv.org/abs/gr-qc/0310027}{{\ttfamily gr-qc/0310027}}].

\bibitem{Myers:1986un}
R.~C. Myers and M.~J. Perry, \emph{{Black Holes in Higher Dimensional
  Space-Times}},
  \href{https://doi.org/10.1016/0003-4916(86)90186-7}{\emph{Annals Phys.}
  {\bfseries 172} (1986) 304}.

\bibitem{Duff:1993ye}
M.~J. Duff and J.~X. Lu, \emph{{Black and super p-branes in diverse
  dimensions}}, \href{https://doi.org/10.1016/0550-3213(94)90586-X}{\emph{Nucl.
  Phys. B} {\bfseries 416} (1994) 301}
  [\href{https://arxiv.org/abs/hep-th/9306052}{{\ttfamily hep-th/9306052}}].

\bibitem{Myers:1998gt}
R.~C. Myers, \emph{{Black holes in higher curvature gravity}}, pp.~121--136.
\newblock 11, 1998.
\newblock \href{https://arxiv.org/abs/gr-qc/9811042}{{\ttfamily
  gr-qc/9811042}}.
\newblock 10.1007/978-94-017-0934-7{${}_{}$}8.

\bibitem{Myers:1987qx}
R.~C. Myers, \emph{{Superstring Gravity and Black Holes}},
  \href{https://doi.org/10.1016/0550-3213(87)90402-0}{\emph{Nucl. Phys. B}
  {\bfseries 289} (1987) 701}.

\bibitem{Bateman:100233}
H.~Bateman and A.~Erdélyi, \emph{{Higher transcendental functions}},
  California Institute of technology. Bateman Manuscript project. McGraw-Hill,
  New York, NY, 1955.

\bibitem{frolov1998black}
V.~Frolov and I.~Novikov, \emph{Black Hole Physics: Basic Concepts and New
  Developments}, Fundamental Theories of Physics. Springer Netherlands, 1998.

\bibitem{Lowe2012}
D.~A. Lowe and A.~Skanata, \emph{Generalized hidden kerr/{CFT}},
  \href{https://doi.org/10.1088/1751-8113/45/47/475401}{\emph{Journal of
  Physics A: Mathematical and Theoretical} {\bfseries 45} (2012) 475401}.

\bibitem{Shi2018}
C.~Shi, J.-d. Zhang and J.~Mei, \emph{Hidden conformal symmetry for vector
  field on various black hole backgrounds},
  \href{https://doi.org/10.1007/JHEP04(2018)001}{\emph{Journal of High Energy
  Physics} {\bfseries 2018} (2018) 1}.

\end{thebibliography}\endgroup
\end{document}